\documentclass[twoside,twocolumn,9pt]{article}
\usepackage{extsizes}
\usepackage[super,sort&compress,comma]{natbib} 
\usepackage[version=3]{mhchem}
\usepackage[left=1.5cm, right=1.5cm, top=1.785cm, bottom=2.0cm]{geometry}
\usepackage{balance}
\usepackage{mathptmx}
\usepackage{sectsty}
\usepackage{graphicx} 
\usepackage{lastpage}
\usepackage[format=plain,justification=justified,singlelinecheck=false,font={stretch=1.125,small,sf},labelfont=bf,labelsep=space]{caption}
\usepackage{float}
\usepackage{fancyhdr}
\usepackage{xurl}   
\usepackage{fnpos}
\usepackage[english]{babel}
\addto{\captionsenglish}{%
  \renewcommand{\refname}{Notes and references}
}
\usepackage{array}
\usepackage{droidsans}
\usepackage{charter}
\usepackage[T1]{fontenc}
\usepackage[usenames,dvipsnames]{xcolor}
\usepackage{setspace}
\usepackage[compact]{titlesec}
\usepackage{hyperref}
\usepackage{url}
\usepackage{epstopdf}
\usepackage{natbib}

\setcitestyle{super,square}
\definecolor{cream}{RGB}{222,217,201}
\definecolor{cream}{RGB}{222,217,201}
\newcolumntype{C}[1]{>{\centering\arraybackslash}m{#1}}

\definecolor{ruzz}{rgb}{1,0,0.5}
\definecolor{ruz}{rgb}{1,0,0.5}
\definecolor{tmod}{RGB}{0,50,150}
\definecolor{tzel}{RGB}{0,200,200}
\definecolor{zel}{RGB}{0,100,50}
\definecolor{fialova}{rgb}{0.5,0,0.5}
\definecolor{white}{rgb}{1.0,1.0,1.0}
\definecolor{tblue}{RGB}{0,0,130}
\definecolor{seda}{RGB}{235,235,235}
\definecolor{sseda}{RGB}{248,248,248}
\definecolor{azur}{RGB}{212,255,230}
\definecolor{pink}{RGB}{242,220,255}
\definecolor{grey}{RGB}{255,240,240}
\definecolor{lavender}{rgb}{0.9, 0.9, 0.98}

\setcitestyle{citesep={, }}

\hypersetup{
	colorlinks = true,
	linkcolor = ruzz,
	citecolor = blue,
	linkbordercolor = red,
	citebordercolor = green,
	urlcolor= fialova
}

\begin{document}

\pagestyle{fancy}
\thispagestyle{plain}
\fancypagestyle{plain}{
\renewcommand{\headrulewidth}{0pt}
}

\makeFNbottom
\makeatletter
\renewcommand\LARGE{\@setfontsize\LARGE{15pt}{17}}
\renewcommand\Large{\@setfontsize\Large{12pt}{14}}
\renewcommand\large{\@setfontsize\large{10pt}{12}}
\renewcommand\footnotesize{\@setfontsize\footnotesize{7pt}{10}}
\makeatother

\renewcommand{\thefootnote}{\fnsymbol{footnote}}
\renewcommand\footnoterule{\vspace*{1pt}%
\color{cream}\hrule width 3.5in height 0.4pt \color{black}\vspace*{5pt}} 
\setcounter{secnumdepth}{5}

\makeatletter 
\renewcommand\@biblabel[1]{#1}            
\renewcommand\@makefntext[1]%
{\noindent\makebox[0pt][r]{\@thefnmark\,}#1}
\makeatother 
\renewcommand{\figurename}{\small{Fig.}~}
\sectionfont{\sffamily\Large}
\subsectionfont{\normalsize}
\subsubsectionfont{\bf}
\setstretch{1.125} 
\setlength{\skip\footins}{0.8cm}
\setlength{\footnotesep}{0.25cm}
\setlength{\jot}{10pt}
\titlespacing*{\section}{0pt}{4pt}{4pt}
\titlespacing*{\subsection}{0pt}{15pt}{1pt}

\fancyfoot{}
\fancyfoot[LO,RE]{\vspace{-7.1pt}\includegraphics[height=9pt]{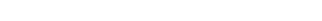}}
\fancyfoot[CO]{\vspace{-7.1pt}\hspace{13.2cm}\includegraphics{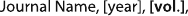}}
\fancyfoot[CE]{\vspace{-7.2pt}\hspace{-14.2cm}\includegraphics{head_foot/RF}}
\fancyfoot[RO]{\footnotesize{\sffamily{1--\pageref{LastPage} ~\textbar  \hspace{2pt}\thepage}}}
\fancyfoot[LE]{\footnotesize{\sffamily{\thepage~\textbar\hspace{3.45cm} 1--\pageref{LastPage}}}}
\fancyhead{}
\renewcommand{\headrulewidth}{0pt} 
\renewcommand{\footrulewidth}{0pt}
\setlength{\arrayrulewidth}{1pt}
\setlength{\columnsep}{6.5mm}
\setlength\bibsep{1pt}

\makeatletter 
\newlength{\figrulesep} 
\setlength{\figrulesep}{0.5\textfloatsep} 

\newcommand{\topfigrule}{\vspace*{-1pt}%
\noindent{\color{cream}\rule[-\figrulesep]{\columnwidth}{1.5pt}} }

\newcommand{\botfigrule}{\vspace*{-2pt}%
\noindent{\color{cream}\rule[\figrulesep]{\columnwidth}{1.5pt}} }

\newcommand{\dblfigrule}{\vspace*{-1pt}%
\noindent{\color{cream}\rule[-\figrulesep]{\textwidth}{1.5pt}} }

\makeatother

\twocolumn[
  \begin{@twocolumnfalse}
{\includegraphics[height=30pt]{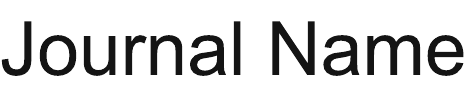}\hfill\raisebox{0pt}[0pt][0pt]{\includegraphics[height=55pt]{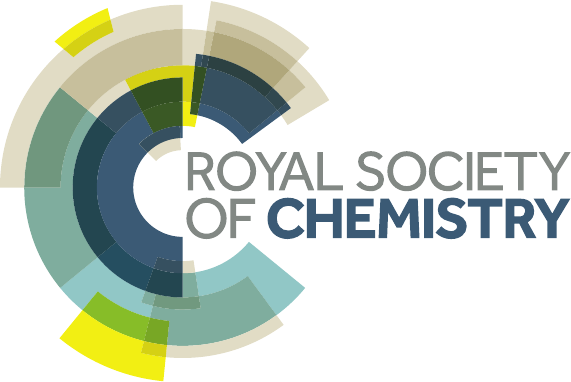}}\\[1ex]
\includegraphics[width=18.5cm]{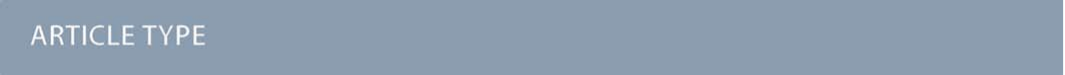}}\par
\vspace{1em}
\sffamily
\begin{tabular}{m{4.5cm} p{13.5cm} }

\includegraphics{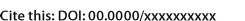} & \noindent\LARGE{\textbf{Stable Thin Clathrate Layers}} \\
\vspace{0.3cm} & \vspace{0.3cm} \\

 & \noindent\large{Eva Posp\' i\v silov\' a$^{a,b,\ast}$} \\



\includegraphics{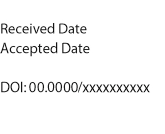} & \noindent\normalsize{We explore
  \emph{elemental} Si, Ge and Sn free-standing slabs in diamond, clathrate and other
  competing structures and evaluate their stability by comparing
  Kohn-Sham total energies at the same atoms/area coverage $\rho$. We
  find that within 3-6 ML range (in units of diamond monolayer), the
  surface energy stabilizes {\textit{clathrate}} thin films against the diamond slabs
  or other competitors lowering internal energy per atom, for all three
  elements. At~1.3-1.6 ML coverages, Si and Ge stable layers are
  adatomic decorations of the puckered honeycomb lattice, forming
  a dumbbell--decorated pattern with rectangular unit cell. Analogical Sn layer is
  unstable to a web-like net of interconnected Sn$_9$ clusters. The
  zoo of~stable thin-layer non-diamond structures is completed by a Sn
  {\textit metallic} bilayer at 2.3 ML coverage.}
\end{tabular}

 \end{@twocolumnfalse} \vspace{0.6cm}

  ]

\renewcommand*\rmdefault{bch}\normalfont\upshape
\rmfamily
\section*{}
\vspace{-1cm}


\footnotetext{\textit{$^{a}$~Institute of Materials and Machine Mechanics, Slovak Academy of Sciences, v.v.i., Dúbravská cesta 9, Bratislava, 84513, Slovakia}}
\footnotetext{\textit{$^{b}$~Institute of Physics, Slovak Academy of Sciences, v.v.i., Dúbravská cesta 11, Bratislava, 84511, Slovakia}}
\footnotetext{\textit{$\ast$~Corresponding author at: Institute of Materials and Machine Mechanics, Slovak Academy of Sciences, v.v.i., Dúbravská cesta 9, Bratislava, 84513, Slovakia. E-mail address: e.pospisilova@savba.sk (E. Pospíšilová).}}





\bigskip

\section{Introduction}
\label{sec:intro}


For bulk Sn, Ge and Si, cubic diamond (Pearson symbol cF8) phase is the ground state structure (GS), lying 23, 27 and 54 meV/at.\cite{pickard2010}, respectively, below its energetically closest sp${}^{3}$ adversary -- type-II clathrate -- according to the basic Density Functional Theory (DFT) setup. The main question of the presented fundamental theoretical DFT study is, whether opening a surface could grant clathrates an~overall energetic advantage over diamond, resulting in \emph{stable} free-standing clathrate thin~films. \\
\indent sp${}^{3}$ slabs exhibit unsaturated -- so called \emph{dangling} -- bonds on~the surface, which can be (partially) satisfied by proper reconstruction. The stability of a thin film, measured by internal energy per atom for the given thickness $\rho \equiv N/S$ (coverage: number of~atoms per area), is thus given by the trade-off between optimal bulk energy and saturation of the surface bonds. The question is, whether the latter effect could significantly favour clathrates. \\
\indent The inquiry is not restricted to clathrates, but spans wide area of known sp${}^{3}$ Column-XIV allotropes and thanks to long Molecular Dynamics (MD) annealings sped up by Machine Learning Force Fields (MLFF) innumerably many other candidates for the stable films as well, including metallic even-layers of Sn. \\
\indent Free-standing thin films minimizing per-atom-energy for the given thickness (coverage) $\rho \equiv N/S$ can be imposed by~structure-neutral substrate such as amorphous or electronically-balanced oxide (\ce{ZrO2}) or graphite wafer. Further stabilization of~the most stable -- \emph{elemental} -- clathrate slabs could be achieved by~lattice-matching substrate, e.g. via \emph{coherent} clathrate-diamond interface\cite{iface}. \\
\indent In the focus here are several-ML (diamond monolayers) thick films, no longer linked to their parent bulk phases by the physical properties. As for the thinnest ones, even interaction between slab's faces comes into play, hence new -- yet unexplored -- features are expected. \\
\indent The paper is organized as follows. In addition to the basic DFT setup applied, the crucial aspects of energy-minimization of~a thin film such as the choice of proper termination, surface reconstruction when cut from the mother bulk phase together with~MD-related approaches starting from random atomic positions or from~the melt are described in~Section \ref{sec:methods}. Section \ref{sec:allo} introducing known parent bulk phases for thin-film production ensues. Section \ref{sec:results} reveals realm of~stability of~clathrate films for~each of~the elements Sn, Ge, Si and compares the associated terminations and reconstructions of~the per-atom-energy minimizing slabs. We go on~to~discuss the possibility of~spontaneous clathrate crystallization from the melt in~Section \ref{sec:discuss} and the impact of van der Waals dispersion interaction in its last Subsection \ref{subsec:vdw}. The most important outputs of our work are summarized in~the final Section \ref{conclusions}. Additional Figures and xyz-files\footnote{xyz-file contains three lattice vectors whose component unit is \AA\, (first three lines), number of atoms (4th line) and fractional coordinates of each atom followed by the atomic number and two times the ordinal number} of \emph{all} stable films are included in~the Supplemental Material (SM) and are referenced as SX.Y from the main text, where the integers X and Y index particular SM Section and Figure, respectively. Our principal results -- internal energy per atom of the most stable films as a~function of the thickness -- are plotted in \autoref{fig:res} and tabulated in the final Section S10 of SM, supplemented with~calculated values of~electronic band gaps of the slabs.

\section{State of the Art}
\label{sec:state-of-the-art}

\indent We shall start this concise review with honeycomb monolayers of Group-IV (G4) elements. Due to the preferred tetrahedral sp${}^{3}$ hybridization of these elements, the only member creating \emph{free-standing} hexagonal monolayers is carbon (graphene)\cite{1atomic_sheets}. Honeycomb monolayers of Pb, Sn, Ge and Si analogous to~graphene -- so called \emph{Xene materials} (X = Si, Ge, Sn and Pb) -- \emph{do} exist, but only on appropriate substrates\cite{epitaxial_stanene}, owing to~the Xene lattice stretching and its strong interactions with the wafer. \\
\indent Notably, high-quality monolayer stanene samples were crystallized as an unusually ultraflat honeycomb lattice on Cu(111) via~Molecular Beam Epitaxy (MBE), fairly chemically inert and stable up to $243\:\mathrm{K}$\cite{epitaxial_stanene}. This observation was indeed unexpected because going down the Group IV sp${}^{3}$ nature of~elements becomes more and more pronounced, reflecting in the buckling, adatomic reconstructions or layer-stacking stabilization mechanisms\cite{stability_groupIV,adatom2,adatom1}. There are also indications\cite{saxena} of free-standing few-layer stanene prepared on liquid hexane by non-equilibrium process of laser ablation, but later in the text (consult \autoref{hexa-square}) we propose another hexagonal polymorph for this structure. \\
\indent Supported (e.g. by Cu(111) binding with $p_{z}$ Sn orbitals, sustaining stanene chemical inactivity) or with suitable chemical functionalization, stanene is regarded as~an ideal platform to~study exotic quantum states of matter; when deposited on~BN, it retains its electronic properties and can act as a quantum spin Hall insulator with~gap as large as $0.3\:\mathrm{eV}$, according to~the theoretical predictions\cite{epitaxial_stanene,db-stanene1,db-stanene3}. Its applications range from~low-power electronics, nanotransistors and high-performance thermoelectrics to~fault-tolerant quantum computation. Extra, superconductivity has been discovered in~few-layer (bilayer and more-layer) stanene atop \ce{PbTe}/\ce{Bi2Te3}/Si(111) sandwich structure\cite{stanene-super}. \\
\indent N.B.: Unlike preceding works\cite{db-stanene1,db-stanene3}, the present paper goes \emph{beyond} stability assessment limited to atomic \& cell relaxations and phonon-frequency evaluation. Our inquiry does not rely on~real phonon frequencies (reported also for dumbbell-stanene); all the free-standing thin films were subjected to long Molecular Dynamics (MD) runs (millions of~fs-steps) at RT and higher ($300-700\:\mathrm{K}$), which led to dumbbell-stanene spontaneous transformation to a web of \ce{Sn9} clusters accompanied by decrease of internal energy per~atom in~the fixed periodic cell, \emph{regardless} of reconstruction type, see \autoref{mono} and \autoref{fig:sn9}. \\
\indent Reconstructed $\sqrt{3}\times\sqrt{3}$ honeycomb monolayer of Ge -- germanene -- was synthesized on the top of $7\times 7$-Au(111) surface\cite{germanene}. Similarly, one-atom-thick Si sheets of $\sqrt{3}\times\sqrt{3}$-reconstructed silicene were deposited on Ag(111)\cite{silicene1}. No experimentally prepared free-standing germanene and silicene are mentioned in the literature. \\ \setcitestyle{square,numbers}
\indent Optimal reconstruction of free-standing germanene was sought after in theoretical (DFT) paper \cite{turkey}, where the so called HDP structure was regarded to be energetically the best. We think that their ``measure of energetics of the phases'' -- maximizing cohesive energy per unit area -- \emph{is not} the relevant thin-film stability criterion and we suggest in~the present work other way to compare the stability of layers: minimization of per-atom energy, $E_{\mathrm{atom}}$, for the given coverage $\rho \equiv N/S$. Wherefore we recalculated all the samples and plotted the results on the corresponding $E = E(\rho)$ graphs. \\ \setcitestyle{super,square}
\indent Energy-minimizing (unreconstructed) bilayer silicene and germanene were theoretically predicted to be composed of~\emph{exactly planar} silicene monolayers in eclipsed (e.g. lonsdaleite) conformation\cite{ge-bilayer1,si-bilayer2,si-bilayer1}. These bilayers are represented by green pentagons marked as ``\texttt{GS-bilayer}'' in \autoref{fig:res} summarizing our chief results. DFT predictions were confirmed by STM experimental images of bilayer Si on~Ag(111)\cite{sibila-ex1,sibila-ex2}. For both bilayer (few-layer) silicene and germanene, the stacking order of~monolayers can induce metal-semiconductor phase transition and can regulate size of the electronic band gap\cite{si-bilayer2,si-bilayer1,ge-bilayer}. Variety of stacking patterns -- majority of which are not available to the bilayer graphene -- are due to the inherent buckling of~monolayer Si and Ge\cite{ge-bilayer1}. \\
\indent We have not investigated impact of stacking sequence on~bilayer stability in the present paper. The scope of this work is limited to $2\times2$, $\sqrt{3}\times\sqrt{3}$, $c(2\times4)$ and $c(2\times8)$ reconstructions of~bilayer (111) bulk cubic diamond cuts with all possible relative placements of adatoms on two mutually interacting slab's faces. \\
\indent Comprehensive theoretical (DFT) survey of atomically thin free-standing Sn films was performed by P. Borlido et al.\cite{SnI}, who discovered the ground state of bilayer Sn. This square metallic -- still \emph{hypothetical} -- structure is supposed to be prepared experimentally on e.g. Ag(100), Au(100), Pt(100), or Al(100) substrates, lying just $149\:\mathrm{meV/at.}$ above~the bulk ground state $\alpha$-Sn (cF8, cubic diamond). \\ \setcitestyle{square,numbers}
\indent Surface energy of various (un)reconstructed planar terminations of \emph{bulk} cubic diamond (cF8) phase of Sn, Ge and Si were discussed in \cite{dia-web,stekolnikov2005,stekolnikov2002,Si-cF8,Ge-cF8,Ge-cF8-2,Sn-cF8}.\setcitestyle{super,square} It was shown by~numerous experimental and theoretical groups that (111)-diamond section of~Sn, Ge and Si is a \emph{cleavage face} possessing the lowest surface energy $\gamma$ among~all the cubic-diamond (Pearson cF8) sections. Note that this statement is valid just after proper surface reconstruction\cite{dia-web,stekolnikov2002}. $5\times5$ Dimer-Adatom-Stacking Fault (DAS), $7\times7$ DAS and $\sqrt{3}\times\sqrt{3}$, $2\times2$, $c(2\times4)$, $c(2\times8)$ adatomic reconstructions were observed on Si and Ge surfaces\cite{zhachuk2}. \\
\indent It was revealed recently by means of DFT that despite $5\times 5$ DAS being the ground-state Si(111) bulk-surface reconstruction, $7\times 7$ high-temperature DAS reconstruction -- most frequently observed in experiments -- freezes in a metastable state below RT upon~cooling\cite{dia-sect,zhachuk3}. As for Ge(111) relaxed bulk surface $c(2\times8)$ adatomic reconstruction is experimentally the most frequent one\cite{stekolnikov2002, zhachuk2}. \\
\indent R. Zhachuk et al. constructed ``strain phase diagrams'': DFT-computed surface energy $\gamma$ of~various reconstructions of~Ge(111) and Si(111)-diamond surfaces as a function of biaxial strain in~the range $\varepsilon \in (-4;\,4)\%$\cite{zhachuk2, zhachuk1}, explaining surface structure transformations during growth of strained Ge(111), Si(111) layers (imposed by the substrate). The calculations were compared with scanning tunneling microscopy data obtained on~stepped Si(111) surfaces grown on Ge(111) substrate and on Ge islands grown on a Si(111)\cite{zhachuk3,zhachuk2,zhachuk1}. \\
\indent It turns out that DAS reconstructions thanks to the presence of~\emph{all} reconstruction elements (each individually) are intrinsically compressive ($\varepsilon < 0$), while purely adatomic reconstructions gain dominance under biaxial tensile strain ($\varepsilon > 0$). Namely Si(111) thin films deposited on Ge(111) substrate display $c(2\times4)$ adatomic surface reconstruction above $\varepsilon \approx 2.5$\% biaxial tensile strain\cite{zhachuk2,zhachuk1}. Local minima of the surface energy for~particular orientations are trade-off between saturation of the dangling bonds and minimization of the surface stress, inherently imposed by the reconstruction units (adatoms, dimers, stacking faults, steps, $\pi$-bonded chains) introducing unusual bond lengths and angles\cite{dia-sect}. \\

\indent There have been a few experiments leading to atomically-thin clathrate layers, deposited on lattice-matching substrates. Our most recent work reports several-atomic-layer thick Sn decagonal clathrate film deposited on quasicrystalline $d$-Al-Ni-Co wafer\cite{Vipin}. The preceding work has shown the versatility of Sn to form quasicrystalline atomic arrangements, growing icosahedral Sn clathrate nanolayer on $i$-Al-Pd-Mn\cite{Morske_dno}. \\
\indent Group of T. Kume deposited atomically-thin Ge type-II clathrate on Ge(111)-diamond surface. The authors hypothesized the existence of thin buffer layer at the interface\cite{oxides,Kume}, but we have managed to construct \emph{coherent} interface between the two\cite{iface}. This kind of coherent clathrate/diamond interface had been formerly displayed in several computational investigations exploiting clever empirical force fields\cite{coloids,Molinero,molinero-mW}, but had never been described in detail (in terms of periodicity, unit cell, atomic bonds), probably owing to quite a large lattice misfit ($\sim11$\%) between diamond and clathrate part of the same G4 element restricting the size of the coherent patches to relatively small islands.\setcitestyle{square,numbers} Our paper \cite{iface} was thus devoted to minimization of that misfit by optimization of materials on~both sides.\setcitestyle{super,square} \\
\indent However, the resulting Sn and Ge configurations in the aforementioned experiments were to large extent governed by the substrate. No experiment states preparation of free-standing clathrate layer on structure-neutral substrate, either liquid, amorphous or electronically-balanced oxide termination. Therefore in the present work, motivated by these studies, the most stable clathrate layers are sought after, \emph{regardless} of substrate. One of~the questions is, whether a combination of square and triangle clathrate tiles (bearing e.g. quasicrystalline symmetry) provides clathrate films with some energetic advantage over e.g. purely triangular (hexagonal) clathrate tilings. \\
\indent Thicker clathrate layers ($\sim \mu\mathrm{m}$) can be used in opto-electronics as p-n junctions, Schottky barriers, LEDs and photocells in the visible spectrum, optical and X-ray detectors, thanks to the tunable electronic band gap in the range $E_{\mathrm{g}}\in[0.6;\,2.25]\:\mathrm{eV}$ by alloying and doping (Na${}_{x}$Si${}_{136}$, Si${}_{34-x}$Ge${}_{x}$, I${}_{8}$Si${}_{46}$, Xe${}_{8}$Si${}_{46}$, Na${}_{x\leq 24}$(Si${}_{1-y}$Ge${}_{y}$)${}_{136}$)\cite{gap2,gap3,gap1,US-patent}. This, however does not apply to the range of coverages investigated in the present work ($\rho \in [1;\,10]\:\mathrm{ML}$, \autoref{fig:res}), since the width of the electronic band gap decreases with~the decreasing thickness of the film. Note that all the DFT-computed (slightly underestimated) band gaps of~our clathrate slabs are just small fractions of $1\:\mathrm{eV}$, mostly below $E_{\mathrm{g}} = 0.5\:\mathrm{eV}$ for all three elements, consult Subsection S10 of SM (outside the absorption rage of visible light, $E_{\mathrm{g}} > 1.7\:\mathrm{eV} \cong$ red light). The band gaps of thin clathrate layers are smaller than that of the corresponding DFT-computed bulk clath-II phase, see Figures~S8.1 and S8.2 (SM). Hence, their possible applications are as yet \emph{unknown}. \\
\indent Besides opto-electronic applications of clathrates, they are also famous thermoelectrics and superconductors, owing to the cage doping, but this paper is aimed solely at \emph{elemental} clathrate thin films, much harder to be experimentally prepared (especially Sn), their structure and symmetry and its impact on slab's stability. \\


\section{Methods \& Definitions}
\label{sec:methods}

\subsection{Density Functional Theory (DFT)}
\label{dft}

\indent For atomic relaxations as well as for Molecular Dynamics (MD) simulations and on-the-fly Machine Learning Force Fields (MLFF) Vienna \emph{ab initio} Simulation Package (VASP)\cite{DFT0} implementation of~the plane-wave Density Functional Theory was employed powered by the Projector Augmented-Wave potentials (PAW)\cite{DFT1} for~core-electron treatment in~Perdew-Wang \texttt{PW91} Generalized Gradient Approximation (GGA)\cite{DFT2} to the exchange-correlation functional describing electron-electron interaction. \\
\indent As a default setup, we applied VASP ``Accurate'' setting (specifying \texttt{ENCUT}, see below, the grids for Fast Fourier Tranform, and the accuracy of the projectors in real space) and default values of energy cutoff for the plane-wave basis set in eV used for~electronic wave-function description, \texttt{ENCUT} (Sn: $103.3\:\mathrm{eV}$, Ge: $173.8\:\mathrm{eV}$, Si: $245.4\:\mathrm{eV}$). Energy was converged on~K-point mesh density (\texttt{KPOINTS}) determining Bloch vectors for~the Brillouin zone sampling to $1\:\mathrm{meV/atom}$ precision. Simulated films infinite in~two dimensions were separated by at~least 15\AA\, vacuum layers in the third dimension to prevent mutual interactions of~the periodic copies. s\textsuperscript{2}p\textsuperscript{2} electrons were treated as valence for each element. All sites in the unit cell along with the unit cell dimensions were relaxed using a conjugate gradient algorithm to minimize energy with an atomic force tolerance of $0.005\:\mathrm{eV/}$\AA\, and a total energy precision of $10^{-5}\:\mathrm{eV}$. \\
\indent Electronic band gaps stated in Subsection S10 of SM were evaluated using the same GGA setup and converged within $0.05\:\mathrm{eV}$-wide window with respect to \texttt{KPOINTS} and smearing width given by parameter \texttt{SIGMA}, which was successively lowered from $0.2\:\mathrm{eV}$ up to $0\:\mathrm{eV}$ while increasing the K-point mesh density. Note that basic PAW GGA setup is well-known to systematically underestimate a bang-gap width. \\
\indent It has been demonstrated\cite{SnI} that despite the spin-orbit coupling \emph{does} have a measurable effect on the formation energy, the energy shift is nearly constant for all 2D Sn structures ($55-59\:\mathrm{meV/at.}$) and as such can be neglected. Another DFT extension, van~der~Waals dispersion interaction (PBE-D3 correction to~DFT), is discussed in Subsection \ref{subsec:vdw}.

\subsection{Ab Initio Molecular Dynamics (AIMD)}
\label{aimd}
\hypertarget{aimd}{}


\indent By ab initio Molecular Dynamics (AIMD) we mean long MD runs -- (tens of) millions of $\sim\mathrm{fs}$-steps -- markedly accelerated by Machine Learning Force Fields (MLFF) as implemented in VASP. \\
\indent AIMD was used within the present work on three occasions. Firstly, AIMD is the simplest way to amorphous bulk and thin films, described in Subsection \ref{amor}, avoiding ingenious construction. Unfortunately, there is no such an automated ab initio way to~clathrates, see Section~\ref{sec:discuss}, where an attempt has been made to recrystallize clathrate slabs from the melt, either in a free-standing slab form or as a Ge slab on an appropriate InN substrate. \\
\indent The third instance was the thin-film configurational-space search. Despite the clathrate-slab notation suggested in Table~\ref{noty}, Subsubsection \ref{clath-termi}, several-atom-thick layers have actually no relation \emph{to any} bulk phase: they are unique identities themselves. Long AIMD runs thus attest to the completeness of~our thinnest-film enumeration, mainly specific Sn metallic even-layers rediscovered this way in Subsection \ref{hexa-square}. AIMD confirms that there are not any other low-energy thin-film atomic arrangements. \\
\indent Namely, for the Sn, Ge and Si thinnest-film production in~the range of coverages $\rho \in [0.85;\,4.0]\:\mathrm{ML}$ with the basic step $\Delta\rho \approx 0.5\:\mathrm{ML}$, $4\times10^{5}$ steps $5\:\mathrm{fs}$ each in the temperature range $T \in [300;\,500]\:\mathrm{K}$ were performed, leading to \emph{no} new phases. As concerns Sn, due to the close competition of square and triangle even-layers in this temperature range and owing to the plethora of Sn phases, Subsection \ref{hexa-square}, we have gradually increased coverage from $2\:\mathrm{ML}$ to $2.5\:\mathrm{ML}$ via random single-atom addition to~the ($23.9$\AA)${}^{3}$ supercell, witnessing a coexistence of the two phases at~approximately $2.29\:\mathrm{ML}$ (for $137-138$ atoms in the cell). \\
\indent Extra, we have executed several Replica Exchange (RE) simulations 20-50 thousand $5\:\mathrm{fs}$-step long starting from random atomic configurations with all three elements, in the temperature range from room up to the respective melting point, again, with \emph{no} new unexpected results. These were either simulations with~fixed boundary conditions (coverage) or complemented by energy cost function scaling either linearly with surface (surface energy) or quadratically with the departure of a \hyperlink{coverage}{coverage} from~its initial value, emulating through the stiffness constant some kind of~elastic penalty for the sample thickening, which we would like to~artificially suppress. \\
\indent RE is a particular example of a delicate trade-off between precision and computational cost, since several simulations are running at~once, calling for as moderate \texttt{KPOINTS}, \texttt{ENCUT} and \texttt{POTIM}, as possible. In the Sn case, we ended with (3 3 1) \texttt{KPOINTS}, while Si and Ge demanded at least 3 specially chosen \texttt{KPOINTS} and even those may not be enough, since Si slabs melted below $1030\:\mathrm{K}$ at~default value of \texttt{ENCUT} and $5\:\mathrm{fs}$ time-step. It is indeed surprising that even single Ba atom addition to our several-dozen-atom periodic cells of free-standing thin-film samples, does not lead to~a~\emph{solo} clathrate cage. We hope that the present work stimulates further research (with enhanced values of all the parameters) towards this goal. \\
\indent N.B.: All samples obtained by means of AIMD, MLFF and RE were finally DFT-relaxed and only energies after DFT relaxation are stated in this text. Hence accuracy of MLFF enters the results -- \autoref{fig:res} -- in \emph{no way}.

\subsection{Atomic Structure, Termination, Thickness, Reconstruction: Definitions}

\subsubsection{Atomic Structure.}

\indent Vast majority of the slabs were carved out from the known bulk phases, described in Section \ref{sec:allo}, by planar sections. The second way to Column-XIV thin films led through the \textit{ab initio} Molecular Dynamics introduced in Subsection \ref{aimd}. Except for~Sn -- a borderline case between metal and semiconductor -- the work deals exclusively with sp${}^{3}$ allotropes where all atoms exhibit more or less tetrahedral environment with~quartet of nearest neighbours forming average bond angle around $109.5^{\circ}$. \\
\indent All the known high-pressure allotropes more than $50\:\mathrm{meV/at.}$ above the respective cubic diamond (cF8) phase, such as BC8, R8, ST12, BT8\cite{haberl}, mC16, bct-5\cite{martonak3}, obtained chiefly by (de)compression, were avoided [see Table S1.1, SM, for phases that were taken into~account], since the main goal is to predict stable slabs with~\emph{lower} internal energy per atom compared to the corresponding diamond slabs of the same thickness. The only exception will be the amorphous phases, due to the intention to crystallize stable clathrate slabs from the melt and owing to their inherent indifference to~the structure of a substrate, in contradistinction to~diamond or clathrates. \\
\indent Extra, atomic configurations directly related to diamond, such as hexagonal diamond (lonsdaleite), so called diamond polytypes, were also excluded from the competition, as \emph{brand new} materials bearing novel properties are in the sights.

\subsubsection{Termination.}

\indent From now on, termination is a particular \emph{orientation} and \emph{position} of~a cutting plane with respect to a bulk crystal, producing planar slab of constant thickness:

\subsubsection{Thickness.}

of a slab is measured by the \emph{coverage} $\rho$ defined by:

\begin{equation}
	\rho \equiv N/S
\end{equation}
\hypertarget{coverage}{}
\vspace{-0.25cm}

\noindent where $N$ stands for the number of atoms per surface area $S$ of~a~periodic cell. Individual cases of planar terminations suitable for~each bulk phase are listed in~the next Section \ref{sec:allo}. \\
\indent For the sake of simple element comparison, coverage $\rho$ will be measured in dimensionless ``monolayer'' (ML) unit, one ML corresponding to $\rho$ value of a single (111) layer of cubic $\alpha$-phase (cF8, diamond) of either Si, Ge or Sn.

\subsubsection{Reconstruction.}

\indent Hereafter, by \emph{reconstruction} we mean addition of atoms of the same kind, \hypertarget{adatom}{\emph{adatoms}}, (i.e. Sn atoms in the Sn-slab case) to the film's surface followed by DFT relaxation of both the atomic positions and lattice parameters. We did not attempt any hydrogenation, whereas ``bare'' surfaces after cell and atomic relaxation will be called anyway ``unreconstructed''. \\
\indent Particular instances of surface reconstruction shall be specified for~each parent bulk phase and for every termination thereof in~the next Section \ref{sec:allo}.

\subsection{Slab Stability Assessment}
\label{assess}
\vspace{0.15cm}                 

\indent Due to the intrinsic positivity of the \hyperlink{gamm}{surface energy} $\gamma$ there is a general trend of decreasing per-atom energy with slab thickness. At~the same time, flat slabs occur at discrete, characteristic values of~{\em coverage} $\rho$ and energy per atom. Hence, we seek after energy-minimizing slab structures for a given \hyperlink{coverage}{coverage} $\rho$, corresponding to certain slab thickness. \\ 
\indent Let's consider two directly constructed points $E(\rho)$, e.g. a couple of (111)-diamond slabs with distinct thicknesses $\rho_{1} \equiv N_{1}/S_{1}$, $\rho_{2} \equiv N_{2}/S_{2}$ with energies per atom $E_{1}$, $E_{2}$, respectively, which may differ in type of surface reconstruction, hence in the value of~\hyperlink{gamm}{surface energy} $\gamma_{1} \neq \gamma_{2}$. An \emph{interpolation} between the points has a meaning of a \emph{terrace} of energy $E$ such that $E_{1} > E > E_{2}$ and coverage $\rho$ such that $\rho_{1} < \rho < \rho_{2}$ between the two slabs, whose contribution to the overall energy-per-atom of the whole system tends to zero for periodic boundary conditions (PBC) tending to~infinity. Because energy, surface and number of atoms per periodic cell are additive quantities satisfying\footnote{Mind that $p_{i} \neq S_{i}/S$, since a \emph{terrace} (step) between slabs of \emph{different} thicknesses is considered here.}:

\vspace{-0.1cm}

\begin{equation}
p_{i} \equiv N_{i}/N,\,\,i\in\{1,\,2\}: \qquad p_{1} + p_{2} = 1
\end{equation}

\vspace{-0.5cm}

\begin{equation}
S = S_{1} + S_{2} = N_{1}\left(\frac{S_{1}}{N_{1}}\right) + N_{2}\left(\frac{S_{2}}{N_{2}}\right), \quad \framebox{$\displaystyle E = p_{1}E_{1} + p_{2}E_{2}$}
\end{equation}

\vspace{0.1cm}

\noindent the \emph{linear} interpolation $E_{\mathrm{atom}}(\rho^{-1}) = E(\rho^{-1})$ follows:

\vspace{-0.25cm}

\begin{equation}
	\frac{S}{N} = p_{1}\left(\frac{S_{1}}{N_{1}}\right) + p_{2}\left(\frac{S_{2}}{N_{2}}\right) \, \Leftrightarrow \, \framebox{$\displaystyle (\rho)^{-1} = p_{1}(\rho_{1})^{-1} + p_{2}(\rho_{2})^{-1}$}
	\label{ave2}
\end{equation}

\vspace{0.1cm}

\noindent corresponding to this particular interpolation on the graphs $E_{\mathrm{atom}}(\rho)$:

\vspace{-0.35cm}

\begin{equation}
\framebox{$\displaystyle E = E_{2} + \frac{\rho_{2} - \rho}{\rho_{2} - \rho_{1}}\,\frac{\rho_{1}}{\rho}(E_{1} - E_{2})$}
\label{rca-erho}        
\end{equation}

\vspace{0.25cm}                             

\noindent The aim of this sort of graphs is to compare thin-film energy per~atom, $E_{\mathrm{atom}}$, of~different atomic configurations, terminations and reconstructions for the common intermediate coverage $\rho$, not associated with any planar slab structure already measured. \\ 
\indent In the simplest case of a constant surface energy $\gamma$, hyperbolic trends (\ref{rca-erho}) of the curves $E_{\mathrm{atom}}(\rho)$ for each structural family of~slabs, cf. Figure \ref{fig:res}, can be derived directly from the definition of the \emph{surface energy}:
   
\vspace{-0.15cm}                 
\hypertarget{gamm}{}                              
\begin{equation}
\gamma \stackrel{\mathrm{def.}}{=} \frac{E_{\mathrm{cell}} - E_{\mathrm{bulk}}N}{2S} \quad \Rightarrow \quad E_{\mathrm{atom}}(\rho) = \frac{E_{\mathrm{cell}}}{N} = 2\,\frac{\gamma}{\rho} +  E_{\mathrm{bulk}}
\label{rho-na-minus-prvu}
\end{equation}

\vspace{0.0cm}

\noindent where $E_{\mathrm{atom}}$ is energy per slab's atom, $E_{\mathrm{cell}}$ is overall energy per periodic cell with the total of $N$ atoms, $E_{\mathrm{bulk}}$ is energy per one atom of the respective bulk phase and $\rho$ is thin-film \hyperlink{coverage}{coverage} (possessing two faces, hence the factor ``2'' in \autoref{rho-na-minus-prvu}). \\
\indent In~the case of~\emph{layered} growth mode with \emph{constant} surface energy $\gamma$ of a thin film and/or in the close vicinity of the scattered points in {Figure} \ref{fig:res}, the hyperbolic dependence (\ref{rho-na-minus-prvu}) describes slabs with (non-)integer number of layers. Each point of~the hyperbolic convex hull can be thus assigned an (in)complete slab; incompleteness meaning a slab with a \emph{terrace}. \\
\indent The free-standing slab setup with PBC (periodic cell extended vertically in the $z$ direction to~include vacuum gap between periodic images of a slab) introduces specific constraints: for~systems with lattice parameters up to~few tens of \AA, it effectively hampers formation of non-planar bulky morphologies (e.g. nanowires, nanoparticles). Importantly, all the slabs in~current simulations are only \emph{metastable} -- stabilized by artificial effect of finite PBC -- whose increase would progressively lead to~spontaneous transformation to nanowires and nanoparticles, respectively, owing to the $\gamma > 0$ of G4 sp${}^{3}$ elements. \\
\indent While this phenomenon actually helps studying slabs as templates for adlayers that might form on~an appropriate substrate by wetting it, we have to keep in mind that our simulations do not fairly assess the tendencies to possible island formation\cite{Vipin}, once the material is deposited on~a substrate. Questions related to \emph{dewetting} are actually highly relevant for sp${}^{3}$-type adlayers\cite{soi}, e.g. silicon on~an insulator. \\


\section{Competing allotropes}
\label{sec:allo}

\subsection{Monolayers}
\label{mono}

\begin{figure}[h!]
	\centering
		\includegraphics[width = 9.0cm]{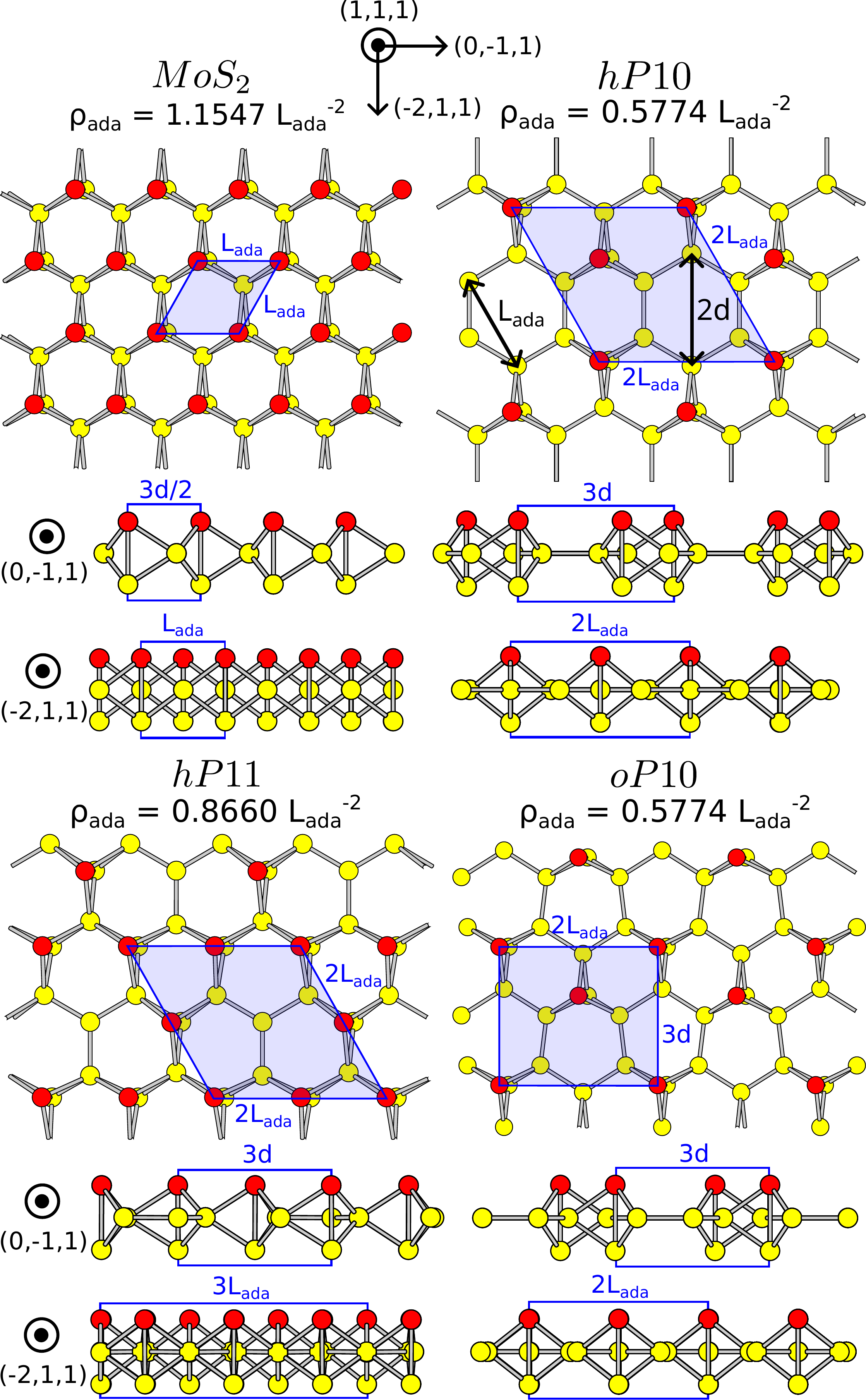}
		\caption{Four reconstructions -- \ce{MoS2}, hP10, hP11, oP10 -- of (111)-diamond Sn, Ge and Si monolayers: top -- front view; middle, bottom -- side views. Solid blue line delimits periodic cell excluding vacuum. Red balls designate adatoms. $\rho_{\mathrm{ada}}$ under Pearson symbol stands for the area-density of~adatoms expressed through~typical length unit $L_{\mathrm{ada}}$: the distance between nearest neighbours on a sublattice of~a~bipartite (111)-diamond lattice; for bulk diamond structure $L_{\mathrm{ada}}(\mathrm{Sn}) = 4.69\:$\AA, $L_{\mathrm{ada}}(\mathrm{Ge}) = 4.08\:$\AA, $L_{\mathrm{ada}}(\mathrm{Si}) = 3.86\:$\AA. Indicated unit cell sizes in universal units of $d$ and $L_{\mathrm{ada}} \approx \sqrt{3}d$ are just for better orientation in various cross sections; here we assume \emph{idealized} planar projections made of regular hexagons with edge length $d$ (i.e. we neglect lattice corrugations and distortions due to the chemical bonds, e.g. with red adatoms).}
		\label{fig:dia111_leporelo}
\end{figure}
\begin{figure}[h!]
	\centering
	\includegraphics[width = 9.0cm]{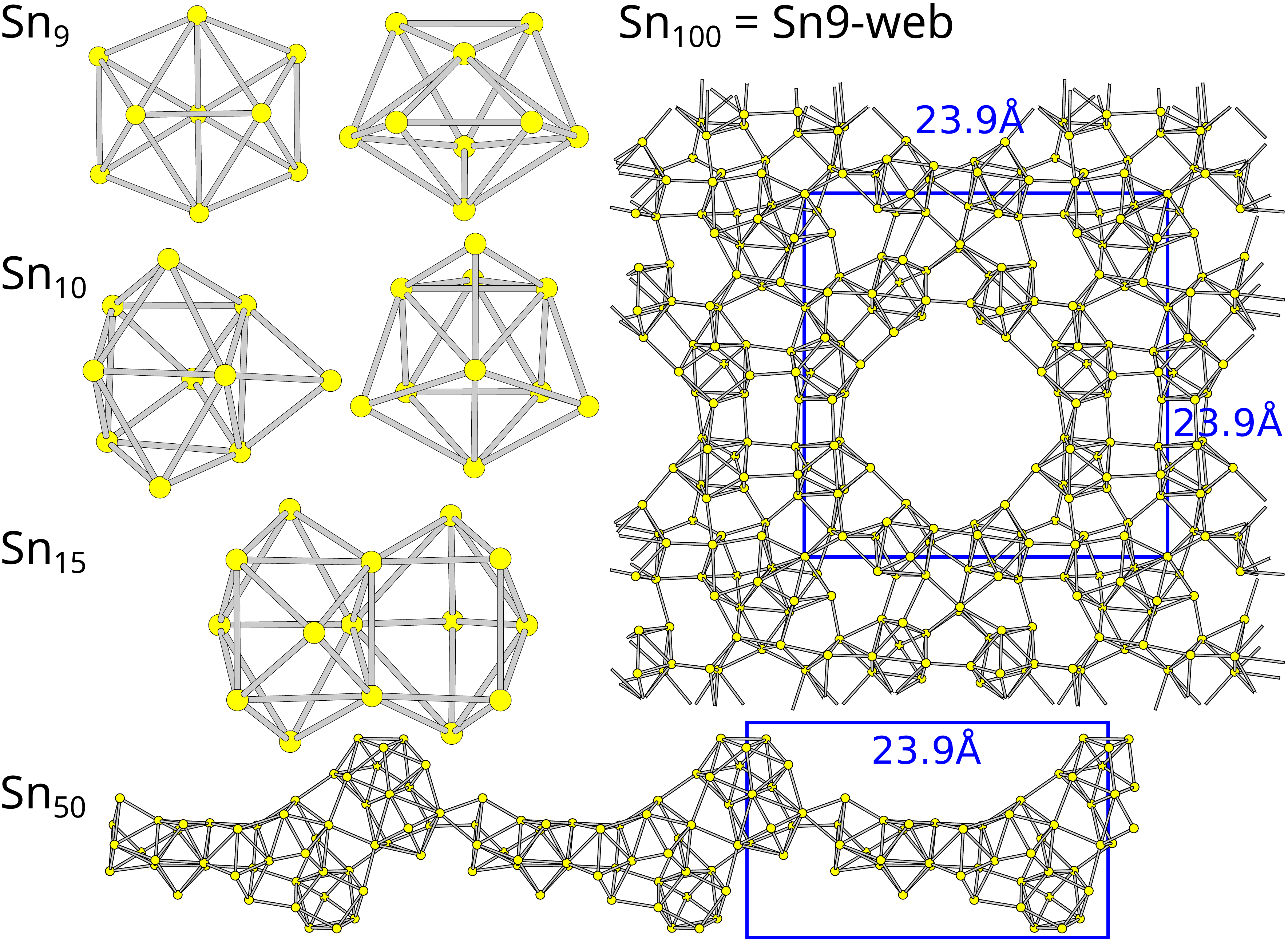}
	\caption{Sn${}_{9}$ and Sn${}_{10}$ clusters (from two different view angles) are fundamental units of threadlike tin structures that form spontaneously for~low coverages $\rho$ (discontinuous layers with holes)\cite{china_particle1, china_particle2}. Sn${}_{15}$ subunit also frequent in Sn clusters comprises two Sn${}_{9}$ basic units sharing a common triangle. Sn${}_{50}$ and Sn${}_{100}$ are examples of our long DFT annealings at~room temperatures with 50 and 100 Sn atoms, respectively, enclosed in~fixed $24\times24\times24\:$ \AA${}^{3}$ cubic container. Sn${}_{100}$ is composed of 8 Sn${}_{9}$ clusters in~a ring. Blue solid borderline delimits periodic motif.}
	\label{fig:sn9}
\end{figure}

\indent Stabilizing reconstruction evidently depends on the layer thickness. Singular point are monolayers, displaying just one atomic layer to reconstruct. However, there are infinitely many ways to~do so. \\
\indent As for Sn, Ge and Si, clear choice\cite{db-stanene1, db-stanene2, turkey} are various reconstructions of (111)-diamond monolayer, so called stanene, germanene and silicene, see Figure \ref{fig:dia111_leporelo}. Because the basic reconstruction unit is \emph{dumbbell}, we will call them hereafter db-monolayers\cite{turkey}. \\
\indent Even though each dumbbell constructed on germanene (silicene) lowers per-atom-energy, MoS${}_{2}$ configuration, corresponding to~the full coverage, has branches with imaginary frequencies in~\textit{ab-initio} phonon spectra indicating mechanical instability\cite{turkey}. The ground state of (Sn, Ge, Si) db-monolayer is therefore non-trivial and requires systematic investigation. \\ 
\indent For \emph{every} admissible dumbbell density per unit area in the given supercell -- from~minimal zero stanene density up to the maximal MoS${}_{2}$ dumbbell density -- \emph{all} dumbbell configurations were tested for the resulting internal energy per atom and plotted as a function of~\hyperlink{coverage}{coverage} $\rho$, see Subsection \ref{assess}. We have chosen $2\times2$ oP10 supercell with~originally 32 stanene atoms (without~adatoms). Adatoms were placed just on one sublattice of the bipartite stanene lattice followed by the systematic elimination of~the symmetrically equivalent adatomic configurations. \\
\indent For both germanene and silicene, oP10 minimizes internal energy per slab's atom, cf. Figure \ref{fig:res}. On the contrary, all the candidate db-monolayers, succumbed in the course of \hyperlink{aimd}{AIMD} simulations, starting from random atomic positions and from hP10 db-stanene as well, to a web of~interconnected Sn$_9$ clusters, Sn9-web, (Figure \ref{fig:sn9}) in the Sn case for the given \hyperlink{coverage}{coverage}, Figure~\ref{fig:res}, accompanied by an internal energy \emph{decrease}. This is our new theoretical prediction, which has not been corroborated by experiment yet. For example, hP10 db-stanene ($\rho = 1.35\:\mathrm{ML}$), 160 atoms per periodic cell, transformed into a web of Sn$_9$ clusters at relatively high temperatures between $559-600\:\mathrm{K}$, lowering the internal energy by $30\:\mathrm{meV/at.}$ in the fixed cell. \\
\indent As it shall be explicitly illustrated by an example of bilayers, Subsection~\ref{hexa-square}, Sn due to its metallic temper favours 6-coordinated atoms resulting in genuine G4 bond-angles equal to~$60^{\circ}$ (Figure~\ref{asi}, typical of a-Ge and a-Si surface\cite{hara2005}) and $109.5^{\circ}$ dominant in a Sn9-web to purely sp${}^{3}$-like coordination of 4, accompanied by ``factitious'' bond angles $54^{\circ}$, $66^{\circ}$, $72^{\circ}$, $129^{\circ}$, enforced by~the arrangements depicted in~Figure~\ref{fig:dia111_leporelo}.


\subsection{Diamond}

\medskip

\indent As a reference for elemental Sn, Ge and Si diamond slabs (no mention of~carbon in this work!), we have chosen (111)-terminated ones. It has been shown by~numerous experimental and theoretical groups that (111)-diamond section of Sn\cite{Sn-cF8}, Ge and Si\cite{stekolnikov2002, stekolnikov2005, Si-cF8, Ge-cF8-2, Ge-cF8, dia-web} is a \emph{cleavage face} possessing the lowest \hyperlink{gamm}{surface energy} $\gamma$ among all the cubic-diamond (cF8) bulk sections. Note that this statement is valid just after proper \emph{bulk} surface reconstruction\cite{stekolnikov2002, dia-web}. \\
\indent Several reconstructions of (111)-diamond surface have been compared: $\sqrt{3}\times\sqrt{3}$, $2\times2$, $c(2\times4)$, $c(2\times8)$\cite{zhachuk1, zhachuk2}, $3\times3$ DAS, $7\times7$ DAS\cite{zhachuk3}, but the variation of the per-atom energies turned out to~be negligible with respect to the clathrate-diamond differences above $3\:\mathrm{ML}$ slab's thickness, cf. Figure \ref{fig:res}. Mind that below, say, $10\:\mathrm{ML}$ surface states or surface atoms cannot be distinguished since practically all atoms belong to the surface\cite{stekolnikov2002}, so the notion of~the \hyperlink{gamm}{surface energy} $\gamma$ is not relevant here. 

\subsection{Clathrates}
\label{clath}

\subsubsection{Clathrate Terminations.}
\label{clath-termi}

\indent We have investigated all planar sections of clathrate types I-IV, leading to equilateral-/isosceles-triangle, square and rectangle in-plane tilings and combinations thereof, Table \ref{noty}. To every such orientation of a section plane with respect to a bulk clathrate there are four distinct terminations\cite{Vipin}, e.g.:

\vspace{-0.0cm}

\begin{equation}
	T_{h} \to T_{c}^{\prime} \to T_{h}^{\prime} \to T_{c}
\label{period}
\end{equation}

\vspace{0.15cm}

\noindent schematically depicts the 4-layer Triangular-tiling period in the ``$z$'' direction perpendicular to the section plane (shared by both isosceles and equilateral triangle, likewise for the square and rectangle tiles) of puckered terminal planes with characteristic ``holes'' -- $T_{h}$, $T_{h}^{\prime}$  -- alternating with flat terminal planes displaying ``caps'' around the tile vertices: $T_{c}^{\prime}$, $T_{c}$, consult Figures S2.1 and S3.1 (SM) for decagonal (iso-tri, rectangle) and dodecagonal tiles (equi-tri, square), respectively. \\
\indent Note that the only exception from the rule (\ref{period}) comprises trio of~(111)-terminations of type-II clathrate where $T_{c}$ planes are missing. These very missing flat planes of clathrate-III 6-caps, Figure S2.1 in~SM, act as a bridge between type-II and type-III clathrates, representing the only difference between the two. \\
\indent Explicit clathrate construction, i.e. atomic decorations of~clathrate tiles mentioned above, was achieved via symmetry-preserving dual transformation\cite{FK1, FK2} of the respective Frank-Kasper (FK) phases. Equally, orientation of section planes was inspired by the well-known FK atomic planes, linked by~dual transformation to the planes of clathrate-cage centers. It turns out that these are the section orientations yielding planar clathrate slabs containing at least 3-coordinated surface atoms, completely without terraces or cages sticking out from~the surface. \\
\indent This, however, is not the whole story about sensible clathrate terminations, since each clathrate cage itself represents an object with~at~least 3-coordinated surface atoms, even when entirely isolated. Adding 5-caps or 6-caps onto puckered $T_{h}$ and $T_{h}^{\prime}$ planes does not change the surface density of unsaturated (so called dangling) bonds. Therefore, we expect that the growth mode of~a~clathrate will be either pure island or mixed island-layer type, depending on the energy scales related to the processes of (1) completion of a surface cage and (2) coalescing cages together\cite{Vipin}.

{{\small
		\begin{table}[h!]
			\centering
			\renewcommand{\arraystretch}{1.3}
			\begin{tabular}[h!]{ l  l  l  l  l }
				\noalign{\hrule height 1.6pt}
				      type      &   Pearson  &   sect.  &   tile                 &  notation        \\
				\hline
				\hline
				      I         &   cP46     &   (100)  &   square               &  Si.square-I-N   \\
				\hline
			          II        &   cF136    &   (110)  &   iso-tri              &  Si.iso-II-N     \\  
			    \hline
			          II        &   cF136    &   (111)  &   equi-tri             &  Si.equi-II-N    \\            
				\hline 
				      III       &   hP40     &   (001)  &   equi-tri             &  Si.equi-III-N   \\
				\hline
				      III       &   hP40     &   (100)  &   rectangle            &  Si.rect-III-N   \\
				\hline
				      IV        &   tP172    &   (001)  &   equi-tri + square    &  Si.dodeca-IV-N  \\ 
				\hline
				      --        &   oP148    &   --     &   iso-tri + rect.      &  Si.deca-N       \\          
				\hline
				\noalign{\hrule height 0.7pt}
			\end{tabular}
			\caption{\label{noty} {\small{\textit{designation of an $N$-monolayer-thick planar slab cut from the parent bulk clathrate, for which Pearson symbol is stated, by planar section with the respective clathrate tile information. Decagonal R2T4 approximant\cite{Vipin} (oP148) does not belong to any canonical clathrate I-IV type}}}}
			\vspace{0.0cm}
\end{table}}}

\subsubsection{Clathrate Nomenclature.}
\label{clath-name}

\indent We shall designate clathrate slabs according to the element, the associated clathrate tile corresponding to particular orientation of a section plane with respect to a bulk clathrate phase and the thickness measured by number of (111)-diamond monolayers $N$, Table \ref{noty}. For the sake of concise and uncluttered notation, unless otherwise stated, the respective $N$ will be rounded to the nearest integer value except for half-integers and optimal surface terminations with energy-minimizing reconstruction, detailed in Section \ref{sec:results}, for the given \hyperlink{coverage}{coverage} $\rho$ are meant.

\medskip

\subsubsection{Clathrate Reconstruction.}
\label{clath-reco}

\indent Here we describe, for the first time, the optimal surface-energy minimizing reconstructions of planar clathrate sections listed in~Table~\ref{noty} for each of the four terminations from Equation (\ref{period}). \\
\indent The basic reconstruction element is again \emph{dumbbell}, but unlike (111)-diamond surface termination, surface adatom S is located \emph{beneath} some group of 3-coordinated surface atoms, whose dangling bonds it should saturate. \\ 
\indent There are three basic positions\cite{Vipin} of a clathrate-surface adatom S (Figures S2.1, S3.1 in SM):

\begin{enumerate}
	\item \underline{Under cage faces aka ``caps'':}
	\begin{itemize}
		\item[1a.] Under 5-gonal cage facets: $T_{c}$, $R_{c}$, $T_{c}^{\prime}$, $R_{c}^{\prime}$.
		\item[1b.] Under 6-gonal cage facets: $T_{c}$, $T_{c}^{\prime}$, $S_{c}$.
	\end{itemize}
	\item \underline{Dumbbell under the 3-coordinated atom} with:
	\begin{itemize}
		\item[2a.] all neighbours 4-coordinated: $T_{c}^{\prime}$, $R_{c}^{\prime}$, $R_{c}$, $S_{c}$.
		\item[2b.] two neighbours 3-coordinated: $R_{h}$, $R_{h}^{\prime}$, $S_{h}$.
		\item[2c.] three neighbours 3-coordinated: $T_{h}^{\prime}$.
	\end{itemize}
	\item \underline{Midedge position:} under the bond of 3-coordinated atoms: $T_{h}$, $R_{h}$, $S_{h}$. \hypertarget{point3}{}
\end{enumerate}

\subsection{Tin-Specific Phases}
\label{hexa-square}

\setcitestyle{square,numbers}
\begin{figure}[h!]
	\centering
	\includegraphics[width = 7.5cm]{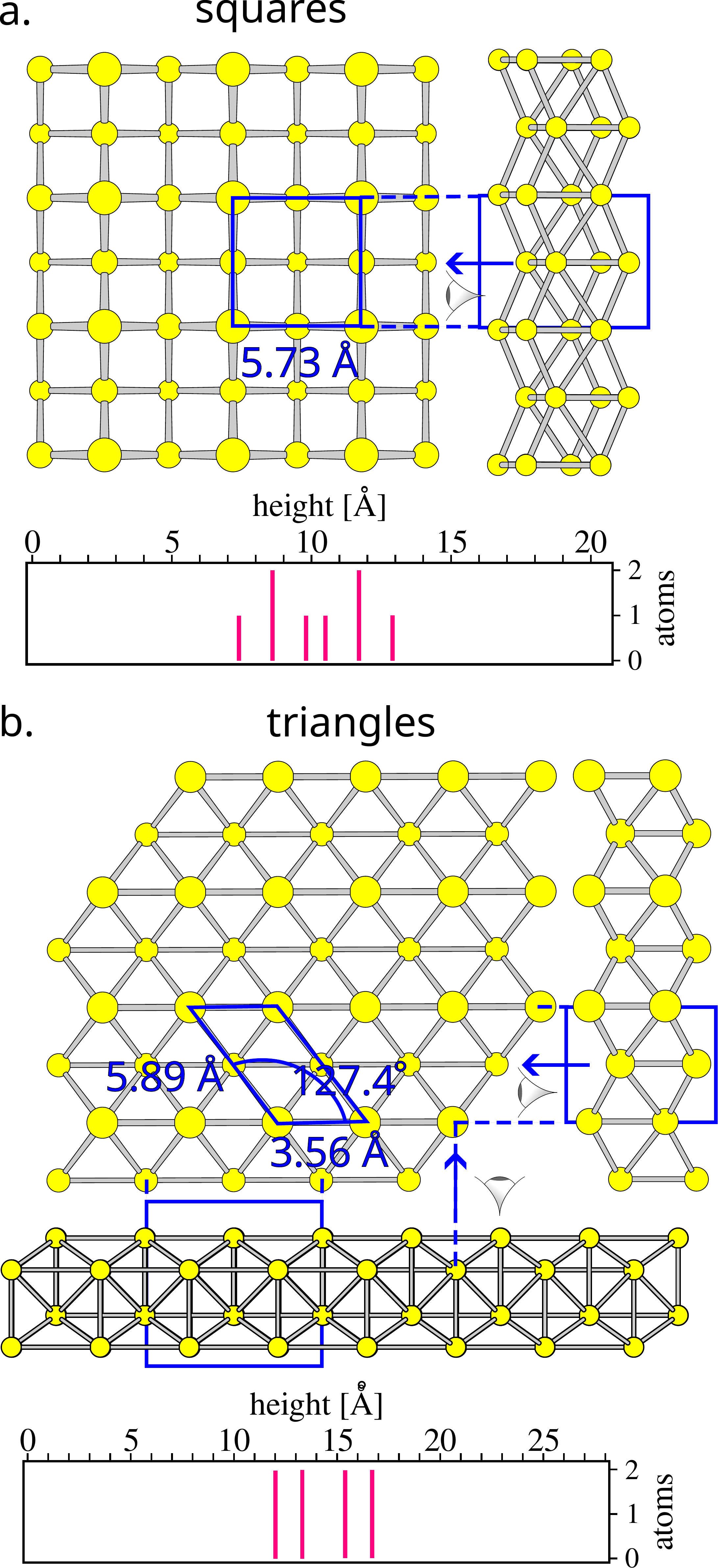}
	\caption{Two Sn thin-film \emph{metallic} phases: square and hexagonal. \underline{Upper panel a.:} Square Sn bilayer: front view and side view, firstly published in \cite{SnI}. \underline{Lower panel b.:} Hexagonal Sn bilayer: front and side view, yet unpublished, discovered via long MD DFT room-temperature runs in~our work. \underline{Bottom panel:} histograms of number of atoms per unit cell (delimited by blue line, excluding vacuum) per~given $z$ coordinate perpendicular to the bilayer plane. Both phases have the front atoms exactly above the bottom atoms of a bilayer in the $z$ direction. Direction of view is indicated by~arrows.}
	\label{fig:hexa}
\end{figure}
\setcitestyle{super}

\indent Unlike Ge and Si, tin owing to its position at the bottom of~Column-XIV lies on the borderline between semiconductors (C, Ge, Si) and metals (Pb), clearly exhibiting metallic properties at~ambient conditions. At~the level of (111)-diamond bilayer ($\rho\approx2$) square and hexagonal metallic Sn films are competing with one another, Figure \ref{fig:hexa}. The couple of metallic phases are designated as ``squares'' and ``triangles'' in result Figure \ref{fig:res}. \\
\indent Whereas square Sn-I phase constitutes the ground state of~Sn bilayer\cite{SnI}, hexagonal counterpart emerges during long MD or Replica Exchange (RE) annealings at room temperature. These observations suggest that hexagonal phase is favoured by (vibrational) entropy, corroborated by numerous defects of~hexagonal polymorphs compared to the relatively perfect square one. All the metallic atomically-thin Sn films showed strong tendency towards the formation of even-layers. \\
\indent Interestingly, both types of metallic bilayers can be viewed as two planar sections of $\gamma$-Sn phase\cite{gama_tin2, Kubiak} in mutually perpendicular directions after symmetry breaking (slight displacement of a~couple of atoms). Besides Sn9-web, Figure \ref{fig:sn9}, hexagonal Sn even-layers, are the second discovery of our DFT RE runs. \\
\indent Despite being presented as stanene 4-layer, in the central Figure~1a\cite{saxena} presumably is depicted triangular Sn 4-layer with~hexagon centers \emph{filled} with Sn atoms, experimentally prepared by non-equilibrium process: pulsed laser ablation of $\beta$-Sn target floating on liquid hexane.

\subsection{Amorphous Phases}
\label{amor}

\subsubsection{Amorphous Tin.}

\indent As opposed to Ge and Si, in the case of bulk Sn there is no such a~local minimum of the potential energy landscape deserving the title \emph{amorphous}. More precisely, the aforementioned minima - even more than one mechanically stable structure - can be readily visited in the DFT universe; the real obstacle, however, resides in~their experimental accessibility. \\
\indent So as to observe a bulk-like amorphous Sn layer by quenched condensation on~a cold substrate, it requires at least 8 atomic \% of some other metal, e.g. Cu, and even then the resulting alloy crystallizes above approximately $20\:\mathrm{K}$\cite{a-Sn1, a-Sn3, a-Sn2}. Another method -- quenched simultaneous condensation of Sn and H on a substrate on nitrogen temperature -- produces bulk-like amorphous semiconducting Sn, stable up to $180\:\mathrm{K}$\cite{a-Sn-H}. The other procedures include for example ion implantation of radioactive Sn isotopes\cite{a-Sn-radio}, or intense electrolytic chemical reaction able to extract selected elements from the common precursor, such as sulphur from SnS${}_{2}$ crystal\cite{a-Sn-chemical}. \\
\indent Possible realization through decompression of high-pressure intermediate obtained by compression of type-II clathrate\cite{martonak1} or $\alpha$-tin\cite{martonak3, popova} seems impracticable for tin, compared to Si and Ge, since analogically DFT-generated metallic Sn disordered liquid ($90\:\mathrm{meV/at.}$ above $\alpha$-Sn) possesses by $10\:\mathrm{meV/at.}$ \emph{lower} internal energy than sp${}^{3}$-amorphous Sn gained by a-Ge rescaling to the typical Sn-Sn distances, Table S1.1, SM. This metallic liquid then readily transforms to various metallic \emph{crystalline} phases of Sn. The phase space of~Sn is much more complicated than that of Si and Ge, and may well hide yet unexplored and even unexpected allotropes\cite{a-Sn-mart}, though.

\medskip

\subsubsection{Amorphous Silicon.}
\label{aasi}

\begin{figure}[h!]
	\centering
	\includegraphics[width = 9.0cm]{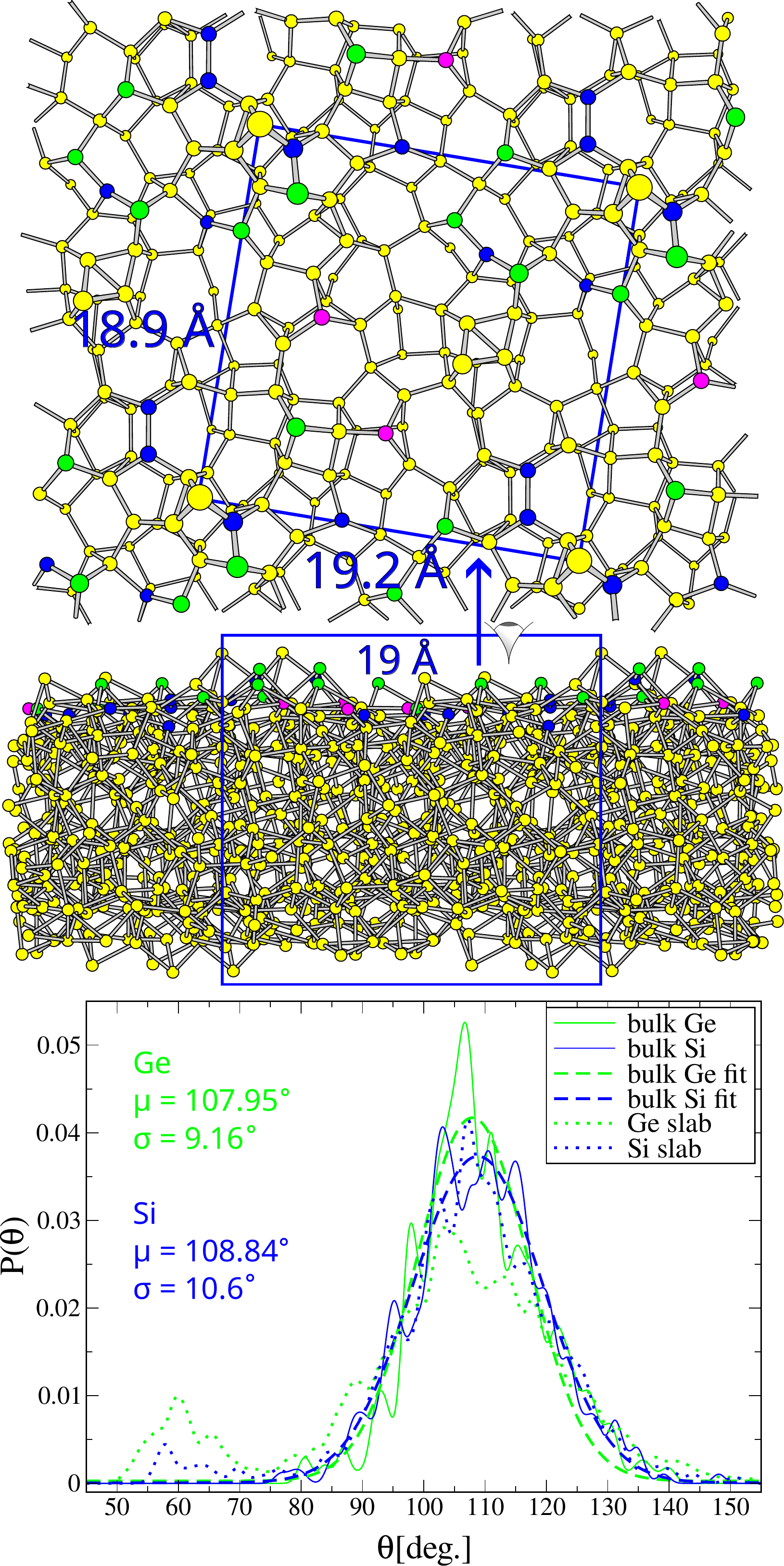}
	\caption{\underline{Top:} Typical surface of a-Si: most of the atoms are 4-coordinated yellow balls, blue balls are sp${}^{2}$-bonded atoms, green balls are p${}^{3}$-bonded and magenta atoms possess two bond angles $\approx100^{\circ}$ and one $\approx120^{\circ}$. There are two atoms with even sharper pyramidal bonding geometry of~$\theta_{\mathrm{ave}} < 80^{\circ}$. Periodic cell without vacuum is delimited by blue line. Size of a ball is proportional to~its distance from the observer. \underline{Middle:} Side view of the corresponding a-Si $4.3\:\mathrm{ML}$-thick slab. On the top panel, just $7$\AA-thick layer from the surface is depicted. \underline{Bottom:} Bond-angle distributions of~the best bulk a-Si and a-Ge without defects, produced by~DFT on-the-fly trained VASP MLFF, fitted by gaussians, compared with thick slabs of~the same material.}
	\label{asi}
\end{figure}

\indent Surprisingly, bulk a-Si can be prepared completely faultless without any coordination defects showing perfect gaussian bond-angle distribution\cite{biswas2004}, Figure \ref{asi}, around (almost ideal) tetrahedral bond-angle $108.6^{\circ}$ with the standard deviation from the narrow interval $9-11^{\circ}$\cite{fortner1989, strubbe2019, pedersen2017, goodnick2019, stich1991, choudhary2005}. This was also the case of~our samples generated using VASP MLFF suite trained on-the-fly during ab initio heating from $300\:\mathrm{K}$ to $900\:\mathrm{K}$ of the quenched melt ($10^4$ $5\:\mathrm{fs}$-steps at $3000\:\mathrm{K}$) followed by~annealing at $900\:\mathrm{K}$\cite{li2020, shodja2014, quench-rate} ($2\times10^{5}$ MLFF $5\:\mathrm{fs}$-steps) and slow cooling ($3\times10^{6}$ MLFF $5\:\mathrm{fs}$-steps) back to~room temperature within~the standard DFT setup, Subsection \ref{dft}. The density of thus prepared bulk a-Si, Figure \ref{asi}, was $98.1\%$ of~$\alpha$-Si (cF8) value, its internal energy being $158\:\mathrm{meV/at.}$ above~Si.cF8, Table~S1.1 (SM), falling into the \hypertarget{bi}{interval} $0.09-0.17\:\mathrm{eV/at.}$ stated in~the literature\cite{pedersen2017, pedersen2020, furukawa2017, custer1994}. \\
\indent The effect of thorough ab initio annealing and relaxation of a-Si \emph{surface} has been validated in solo paper\cite{hara2005}. It has been proven that DFT treatment leads to~the significant relaxation, namely to the p${}^{3}$ and sp${}^{2}$ \hypertarget{hybb}{hybridizations} of 3-coordinated surface atoms, with~average bond angles $\theta_{\mathrm{ave}}$ of $99.4^{\circ}$ and $119.1^{\circ}$, respectively, quite distinct from $108^{\circ}$ $\theta_{\mathrm{ave}}$ of 3-fold bulk atoms. This effect is sizable, since 3-fold atoms comprise $\sim 22\%$ of~the surface; the rest consisting of $\sim 74\%$ sp${}^{3}$- and $4\%$ 5-coordination. Besides~characteristic hybridizations, another intrinsic surface feature are \hypertarget{3ring}{3-member rings} accompanied by small $50^{\circ}-60^{\circ}$ bond-angle peak, not found in the bulk a-Si distribution\cite{hara2005}, approved by our DFT results, Figure \ref{asi}. \\
\indent The a-Si surface energy is supposed to be $\gamma = 1.05\pm0.14\:\mathrm{J/m^{2}}$\cite{hara2005}, or $1.34\:\mathrm{J/m^{2}}$\cite{shodja2014}. The latter paper\cite{shodja2014} states somewhat large $\gamma$ -- even larger than quoted experimental values $1.14\:\mathrm{J/m^{2}}$\cite{gamma-111-1}, $1.23\:\mathrm{J/m^{2}}$\cite{gamma-111-2}, $1.24\:\mathrm{J/m^{2}}$\cite{gamma-111-3}, of ideal (111) diamond termination -- but \emph{without}~proper annealing after~a~tearing test. ($1\:\mathrm{erg/cm^{2}} = \mathrm{mJ/m^{2}}$ and $1\:\mathrm{eV/}$\AA${}^{2} = 16.02176634\:\mathrm{J/m^{2}}$) According to Materials Project\cite{dia-web}, the surface energy of reconstructed (111) Si.cF8 surface is $1.30\:\mathrm{J/m^{2}}$. \\
\indent Our DFT computed (111) Si.cF8 unreconstructed surface has energy $1.57\:\mathrm{J/m^{2}}$, its $c(2\times4)$ reconstruction\cite{zhachuk1} has surface energy $\gamma = 1.27\:\mathrm{J/m^{2}}$, in good agreement with literature. Our a-Si slabs with~coverages ranging from $1.75-4.98\:\mathrm{ML}$ (for~unit conversion: $1\:\mathrm{ML} = 0.15469\:\mathrm{at./}$\AA${}^{2}$ is the coverage of one Si diamond monolayer, $1\:\mathrm{ML}$) possess energies from the interval $0.77-0.87\:\mathrm{J/m^{2}}$ or $48-54\:\mathrm{meV/}$\AA${}^{2}$: as expected lower than (111)-cF8 termination; $\gamma$ of \emph{partially} crystallized (to cF8) a-Si samples attacking lower bound. Mind that \emph{computed} value of a-Si $\gamma$ strongly depends on the \emph{bulk} a-Si energy via~Equation \ref{rho-na-minus-prvu} ($E_{\mathrm{bulk}}$ term), which, according to~the literature, is from the relatively \hyperlink{bi}{broad interval}, see above. For further comparison, surface energy of~the best reconstructed Si clath-II slabs, Figure \ref{fig:restruct}, is $0.865\:\mathrm{J/m^{2}}$, on~the a-Si $\gamma$ \emph{upper} limit.

\medskip

\subsubsection{Amorphous Germanium.}
\label{aage}

\begin{figure}[h!]
	\centering
	\includegraphics[width = 9.0cm]{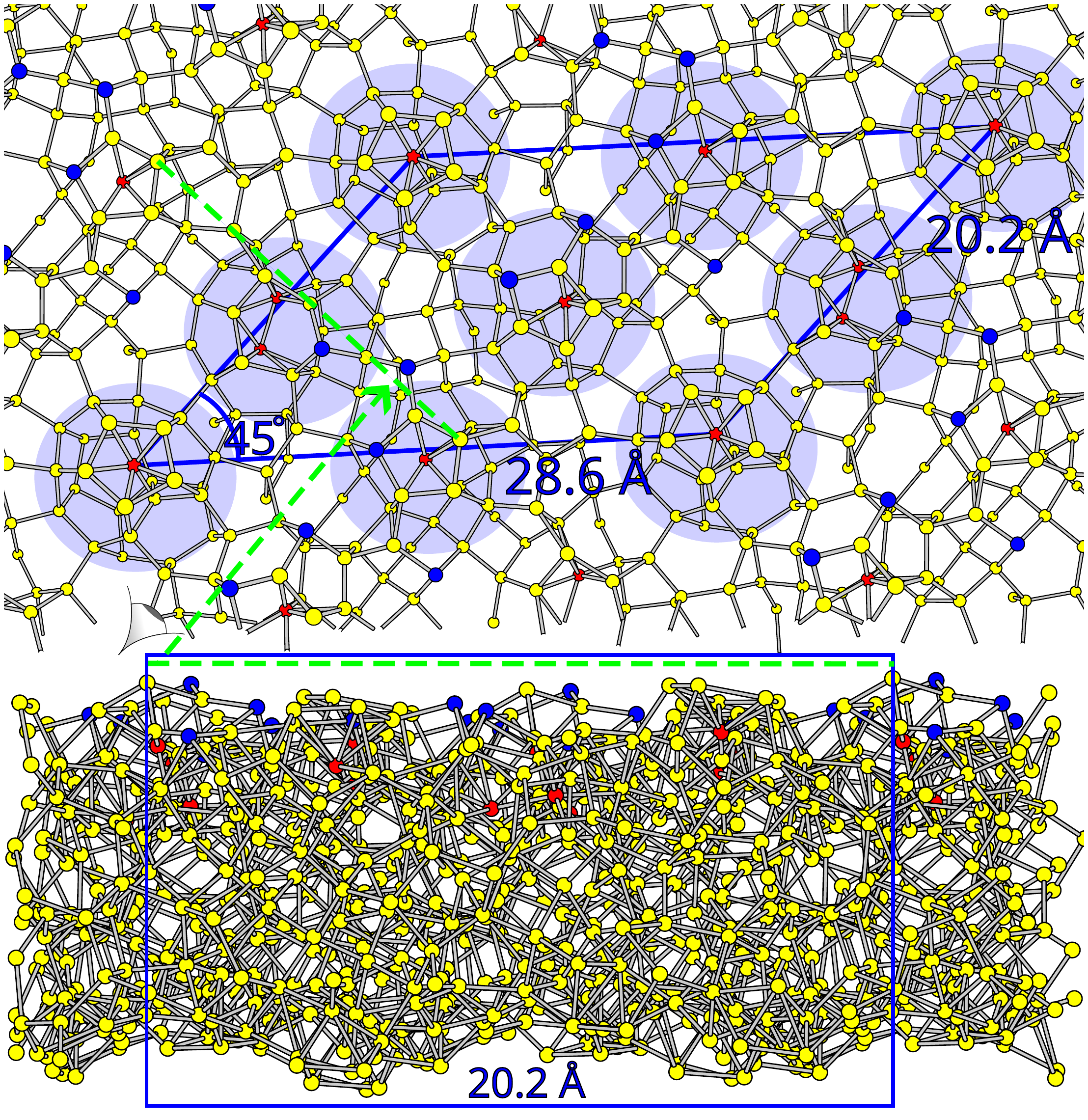}
	\caption{\underline{Top:} Typical a-Ge surface obtained by VASP on-the-fly MLFF from the melt by~repeated annealing at $646\:\mathrm{K}$ and slow quench to $300\:\mathrm{K}$. Periodic cell without vacuum is delimited by blue line. Most of the surface atoms are (approx. sp${}^{3}$-) 4-coordinated: yellow balls. Red balls: over-coordinated adatoms. Blue balls: dangling bonds, mainly associated with 5- and 6-caps. \underline{Bottom:} $4.2\:\mathrm{ML}$-thick slab, whose surface is plotted above. On the top panel, just $8$\AA-thick layer from the surface is depicted.}
	\label{age}
\end{figure}

\indent Despite structural properties of bulk a-Ge being the same as those of a-Si, cf. Figure~\ref{asi}, and despite bulk a-Ge constructed directly via a-Si rescaling to the characteristic Ge-Ge distances succeeded by~atomic relaxation, we have observed for the first time that a-Ge \emph{surface} features are completely different, Figure~\ref{age}. Note that elemental Ge thin films were in fashion in 1970s and 1980s, when DFT was still in its infancy, hence all we know about~structural units of a-Ge (thin film) surface comes from~DFT simulations of~our own. \\
\indent The density of bulk a-Ge comprises 96.68\% of the Ge.cF8 value and is by $128\:\mathrm{meV/at.}$ unstable at $0\:\mathrm{K}$ to diamond, Table S1.1, SM. a-Ge slabs were prepared by long annealing slightly below $T_{\mathrm{c}} = 650\:\mathrm{K}$, just as bulk a-Ge. \\
\indent The surface energy of a-Ge slabs with characteristic 5-caps and 6-caps on the surface, Figure \ref{age}, lies in the interval $(0.45-0.55)\:\mathrm{J/m^{2}}$ or equivalently $(28-34)\:\mathrm{meV/}$\AA${}^{2}$, approximately one half of unreconstructed (111) diamond, $1.07\:\mathrm{J/m^{2}}$ (Materials Project states the value $1.11\:\mathrm{J/m^{2}}$\cite{dia-web}). These a-Ge clathrate-surface elements manifest themselves in~the bond-angle distribution as a strong peak around~$60^{\circ}$, Figure \ref{asi}, even stronger than that of~a-Si surface of different origin. \\
\indent For further comparison, surface energy of (111) diamond can be lowered to $0.865\:\mathrm{J/m^{2}}$ by~$\sqrt{3}\times\sqrt{3}$ reconstruction, which is still much higher than a-Ge values. Surface energy of~the best clathrate reconstructions, 5-caps and 6-caps with one adatom, consult Figures S2.1 and S3.1 (SM) and Figure \ref{fig:restruct}, are $0.734\:\mathrm{J/m^{2}}$ and $0.559\:\mathrm{J/m^{2}}$, respectively, again attacking upper threshold of~a-Ge range, like in the a-Si case (but substantially lower than $\gamma$ of~reconstructed Ge diamond surface). 

\subsection{Chiral Framework Structures}
\label{cfs}

\medskip

\indent To the best of our knowledge, the only sp${}^{3}$ G4 allotropes possessing \emph{lower} internal energy per~atom than type-II clathrate \emph{not} structurally related to diamond are Chiral Framework Structures, CFS, or spirals: CFS-5\cite{pickard2015}, CFS-6\cite{pickard2010} or NGS\cite{conesa}, T12\cite{T12}, see also Table S1.1, SM. \\
\indent All three structures are fully sp${}^{3}$-hybridized, displaying main peak of the bond-angle distribution around $109.5^{\circ}$ and do not contain any imaginary frequences supporting dynamical stability of these allotropes at ambient pressure\cite{pickard2015, T12}. \\
\indent Lattice Gas Monte Carlo (LGMC) algorithm was deployed to~seek after the best planar terminations for thin films. An~attractive toy potential was applied that favored two CN=3 surface atoms over a couple of CN=2, CN=4 atoms: thus simulations tend to~produce surfaces similar to the unreconstructed (111)-diamond (or clathrate). Simulation itself was a multiply repeated cooling while recording the lowest-energy configurations (by our own program). \\
\indent LGMC research of CFS-6 ended in two surface terminations normal to (120) and (001) directions [(001) is the axis of the main spiral with~hexagonal cross section], subjected to subsequent reconstructions for several Si, Ge and Sn film thicknesses. \\
\indent For both sections two surface reconstructions were attempted: \hyperlink{adatom}{adatoms} were firstly placed below each bond joining a couple of unsaturated atoms -- (001)-1, (120)-1 -- and secondly to the valley between them -- (001)-2, (120)-2. As concerns thicknesses above $3.5\:\mathrm{ML}$ (Si), relaxations have just effect on the slab's \emph{surface}; the interior remains spiral-like and does not change its phase upon relaxation\cite{diss}. \\
\indent Atomic and cell relaxations left (120) (un)reconstructed surface intact, while surface of~(001) termination has changed substantially and manifested~3- and 4-coordinated atoms only. As for (001)-1 and (001)-2 reconstructions, adatoms move towards the spiral's main axis, saturating as many dangling bonds of the surrounding spiral as possible. \\
\indent As expected from the already probed clathrate-surface reconstructions, Subsubsection \ref{clath-reco}, the only successful attempt was (120)-1 for Ge and Sn, resembling midedge reconstruction of~clath-II T${}_{h}$ tiles, cf. Figure S2.1, S3.1 (SM), which was forbidden for Si, Figure \ref{fig:res}. Si CFS-6, which is energetically degenerated with clath-II bulk, can not be cut and reconstructed \emph{effectively}, therefore only unreconstructed Si.CFS surfaces were included in~Figure~\ref{fig:res} summarizing the chief results. \\
\indent (120) termination is clearly better than (001) owing to the smaller area-density of~the coordination defects (dangling bonds) and the absence of 2-coordinated atoms, but both are worse compared to the best clath-II planar sections, Figure \ref{fig:restruct}. Clathrates exhibit lower per-atom energy and lower \hyperlink{gamm}{surface energy} $\gamma$ for all three elements and thicknesses under question. \\
\indent As for CFS-5, there are two equivalent sections in directions (100) and (010), with just 3- and 4-coordinated atoms. In this case, 3-fold atoms are arranged in 4-atom chains per periodic box, exhibiting highest area-density of coordination defects among all CFS sections already mentioned. Several solo-adatom reconstructions were attempted, none of~them being successful at lowering energy-per-atom or even surface energy $\gamma$, so compositions of several such adatoms seem to be also useless (absent in~Figure~\ref{fig:res}). \\
\indent Finally, natural planar termination of T12 normal to (001) direction with exclusively 3- and 4-coordinated surface atoms is the same as (120) cut of CFS-6, since both structures are stackings of~equal 3-atom thick layers, having almost the same bulk energy per atom, wherefore the reconstructions of T12 were not executed. \\


\section{Results}
\label{sec:results}

\indent Our principal results are \hyperlink{coverage}{coverage} ranges for which clathrate slabs are the ground state with $\Delta E(\rho)\equiv E_{\mathrm{clath}}(\rho)-E_{\mathrm{ref}}(\rho)<0$; where $E_{\mathrm{ref}}(\rho)$ is energy of the closest competing structure, usually diamond. They are summarized in Table \ref{tab:sum}, Figure \ref{fig:res} (see also Figures~S4.1, S4.2 and S4.3 in SM) and the surface reconstruction of per-atom-energy minimizing clathrate slabs corresponding to~$\rho_{\mathrm{max}}$, where $\Delta E$ is maximal, are depicted in~Figure~\ref{fig:restruct}. xyz-files of all stable thin films except for diamond are given in~Section~S9 (SM). Finally, in~Subsection~\ref{ele-res} a possible mechanism of~electronic stabilization of~clathrate thin films with~respect to~diamond ones is advocated.

\begin{table}
	\begin{tabular}{cccccc}
		element&  clath-II    &$\rho_0$& $\rho_{\mathrm{max}}$ & $\rho_1$ & $\Delta E_{\mathrm{max}}$   \\
		&  surface     & \multicolumn{3}{c}{ML} & meV/atom   \\
		\hline
		Sn     &    (111)     & 3.2  &   3.2  &   10.0    &  -35   \\
		Ge     &  (111)\&(110)& 2.5  &   3.4  &    7.4    &  -34   \\
		Si     &    (110)     & 2.5  &   3.3  &    5.3    &  -44   \\  
	\end{tabular}
\caption{\label{tab:sum}{\small{\textit{
	coverage intervals at which the ground-state slab is a clathrate (always type-II): columns $\rho_0$ and $\rho_1$ are endpoints of the clathrate stability intervals expressed in (111)-diamond-monolayer units (ML). $\Delta E_{\mathrm{max}}$ is the ``maximum energy gain'' relative to competitors at $\rho_{\mathrm{max}}$
}}}}
\end{table}

\subsection{Tin}
\label{Sn-res}

\begin{figure*}[h!]
	\centering
	\includegraphics[width = 18.0cm]{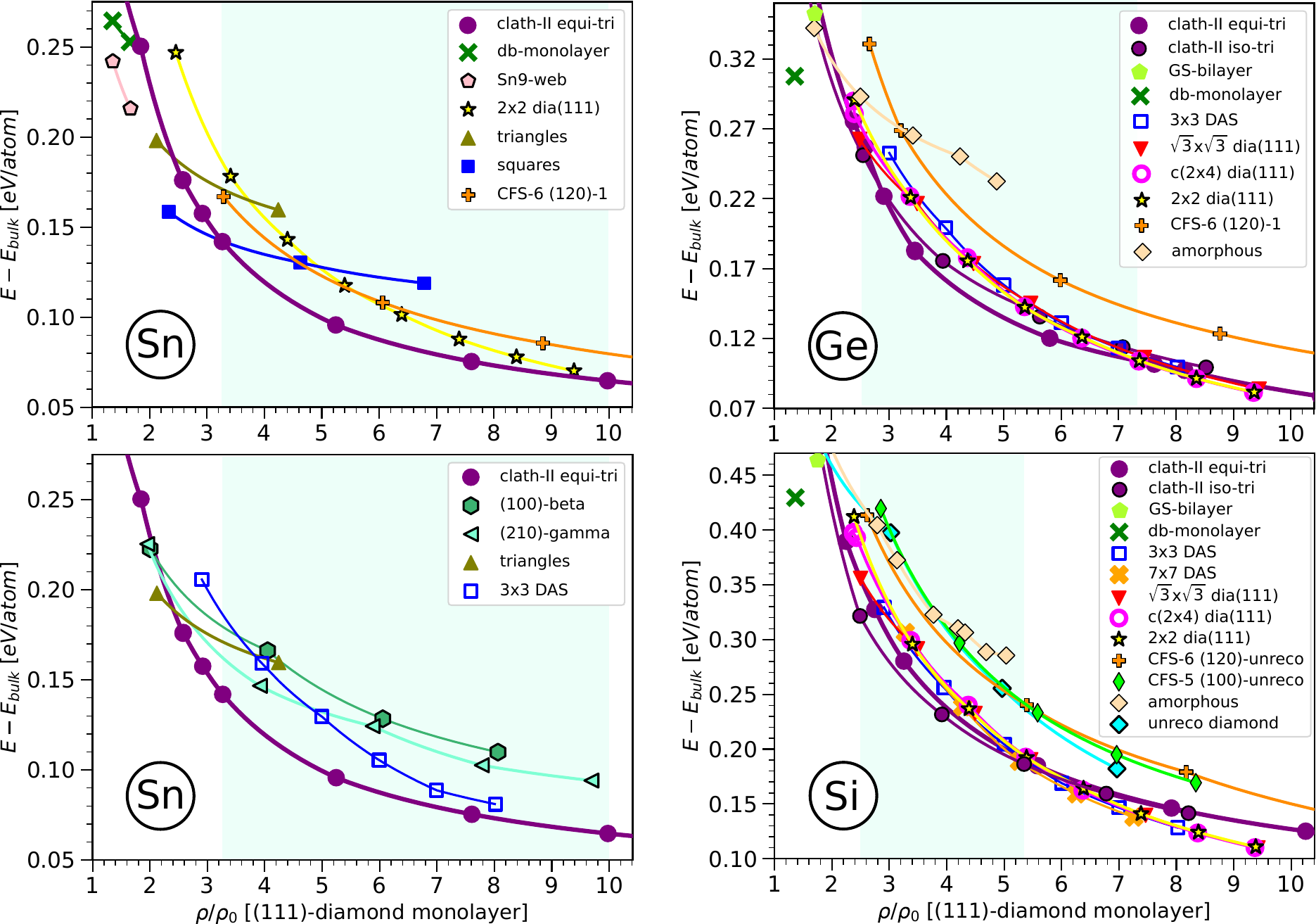}
	\caption{Shaded green area: stability region of Sn, Ge, Si clathrate thin films. DFT internal energy with respect to the cF8 bulk phase as a function of~coverage is plotted in each case.}
	\label{fig:res}
\end{figure*}

\begin{figure*}[h!]
	\centering
	\includegraphics[width = 18.0cm]{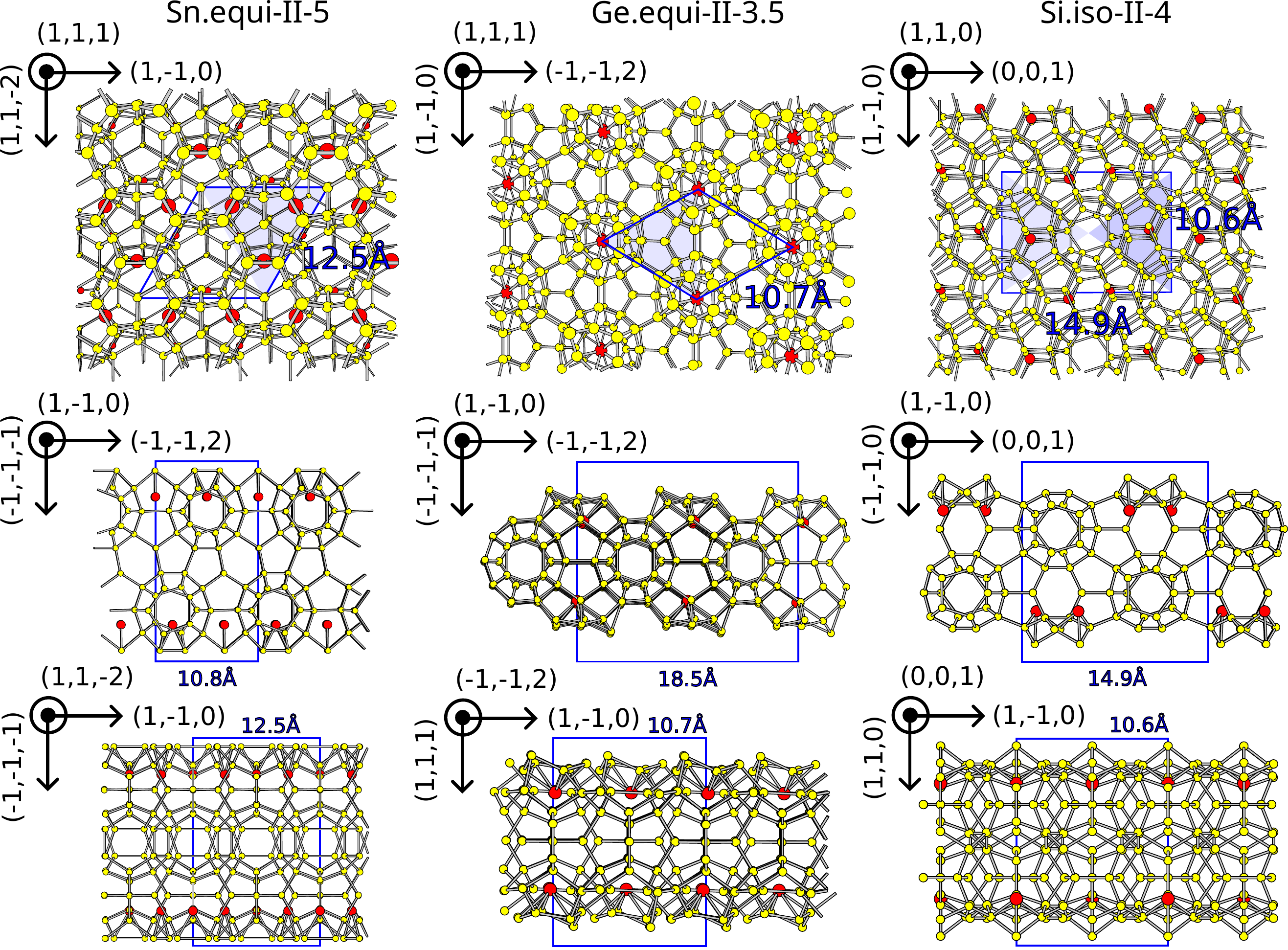}
	\caption{The structure of the most stable free-standing Sn (left panels), Ge (middle panels), Si (right panels) clathrate type-II thin films: their termination and reconstruction as a function of~a chemical element. Shaded areas correspond to elementary-triangle clathrate tiles (isosceles for~Si, equilateral for~Ge and Sn). Periodic cell without~vacuum is highlighted by solid blue line. Surface adatoms are bigger red balls. The integer in the sample's label designates number of (111)-diamond ML in~thickness (coverage $\rho$). Top: front view; middle, bottom: side views of the same slab. View directions are provided in parentheses.}
	\label{fig:restruct}
\end{figure*}

\indent For the lowest coverages below $2\:\mathrm{ML}$, discontinuous structures are formed, depending on the size of a supercell. At $\rho \approx 1.7\:\mathrm{ML}$, thread-like arrangements of Sn${}_{9}$, Sn${}_{10}$ and Sn${}_{15}$ units\cite{china_particle1, china_particle2} turn into a Sn9-web, Figures \ref{fig:sn9}, \ref{fig:res}. These small Sn ``houses'' seem to~be the energy-minimizing interface between~metallic Sn bilayers with~surrounding vacuum. \\
\indent Around $\rho \approx 2\:\mathrm{ML}$, uneven match among hexagonal and square metallic bilayers happens, Figure \ref{fig:hexa}, designated as ``triangles'' and ``squares'' in Figure \ref{fig:res}, respectively, squares being the clear winner by~$\approx 30\:\mathrm{meV/at.}$, Subsection \ref{hexa-square}. \\
\indent Finally, above $\rho \approx 3\:\mathrm{ML}$ up to $\rho \approx 10\:\mathrm{ML}$, clathrate slabs get a word in, Table \ref{tab:sum}, Figure \ref{fig:res}. For more than $\rho \approx 10.4\:\mathrm{ML}$, approaching the bulk limit, thin-film clathrates are surpassed by reconstructed thick diamond layers. \\
\indent Internal-energy-minimizing clathrate films are (111)-cut of~type-II bulk showing equilateral-triangle $T_{h}$ tiling in the section plane, Subsubsection \ref{clath-termi}, reconstructed equally by adatoms at~the midedge positions (in~the middle of each tile's side), samples Sn.equi-II-N, $N \in \{3;\,5;\,7.5;\,10\}$, Section S9.1 (SM) and Figure \ref{fig:restruct}. Owing to~its metallic behaviour, Sn favours the highest area-density of~adatoms from~amid Group~IV; this particular reconstruction presenting such an example, cf. Figure S3.1 (SM), resulting in~1.5 adatom per~triangle tile -- the largest admissible clathrate adatomic surface density. \\
\indent As concerns CFS, Subsection \ref{cfs}, the best reconstructed surface, CFS-6 (120)-1, possesses both higher area-density of defects (in the form of 3-coordinated atoms) as well as greater bulk per-atom-energy, thus dropping out of the game, Figure \ref{fig:res}. 

\subsection{Germanium}
\label{Ge-res}

\indent The most stable Si and Ge monolayer is oP10-reconstructed germanene and silicene, Subsection \ref{mono}, Figure \ref{fig:res}. Unlike metallic Sn, there is no known (meta)stable Ge or Si \emph{bilayer}. Above~$\rho \approx 2.5\:\mathrm{ML}$ clathrates clearly reign stable-thin-film configurational space, starting with~Ge.iso-II-2.5 (S9.2.2 in SM) one-cage thick isosceles-triangle (110)-cut of type-II bulk composed of~combined $T_{h} + T_{h}^{\prime}$ tiles, Figure S2.1 of SM, reconstructed by~solo adatom per~$T_{h}$ tile (i.e. without midedge adatoms in $T_{h}^{\prime}$ tiles). \\
\indent Next, in the thickness range $\rho \in[2.5;\,7.4]\:\mathrm{ML}$, Table \ref{tab:sum}, just as in~the case of Sn, (111)-cut clath-II films designated as Ge.equi-II-N, $N \in \{3;\,3.5;\,6\}$, Section S9.2 (SM), are by far most stable. This time, however, the best termination are markedly \emph{deformed} $T_{c}^{\prime}$ 6-caps, Subsubsection \ref{clath-termi}, with one adatom each. The net result of thorough \hyperlink{aimd}{AIMD} annealing and subsequent relaxation is transition of 6-cap to effectively 5-cap plus one atom, Figure~\ref{fig:restruct}, for all the Ge.equi-II-N samples. The configurational space above~$\rho \approx 7\:\mathrm{ML}$ is obviously a bulk limit, where diamond films become dominant, regardless of reconstruction, Figure~\ref{fig:res}. \\
\indent As CFS layers are concerned, the concluding remark from~the preceding Subsection~\ref{Sn-res} holds, the best CFS-6 (120)-1 moving even further away from the convex hull compared to Sn.CFS-6.

\subsection{Silicon}

\indent The area-density of adatoms gradually decreases in the sequence $\mathrm{Sn} \to \mathrm{Ge} \to \mathrm{Si}$. From the G4, Si is most sensitive to~adatom addition, unequivocally favouring isosceles-triangle (110)-cut of~type-II bulk composed of combined $T_{h} + T_{h}^{\prime}$ tiles, Figure~S2.1 of SM, in~the whole range of Si clathrate-film stability: $\rho \in[2.5;\,5]\:\mathrm{ML}$, Table \ref{tab:sum}, resulting in samples labeled Si.iso-II-N at~discrete thicknesses $N \in \{2.5;\,4;\,5\}$, Section S9.3 (SM). Like in~the Ge case, they are reconstructed by solo adatom in the middle of $T_{h}$ tile (i.e. without adatoms at midedge $T_{h}^{\prime}$ positions). \\
\indent According to Figures~S4.1, S4.2, S4.3 (SM), clathrate films are already winning by cheaper surface termination \emph{before} surface reconstruction, but proper reconstructions render them another important advantage over the diamond slabs. \\
\indent As expected, for Si, the best CFS surface is \emph{un}reconstructed, Figure \ref{fig:res}, almost overlaying high-energy amorphous $E(\rho)$ curve far from the overall convex envelope. \\ 
\indent Note that despite the lower surface energy $\gamma$ for both a-Ge and a-Si, Subsubsections \ref{aage} and \ref{aasi}, respectively, the stability of~clathrate (and even diamond) slabs over amorphous ones, Table \ref{fig:res}, is secured by their \emph{bulk} energy, Table S1.1 (SM).

\subsection{Electronic Stabilization of Clathrate Films}
\label{ele-res}

\indent Electronic densities of states (eDOS) of the most stable Sn, Ge and Si clathrate slabs are displayed in Figure S8.2 (SM). Clathrates definitely enhance the tendency of the sp${}^{3}$ elements towards bandgap formation -- this is clearly visible already in~the bulk, see Figure S8.1 (SM). \\
\indent Trends are evident both with respect to clathrate/diamond competition and as a function of \hyperlink{coverage}{coverage}, because the thinnest $2.5\:\mathrm{ML}$-slabs (samples Sn.equi-II-2.5, Ge.iso-II-2.5, Si.iso-II-2.5) exhibit distinct eDOS as well as $\sim3.5\:\mathrm{ML}$-thick ones (samples Ge.equi-II-3.5, Si.equi-III-3.5), whereas eDOS of thicker slabs approaches the bulk limit. \\
\indent Comparing clathrate vs. diamond, ultrathin slabs confirm and enhance clathrate tendency for the gap formation, while the reconstructed diamond slabs tend to fill the electronic states just above or even across the Fermi energy. This trend must be the substantial reason for the enhanced clathrate stability. \\
\indent Looking at the slab width variation, it is obvious that $2.5\:\mathrm{ML}$-slab eDOS stand out loosing any correspondence with the bulk. $3.5\:\mathrm{ML}$-thick Ge and Si slabs are distinguished by their bandgaps reaching the bulk-bandgap width, and are broader than bandgaps of~thicker slabs of the same kind. These results are further substantiated by calculated values of electronic band gaps of the most stable slabs, Subsection S10 of SM, showing that clathrate thin films exhibit wider gaps (of the order of $\sim 0.1\:\mathrm{eV}$ in the thickness range considered, $\rho \in [1;\,10]\:\mathrm{ML}$) compared to the diamond ones of~the same thickness.


\medskip

\section{Discussion}
\label{sec:discuss}

\indent Here we attempt to crystallize the most stable clathrate thin films from the melt to directly prove the chief results of our work, Figures \ref{fig:res} and \ref{fig:restruct}. The highest expectations are from the a-Ge slabs, owing to the promising spontaneously reconstructed surface -- 5-caps and 6-caps with one or two \hyperlink{adatom}{adatoms}, Figure \ref{age} -- that might serve as a condensation centers for clathrate cages.

\subsection{Crystallization of Free-Standing Molten Si Slab}

\indent a-Si slabs, energetically unstable at $0\:\mathrm{K}$ with respect to both diamond and clathrate slabs for any \hyperlink{coverage}{coverage} $\rho$, Figure \ref{fig:res}, crystallize \emph{exothermically} into diamond after annealing at~$T_{\mathrm{c}}\in[900-1100]\:\mathrm{K}$; the thicker slab the higher annealing temperature (due to the higher cohesive energy; $T_{\mathrm{c}} = 1100\:\mathrm{K}$ constituting the bulk limit). All of our a-Si slabs were prepared either by long \hyperlink{aimd}{DFT MD} (on-the-fly VASP MLFF) annealing slightly below $T_{\mathrm{c}}$ or by Replica Exchange method covering the temperature range $[400\:\mathrm{K};\,T_{\mathrm{c}}]$. \\
\indent In particular, partial crystallization of $3.2\:\mathrm{ML}$-thick a-Si slab into~cF8 was accompanied by $36\:\mathrm{meV/at.}$ energy release while annealing at $900\:\mathrm{K}$ and cooling down to~$300\:\mathrm{K}$. a-Si slabs of~coverages $\rho = 3.9\:\mathrm{ML}$ and $4.4\:\mathrm{ML}$ lowered their internal energies by~$28\:\mathrm{meV/at.}$ and $30\:\mathrm{meV/at.}$ during the same procedure, respectively. Crystallization of~$5.0\:\mathrm{ML}$-thick slab required annealing temperature at least $1000\:\mathrm{K}$. The exothermic change of~a-Si phase is an evidence that amorphous slabs can not compete with more stable diamond ones at $0\:\mathrm{K}$ at any coverage.

\subsection{Crystallization of Free-Standing Molten Ge Slab}



\indent Unlike a-Si bulk surface containing exclusively p${}^{3}$-, sp${}^{2}$- and sp${}^{3}$-bonded atoms, Figure \ref{asi}, a-Ge surface displays characteristic 5-caps and 6-caps with one or two \hyperlink{adatom}{adatoms}, Figure \ref{age}, which can be regarded as nuclei of clathrate cages. Therefore we attempted to crystallize one of the most stable free-standing clathrate films -- Ge.equi-II-3.5 -- from the melt in fixed-volume simulation, Figure S6.1 (SM). Initial atomic configuration \emph{before} melting can be found in Subsubsection S9.2.4 (SM). \\
\indent Slow cooling down of $3000\:\mathrm{K}$-melt to room temperature in~$2\times 10^{4}$ MD $\Delta t = 5\:\mathrm{fs}$-steps produced (from both faces) symmetrical disordered slab composed mainly of eclipsed hexagons and pentagons, $120\:\mathrm{meV/at.}$ above original clathrate Ge.equi-II-3.5 48-atom sample. \\
\indent Subsequent annealing $520\:\mathrm{K} \to 350\:\mathrm{K}$ in $4\times 10^{6}$ MLFF $\Delta t = 5\:\mathrm{fs}$-steps resulted in asymmetrical diamond film with dumbbell reconstruction from one side and kind of warts from the other, Figure S6.1 (SM). This diamond slab, not surprisingly, was still $45\:\mathrm{meV/at.}$ above initial clathrate Ge.equi-II-3.5 sample. \\
\indent A couple of long zero-pressure MLFF annealings at $450\:\mathrm{K}$ (or $400\:\mathrm{K}$) and subsequent slow cooling to room temperature -- $\sim 10^{6}$ steps $5\:\mathrm{fs}$ each -- of 192-atom sample [$2\times2\times1$ supercell of 48-atom Ge.equi-II-3.5 slab] in fixed periodic cell produced several characteristic 5-, 6- and even 7-cap features on the slab's surface like in~Figure~\ref{age}, but with no vital signs of a clathrate cage. Phase transition to~the original clathrate slab (before melting) was prohibited by~a~kinetic barrier, as the slab got stuck in amorphous configuration. The same holds for~240-atom sample shown in~Figure~\ref{age}, starting from random atomic positions, annealed millions of $1\:\mathrm{fs}$-steps at fixed $650\:\mathrm{K}$ temperature (of bulk a-Ge to cF8-Ge transition) and cooled to room temperature during $\sim 10^{4}$ steps (ought to be longer). \\
\indent There was just one purely MD simulation (before~the advent of~MLFF in~VASP) with 150-atom slab at fixed temperature $400\:\mathrm{K}$, half million of $5\:\mathrm{fs}$-steps long, starting from random atomic positions, that produced single reconstructed clathrate cage. \\
\indent Since transformation of the-best-clathrate-slab melt into a diamond one (cubic, not lonsdaleite) was well reproducible in many Si and Ge cases of~distinct thicknesses, we assume that clathrates are kinetically forbidden. Next we shall try to crystallize them on~specially designed clathrate-matching substrates\cite{iface}.

\subsection{Crystallization of Molten Ge Slab on InN}



\indent Even more interesting is the (re)crystallization of the same molten Ge.equi-II-3.5 layer on coherent InN binary-diamond substrate. The coherency of such an interface was the objective of~article\cite{iface}, where it has been shown that InN wafer gives Ge type-II or type-III clathrate growth fundamental advantage: its bulk-lattice misfit equals 2.3\% instead of 12.6\% misfit of Ge diamond film on~the same InN substrate. \\
\indent Recrystallization of Ge.equi-II-3.5 on InN led unexpectedly to~a~thin disordered Ge film connected with InN via a crystalline layer of~regular 5-gons, Figure S6.2 (SM) [see Subsection S9.4.1 for~initial xyz-file], $54\:\mathrm{meV/at.}$ above the original clathrate-diamond interface. The final result is probably due to the very 5-gonal buffer layer, which, as opposed to the coherent clathrate-diamond interface, saturates \emph{all} the sp${}^{3}$ bonds from the diamond side. As has been already pointed out\cite{iface}, clathrate is able to saturate at~most 6 out~of~9 dangling bonds on the diamond surface, Figure S6.3. Even more can be observed: not just all bonds of the substrate are saturated, but all bond angles with vertices at the substrate atoms are around the ideal tetrahedral angle ($\approx 109.5^{\circ}$), Figures S6.5, S6.8. Ge melt constitutes an intermediate situation between~Ge clathrate and recrystallized Ge layer from the point of~view of~bond angles; all N bonds are saturated even by the melt, Figures S6.4, S6.7. \\
\indent The statement above holds for~the nitrogen termination of~InN substrate (from~both sides), the more reactive element from~the couple (In, N). If the substrate were terminated by In, it would not grant any benefit to~an overlaying Ge clathrate at all. The same would be perhaps true for other electronically balanced substrates such as graphite or \ce{ZrO2}, were it not for the covalent bonds e.g. between graphite and Ge-clathrate-film adatoms, already witnessed in DFT relaxations of our own. \\
\indent When exactly the same experiment was carried out with thicker Ge layer, Ge.equi-II-6, the same 5-gonal buffer monolayer linked InN substrate with adjacent cubic-diamond Ge thin film, Figure S6.6 (SM), with some disordered mess on its outer surface. Thus it can be concluded that it is the very Ge buffer layer playing first fiddle on a N-terminated substrate. Mind, on the other hand, that the final diamond layer is by $38\:\mathrm{meV/at.}$ worse compared to~the original clathrate-diamond setup, see Subsubsection S9.4.2 for~the initial xyz-file, so the reason for diamond Ge output is kinetic, as well.

\subsection{van der Waals Dispersion Correction}
\label{subsec:vdw}

\indent Karttunen et al. demonstrated that for silicon the dispersion interaction neglected in~original PBE functional not only adds notable binding contribution of $\sim 0.3\:\mathrm{eV/at.}$ to~the bulk systems, but also impacts considerably on energy differences due to~the approximately linear dependence of the dispersion energy on~the {\em atomic density}\cite{kart2017}. The same picture emerges also for~germanium\cite{tran}. Both cited sources point out effectiveness of~the Grimme's et al. dispersion correction implemented along PBE functional (``PBE-D3'') in VASP\cite{grimme1, grimme2, grimme} in comparison with other ways of treating dispersion within the DFT framework. Thus the dispersion correction practically doubles $\Delta E$ between bulk clathrate and cubic diamond structure, from~$50$ to~$100\:\mathrm{meV/at.}$ for Si, and from $30$ to~$90\:\mathrm{meV/at.}$ for Ge, concerning cubic type-II clathrate (Pearson cF136), owing to the somewhat higher density of~the diamond phase\cite{kart2017}. \\
\indent However, the situation is different in the case of tin. According to~references\cite{tran, blaha} binding energy of cubic-diamond $\alpha$-Sn is described accurately by PBE {\em without} dispersion correction, while PBE-D3 leads to overbinding by a clear margin. Added to~this, PBE-D3 overestimates dramatically dispersion energy of~the metallic $\beta$-Sn \emph{reverting} $\Delta E$ between $\alpha$ and $\beta$ from~$42\:\mathrm{meV/at.}$ in~favour of $\alpha$ to $43\:\mathrm{meV/at.}$ in favour of $\beta$(!). Since vibrational free energy privileges $\beta$ (and we checked that the dispersion correction does not shift the phonon frequencies markedly), PBE-D3 contradicts the experimental fact the stable low-temperature phase being $\alpha$ up to approximately room temperature. \\
\indent In Figure S7.1 (SM), the effect of PBE-D3 correction on~our results for Si and Ge is displayed (\texttt{IVDW=12} in VASP), but we refrain from doing the same for Sn. \\
\indent Effect of a long-range dispersion interaction in general increases with the thickness of~the slab culminating in bulk, so the thinnest films in the focus of the present study are afflicted to much lesser extent. Nevertheless -- as expected, the range of~clathrate-film stability has been considerably retrenched by~dispersion correction for both Ge and Si. \\
\indent The original interval of Ge clath-II stability, $\rho \in (2.5;\,7.4)\:\mathrm{ML}$, Figure \ref{fig:res}, reduced to~$\rho \in (3;\,5.5)\:\mathrm{ML}$, Figure S7.1, leaving single (out of 4 stable slabs without vdW) directly constructed thin film Ge.equi-II-3.5 of thickness (coverage) $\rho \approx 3.5\:\mathrm{ML}$ on the overall convex hull: its energetic advantage over the diamond slab of the same coverage diminished from $34\:\mathrm{meV/at.}$ to~$24\:\mathrm{meV/at.}$ (both below the diamond slab). \\
\indent The original interval of Si clath-II stability, $\rho \in (2.5;\,5.3)\:\mathrm{ML}$, Figure \ref{fig:res}, reduced to~$\rho \in (2.5;\,4.0)\:\mathrm{ML}$, Figure S7.1, leaving solo (out of 3 stable slabs without vdW) directly constructed thin film Si.iso-II-2.5 of~thickness (coverage) $\rho \approx 2.5\:\mathrm{ML}$ on the overall convex envelope: its energetic benefit over the diamond slab of~the same coverage lessened from $44\:\mathrm{meV/at.}$ to~$23\:\mathrm{meV/at.}$ (both with~respect to the diamond slab) as well. \\


\section{Conclusions}
\label{conclusions}
This work exposes intervals of stability of several-ML thick clathrate layers, namely  $[2.5;\,7.4]\:\mathrm{ML}$ for Ge, $[2.5;\,5.3]\:\mathrm{ML}$ for~Si and $[3.4;\,10.0]\:\mathrm{ML}$ for Sn, where ML stands for the unit thickness of (111)-diamond monolayer of the given element. Column-XIV clathrate slabs are stabilized by means of the surface energy in two respects: cheaper terminations and higher energy gain by~reconstructions compared to the diamond films, meaning that clathrate films possess lower surface density of unsaturated dangling bonds than any known termination of sp${}^{3}$ allotrope and can be reconstructed most effectively. Within these thickness ranges lower surface energy reflects in~lower internal energy per~atom, which is taken as a stability measure. \\
\indent Natural planar clathrate terminations, as well as the recipe for~their reconstructions have been described for the first time, being superior to their diamond counterparts from the point of~view of~the surface energy. \\
\indent Combined (do)decagonal rectangle(square)-triangle clathrate tilings were not efficient in~lowering internal energy per atom, with respect to the pure triangular ones. \\
\indent Apart from clathrate thin films, we have discovered several Sn nanolayer phases during long MD runs at room temperature ($200-400\:\mathrm{K}$), namely Sn${}_{9}$-web phase and hexagonal even-layers. In the latter case, the effect of entropy is playing decisive role in~the layer stabilization since strongly defective triangular phase emerged at finite temperature and was presumably prepared during non-equilibrium processes\cite{saxena}, clearly far away ($30\:\mathrm{meV/at.}$) from the ground-state. \\
\indent We hope that the present work instigates future experimental preparation of the suggested elemental stable clathrate thin films even in quasi-free-standing form on electronically balanced and/or amorphous substrates, whose novel unique properties would not be governed by the above-mentioned peculiar and often very complex intermetallic wafers (such as $i$-Al-Pd-Mn or $d$-Al-Ni-Co). Such a substrate should somehow disqualify the main rival - diamond - with respect to clathrates or could at least serve as a structure-neutral medium. The logical consequence would then be an investigation of influence of finite temperature on stability of clathrate thin layers. \\


\section*{Conflicts of interest}
The author declares that she has no known competing financial interests or personal relationships that could have appeared to~influence the work reported in this paper. \\


\section*{Data availability}
The data supporting this article have been included as part of the Supplementary Information. \\


\section*{Acknowledgements}

I acknowledge support from grants VEGA 2/0144/21, APVV-19-0369 and APVV-20-0124. I also acknowledge opportunity to~write this article and finish the calculations after the defense of~my Dissertation Thesis\cite{diss} entitled ``Search for Stable Clathrate Films within~Elements from~Column~XIV of~the Periodic Table'', under supervisor Dr. Marek Mihalkovič, containing vast majority of~material for this article, at~the Institute of Materials and Machine Mechanics in~Bratislava. Calculations were performed at the Computing Center of the Slovak Academy of Sciences using the supercomputing infrastructure acquired under Projects ITMS 26230120002 and 26210120002 and EuroHPC grant no. 101101903 (supercomputer Devana). Part of the research results was obtained using the computational resources procured in the national project National competence centre for high performance computing (project code: 311070AKF2) funded by European Regional Development Fund, EU Structural Funds Informatization of society, Operational Program Integrated Infrastructure. \\




\balance

\renewcommand\refname{References}

\bibliography{Bibliography} 
\bibliographystyle{unsrtnat}

\end{document}


\pagestyle{empty}
\definecolor{ruz}{rgb}{1,0,0.5}
\definecolor{tmod}{RGB}{0,50,150}
\definecolor{tzel}{RGB}{0,200,200}
\definecolor{zel}{RGB}{0,100,50}
\definecolor{fialova}{rgb}{0.5,0,0.5}
\definecolor{white}{rgb}{1.0,1.0,1.0}
\definecolor{tblue}{RGB}{0,0,130}
\definecolor{seda}{RGB}{235,235,235}
\definecolor{sseda}{RGB}{248,248,248}
\definecolor{azur}{RGB}{212,255,230}
\definecolor{pink}{RGB}{242,220,255}
\definecolor{grey}{RGB}{255,240,240}
\definecolor{lavender}{rgb}{0.9, 0.9, 0.98}

\hyphenation{Di-me-ri-sa-ti-on alu-mi-no-si-li-ca-tes}

\numberwithin{equation}{section}
\numberwithin{figure}{section}
\numberwithin{table}{section}












\thispagestyle{empty}
\hypertarget{tocc}{}

\begin{center}
	\noindent {\Huge{\textbf{Stable Thin Clathrate Layers}}} \\
\end{center}

\smallskip

\begin{center}
	\noindent {\large{\textbf{Eva Posp\'{i}\v{s}ilov\'{a}}}} \\
\end{center}

\smallskip

\begin{center}
\noindent {\Huge{\textbf{Supplemental Material}}} \\
\end{center}

\vspace{1.0cm}

\noindent \underline{Note:} Invisible hyperlinks (white places) at the corners of \emph{every} page of this Supplemental Material (SM) bring you back to this Table of Contents (TOC). \\

\vspace{0.5cm}

\indent 
\noindent \hspace{-0.76cm} {\Large{\textbf{Table of Contents}}} \\
\label{sec:reco}
\vspace{0.5cm}

\noindent \hyperlink{s1}{\large{\textbf{S1. Comparison of Column-XIV Bulk Energies}}} \\

\noindent \hyperlink{s2}{\large{\textbf{S2. Surface Reconstruction of Decagonal Clathrate Tiles}}} \\

\noindent \hyperlink{s3}{\large{\textbf{S3. Surface Reconstruction of Dodecagonal Clathrate Tiles}}} \\

\noindent \hyperlink{s4}{\large{\textbf{S4. Reconstructed vs. Unreconstructed Surfaces}}} \\

\noindent \hyperlink{s5}{\large{\textbf{S5. Energy-Comparison of Various Types of Clathrate Slabs}}} \\

\noindent \hyperlink{s6}{\large{\textbf{S6. Crystallization of Clathrate Slabs from the Melt}}} \\

\noindent \hyperlink{s7}{\large{\textbf{S7. van der Waals Dispersion Correction of G4 Slabs}}} \\

\noindent \hyperlink{s8}{\large{\textbf{S8. Electronic Density of States: Clathrate Slabs \& Bulk}}} \\

\noindent \hyperlink{s9}{\large{\textbf{S9. xyz Files}}} \\

\smallskip

\indent\quad \hyperlink{s9.1}{\large{\textbf{S9.1 Tin-Slab xyz Files}}} \\

\indent\quad\quad \indent \hyperlink{s9.1.1}{\textbf{S9.1.1 Sn9-web}} \\

\indent\quad\quad \indent \hyperlink{s9.1.2}{\textbf{S9.1.2 triangles}} \\

\indent\quad\quad \indent \hyperlink{s9.1.3}{\textbf{S9.1.3 squares}} \\

\indent\quad\quad \indent \hyperlink{s9.1.4}{\textbf{S9.1.4 Sn.equi-II-3}} \\

\indent\quad\quad \indent \hyperlink{s9.1.5}{\textbf{S9.1.5 Sn.equi-II-5}} \\

\indent\quad\quad \indent \hyperlink{s9.1.6}{\textbf{S9.1.6 Sn.equi-II-7.5}} \\

\indent\quad\quad \indent \hyperlink{s9.1.7}{\textbf{S9.1.7 Sn.equi-II-10}} \\

\smallskip

\indent\quad \hyperlink{s9.2}{\large{\textbf{S9.2 Germanium-Slab xyz Files}}} \\

\indent\quad\quad \indent \hyperlink{s9.2.1}{\textbf{S9.2.1 Ge.db-monolayer (oP10)}} \\

\indent\quad\quad \indent \hyperlink{s9.2.2}{\textbf{S9.2.2 Ge.iso-II-2.5}} \\

\indent\quad\quad \indent \hyperlink{s9.2.3}{\textbf{S9.2.3 Ge.equi-II-3}} \\

\indent\quad\quad \indent \hyperlink{s9.2.4}{\textbf{S9.2.4 Ge.equi-II-3.5}} \\

\indent\quad\quad \indent \hyperlink{s9.2.5}{\textbf{S9.2.5 Ge.iso-II-4}} \\

\indent\quad\quad \indent \hyperlink{s9.2.6}{\textbf{S9.2.6 Ge.equi-II-6}} \\

\smallskip

\indent\quad \hyperlink{s9.3}{\large{\textbf{S9.3 Silicon-Slab xyz Files}}} \\

\indent\quad\quad \indent \hyperlink{s9.3.1}{\textbf{S9.3.1 Si.db-monolayer (oP10)}} \\

\indent\quad\quad \indent \hyperlink{s9.3.2}{\textbf{S9.3.2 Si.iso-II-2.5}} \\

\indent\quad\quad \indent \hyperlink{s9.3.3}{\textbf{S9.3.3 Si.equi-II-3}} \\

\indent\quad\quad \indent \hyperlink{s9.3.4}{\textbf{S9.3.4 Si.iso-II-4}} \\

\indent\quad\quad \indent \hyperlink{s9.3.5}{\textbf{S9.3.5 Si.iso-II-5}} \\

\smallskip

\indent\quad \hyperlink{s9.4}{\large{\textbf{S9.4 Ge-Clathrate/InN}}} \\

\indent\quad\quad \indent \hyperlink{s9.4.1}{\textbf{S9.4.1 Thinner Ge Clathrate Film on InN}} \\

\indent\quad\quad \indent \hyperlink{s9.4.2}{\textbf{S9.4.2 Thicker Ge Clathrate Film on InN}} \\

\noindent \hyperlink{s10}{\large{\textbf{S10. Tabulated $\bm{E(\rho)}$ Dependencies}}} \\

\smallskip

\indent\quad \hyperlink{s10.1}{\large{\textbf{S10.1 $\bm{E(\rho)}$ of Sn}}} \\

\indent\quad \hyperlink{s10.2}{\large{\textbf{S10.2 $\bm{E(\rho)}$ of Ge}}} \\

\indent\quad \hyperlink{s10.3}{\large{\textbf{S10.3 $\bm{E(\rho)}$ of Si}}} \\

\vspace{0.5cm}

\noindent \underline{Note:} Invisible hyperlinks (white places) at the corners of \emph{every} page of this Supplemental Material (SM) bring you back to this Table of Contents (TOC).


\newpage
\hypertarget{s1}{}
\noindent {\Large{\textbf{S1. Comparison of Column-XIV Bulk Energies}}}
\label{sec:energy-comp}

\vspace{1.0cm}

{{\small
		\begin{table}[h!]
			\centering
			\renewcommand{\arraystretch}{1.3}
			\begin{tabular}[h!]{!{\vrule width 1.7pt} C{2.0cm} | C{1.7cm} | C{1.7cm} | C{1.7cm} | C{1.7cm} | C{1.7cm} !{\vrule width 1.7pt}}
				\noalign{\hrule height 1.6pt}
				{\cellcolor{pink}{Config.}} & {\cellcolor{azur}{C}} &  {\cellcolor{azur}{Si}} &  {\cellcolor{azur}{Ge}} & {\cellcolor{azur}{Sn}} &  {\cellcolor{azur}{Pb}}  \\
				\hline
				\hline
				fcc      & 4573    & 608    & 331    & 75      & \textcolor{red}{\textbf{-3.598}} \\
				\hline
				bcc      & 4367    & 587    & 341    & 86      & 42     \\
				\hline
				graphite & \textcolor{red}{\textbf{-9.231}}    & 640 & 602 & 570 & 719 \\
				\hline
				hP1      & 971     & 352    & 243    & 45      & 96     \\
				\hline
				tI4      &  cF8    & 328    & 228    & 46      & 128    \\
				\hline
				cF8      & 131     & \textcolor{red}{\textbf{-5.420}} & \textcolor{red}{\textbf{-4.516}} & \textcolor{red}{\textbf{-3.863}} & {\cellcolor{seda}{251}} \\
				\hline
				lonsdaleite   & 155     & 11     & 18     & 16     & {\cellcolor{seda}{292}} \\
				\hline
				clathI   & 239     & 64     & 33     & 28      & {\cellcolor{seda}{279}} \\
				\hline
				clathII  & 205     & 54     & 27     & 23      & {\cellcolor{seda}{280}} \\
				\hline
				clathIII & 260     & 81     & 48     & 40      & {\cellcolor{seda}{288}} \\
				\hline
				clathIV  & 243     & 69     & 38     & 31      & {\cellcolor{seda}{276*}} \\
				\hline
				deca     & 241     & 71     & 41     & 35      & {\cellcolor{seda}{299*}} \\
				\hline
				icosa    & 288     & 92     & 59     & 49      & {\cellcolor{seda}{274*}} \\
				\hline
				CFS-6    & 241     & 55     & 35     & 30      & {\cellcolor{seda}{286}} \\
				\hline
				CFS-5    & 263     & 41     & 39     & 36      & {\cellcolor{seda}{289}} \\
				\hline
				T12      & 238     & 45     & 36     & 32      & {\cellcolor{seda}{278}} \\
				\hline
				amorph   & --      & 158    & 128    & 91--101 & -- \\
				\hline
				\noalign{\hrule height 1.2pt}
			\end{tabular}
			\begin{tabular}[h!]{p{16.0cm}}
				\vspace{-0.0cm}
				\label{comp-energy}Table S1.1: {\small{Bulk energies of famous Column-XIV allotropes, that were taken into slab internal-energy competition, see Section 4 in the main text. Internal energies of 17 bulk configurations of~G4 elements after~cell and atomic relaxations starting from sound initial atomic positions and lattice parameters. cF8 stands for the cubic diamond phase; lonsdaleite (wurtzite) is the hexagonal diamond polytype with eclipsed bond conformations. ``deca'' stands for~``decagonal clathrate approximant'', \emph{Pearson symbol} oP392. ``icosa'' means ``icosahedral clathrate approximant'', cI920. „clathI'' (cP46), ``clathII'' (cF136), ``clathIII'' (hP40) and ``clathIV'' (tP172) denotes type I, II, III and IV clathrate, respectively. ``CFS'' means Chiral Framework Structures (including T12), see Subsection 4.6 (of the main text). ``amorph'' denotes sp${}^{3}$-amorphous phases of Ge and Si, visit Subsection 4.5; for Sn the lower bound corresponds to a-metal, whereas upper bound to~an~abstract a-sp${}^{3}$ Sn obtained by a-Ge rescaling to the typical Sn interatomic distances. Kohn-Sham energies in~$\mathrm{eV/atom}$ of~the most stable (GS) structures are \textcolor{red}{\textbf{highlighted by red}}. GS energies have been subtracted from~the remaining energies and the corresponding differences $E - E_{\mathrm{GS}}$ are stated in~units of~$\mathrm{meV/atom}$ for~easier comparison of~\emph{distinct} elements. * -- amorphized or strongly deformed during relaxation process. Without~former annealing or external strain. Converged on K-points with~default VASP DFT PAW GGA (PW91) ENCUT-s.}}
			\end{tabular}
\end{table}}}


\newpage
\hypertarget{s2}{}
\noindent {\Large{\textbf{S2. Surface Reconstruction of Decagonal Tiles}}}
\label{sec:reco-deca}

\vspace{0.2cm}

\hypertarget{s21}{}
\begin{figure}[h!]
	\label{comp-deca}
	\centering
	\hspace{0.0cm}\includegraphics[width = 13.7cm]{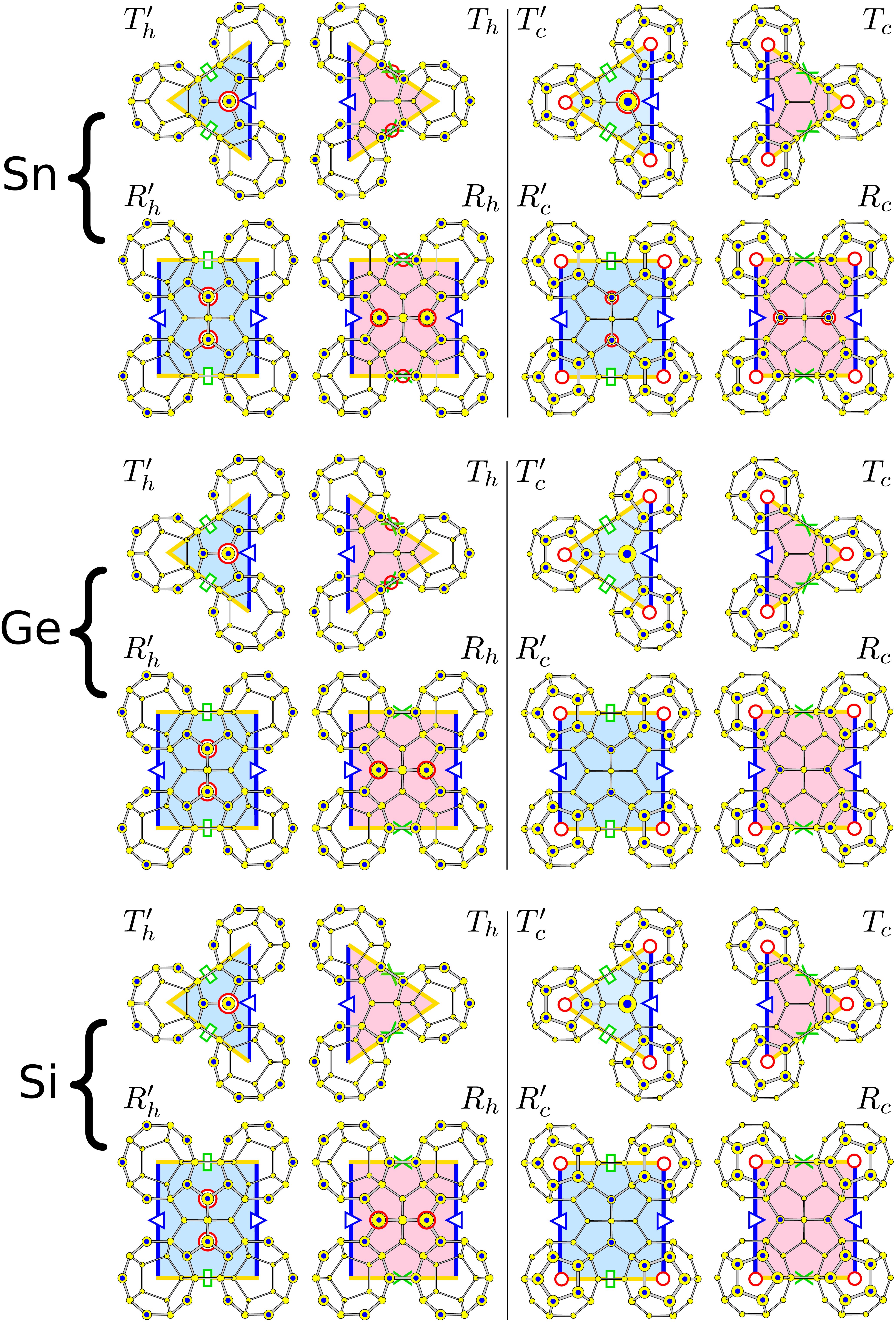}
	\begin{tabular}[h!]{p{16.0cm}}
		\vspace{-0.15cm}
		Figure S2.1: {\small{Reconstruction of \emph{decagonal} Sn, Ge, Si clathrate tiles: isosceles triangle (T) and rectangle (R) terminated by cage ``caps'' (c) and ``holes'' (h). Adatoms (beneath the original atoms, inside the film) highlighted by red rings. 3-coord. atoms before the reconstruction marked with central blue dots, the rest are 4-coord. atoms. Side lengths distinguished by blue (longer) and yellow (shorter) color. Sides that fit together marked with~green crosses, rectangles.}}
	\end{tabular}
\end{figure}


\newpage
\hypertarget{s3}{}
\noindent {\Large{\textbf{S3. Surface Reconstruction of Dodecagonal Tiles}}} \\
\label{sec:reco-dodeca}

\hypertarget{s31}{}
\begin{figure}[h!]
	\centering
	\hspace{0.0cm}\includegraphics[width = 15.8cm]{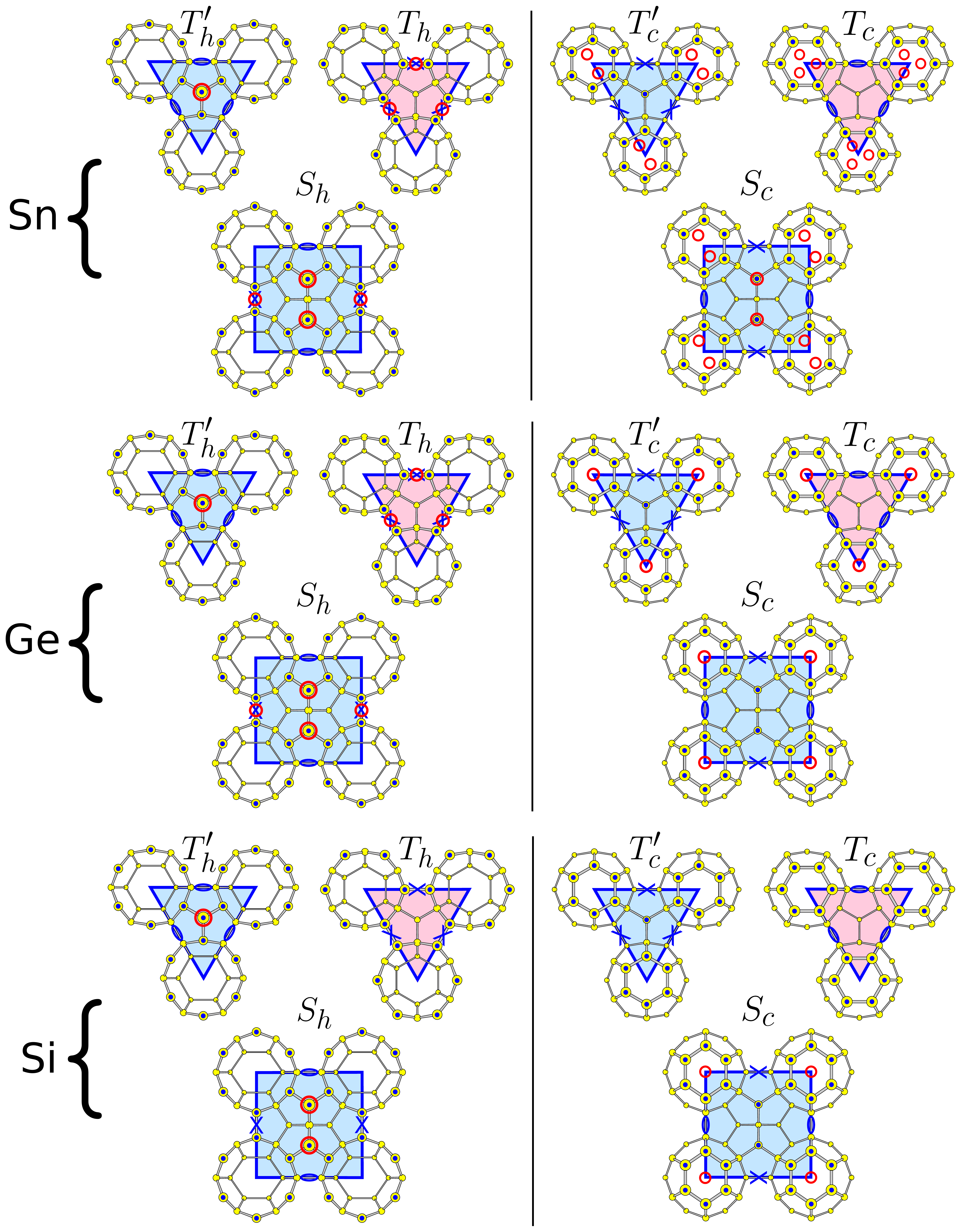}
	\begin{tabular}[h!]{p{16.0cm}}
		\vspace{-0.35cm}
		Figure S3.1: {\small{Optimal reconstruction of \emph{dodecagonal} Sn, Ge and Si clathrate tiles: equilateral triangle (T) and square (S) terminated by cage ``caps'' (c) and ``holes'' (h). For the explanation of the symbols and color scheme, consult Figure S2.1. Note that placing the ``caps'' on the top of the ``holes'' does not change the number of unsaturated 3-coordinated atoms (distinguished by blue central dots); this holds also for the decagonal cases, see Figure S2.1.}}
	\end{tabular}
\end{figure}


\newpage
\hypertarget{s4}{}
\noindent {\Large{\textbf{S4. Reconstructed vs. Unreconstructed Surfaces}}} \\
\label{sec:reco-unreco}

Figures S4.1 and S4.3 below implies that Sn and Si clathrate slabs possess clear energetic advantage over the corresponding (111)-diamond films of the same coverage already \emph{before} surface reconstruction (addition of Si atoms to the slab's surface followed by atomic-position DFT relaxation). However, the appropriate surface reconstruction is responsible for further energy decrease of clathrates with respect to the diamond rivals. \\
\indent Notably, as can be observed in Figure S4.2, the opposite situation happens for Ge, where clathrate's internal-energy advantage over diamond is markedly more pronounced for the \emph{unreconstructed} slabs. \\
\indent In all the graphs below, the region of clathrate film stability is highlighted by green shadow, compare with Figure~6 of the main text. Reconstructed slabs correspond to~the~unreconstructed structures (``reconstructed'' = ``unreconstructed'' + adatoms) in each case. As for the reconstructed ones, the most stable clathrate and diamond films for each coverage $\rho$ from Figure~6 of~the~main text were chosen for comparison, confer Results section. \\

\vspace{0.5cm}

\begin{figure}[h!]
	\centering
	\hspace{0.0cm}\includegraphics[width = 10.0cm]{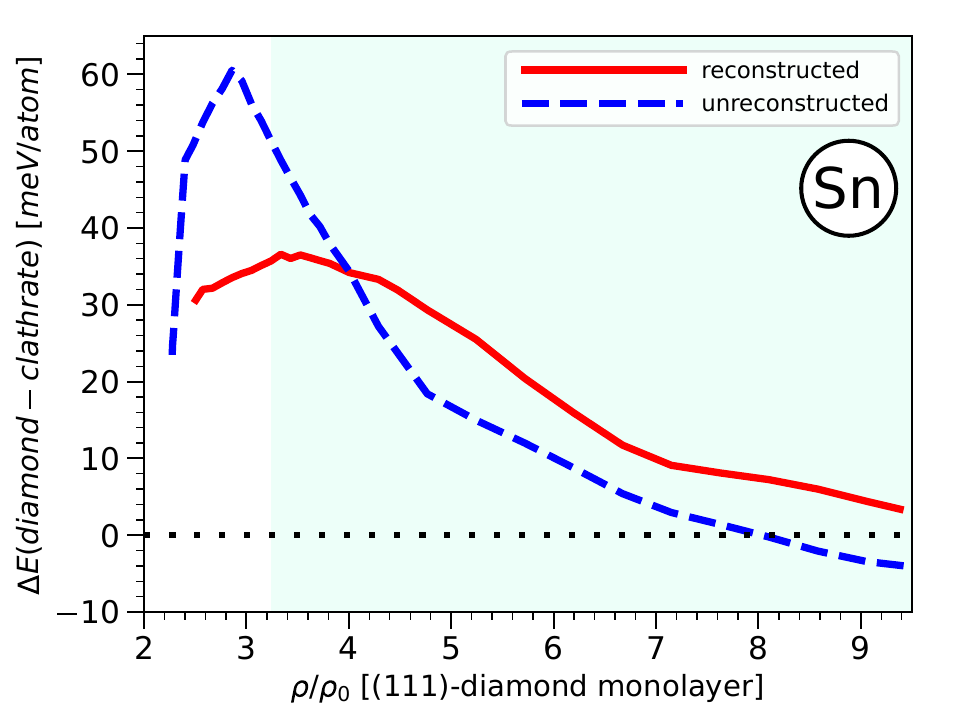}
	\begin{tabular}[h!]{p{10.0cm}}
		\vspace{-0.25cm}
		\label{reko-nereko-Sn}Figure S4.1: {\small{Comparison of per-atom energy difference between the most stable Sn clathrate slabs and Sn (111)-diamond ones for reconstructed and unreconstructed case.}}
	\end{tabular}
\end{figure}

\begin{figure}[h!]
	\centering
	\hspace{0.0cm}\includegraphics[width = 10.0cm]{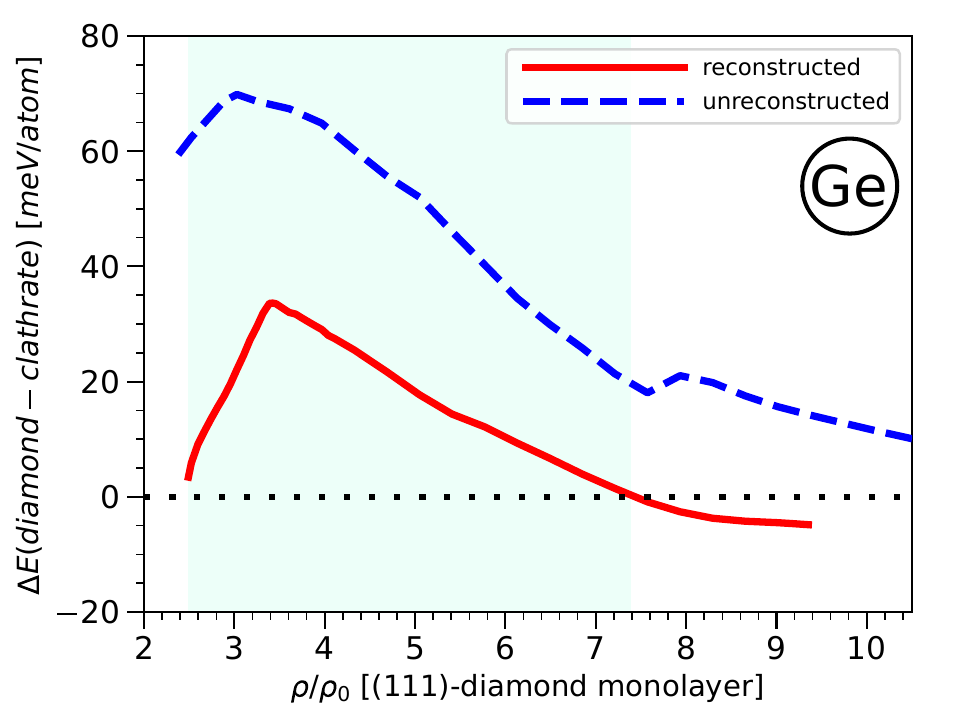}
	\begin{tabular}[h!]{p{10.0cm}}
		\vspace{-0.25cm}
		\label{reko-nereko-Ge}Figure S4.2: {\small{Comparison of per-atom energy difference between the most stable Ge clathrate slabs and Ge (111)-diamond ones for reconstructed and unreconstructed case.}}
	\end{tabular}
\end{figure}

\begin{figure}[h!]
	\centering
	\hspace{0.0cm}\includegraphics[width = 10.0cm]{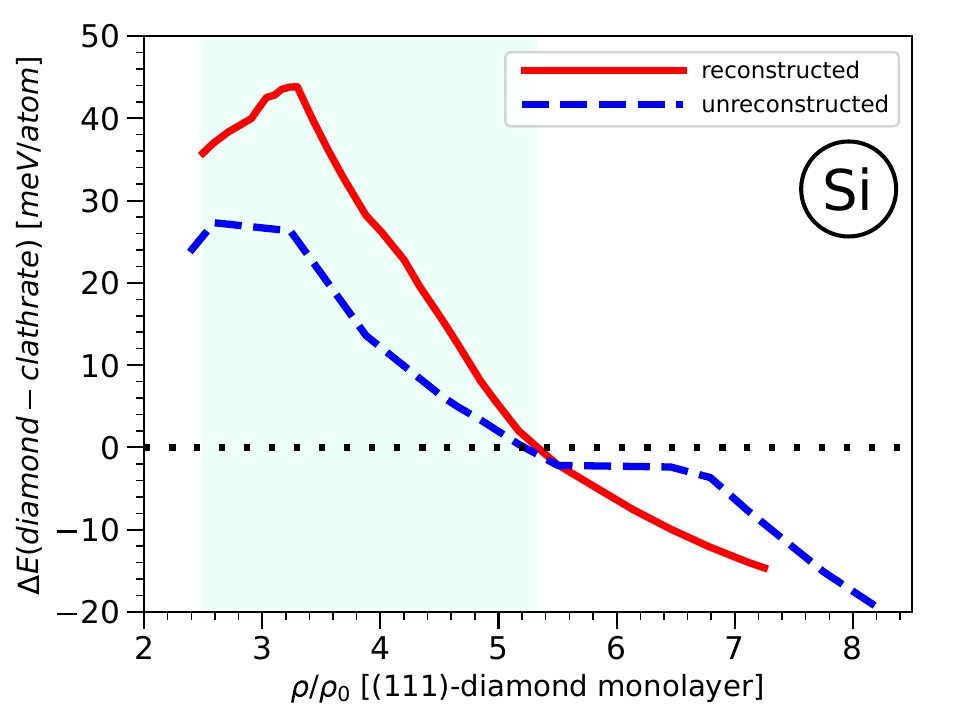}
	\begin{tabular}[h!]{p{10.0cm}}
		\vspace{-0.25cm}
		\label{reko-nereko-Si}Figure S4.3: {\small{Comparison of per-atom energy difference between the most stable Si clathrate slabs and Si (111)-diamond ones for reconstructed and unreconstructed case.}}
	\end{tabular}
\end{figure}


\clearpage

\hypertarget{s5}{}
\noindent {\Large{\textbf{S5. Comparison of Various Types of Clathrate Slabs}}}
\label{sec:clath-comparo}

\begin{figure}[h!]
	\centering
	\hspace{0.0cm}\includegraphics[width = 10.0cm]{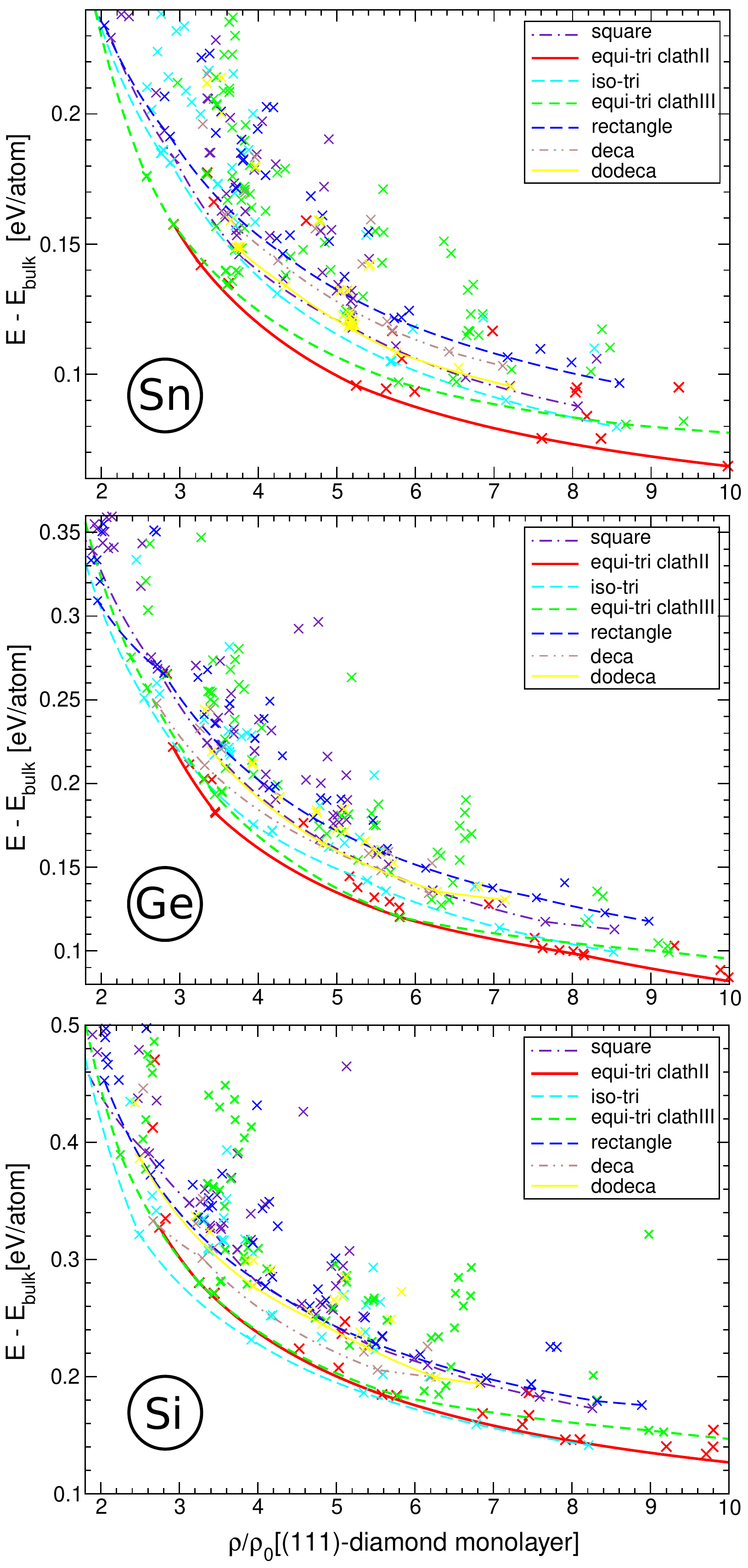}
	\begin{tabular}[h!]{p{16.0cm}}
		\vspace{-0.25cm}
		\label{comp-tiles-spol}Figure S5.1: {\small{Stability comparison (DFT energy per atom at $0\:\mathrm{K}$) of~seven types of Sn, Ge and Si clathrate thin films: clath-I square, clath-II equilateral triangle, clath-II isosceles triangle, clath-III equilateral triangle and rectangle, decagonal and clath-IV dodecagonal tiling reconstructed according to Figures \hyperlink{s21}{S2.1}, \hyperlink{s31}{S3.1}. Each cross represents one attempt for clath-tile reconstruction.}}
	\end{tabular}
\end{figure}


\newpage
\hypertarget{s6}{}
\noindent {\Large{\textbf{S6. Crystallization of Clathrate Slabs from the Melt}}}
\label{sec:crystallize}

\begin{figure}[h!]
	\vspace{-3.2cm}
	\includegraphics[width=15.8cm]{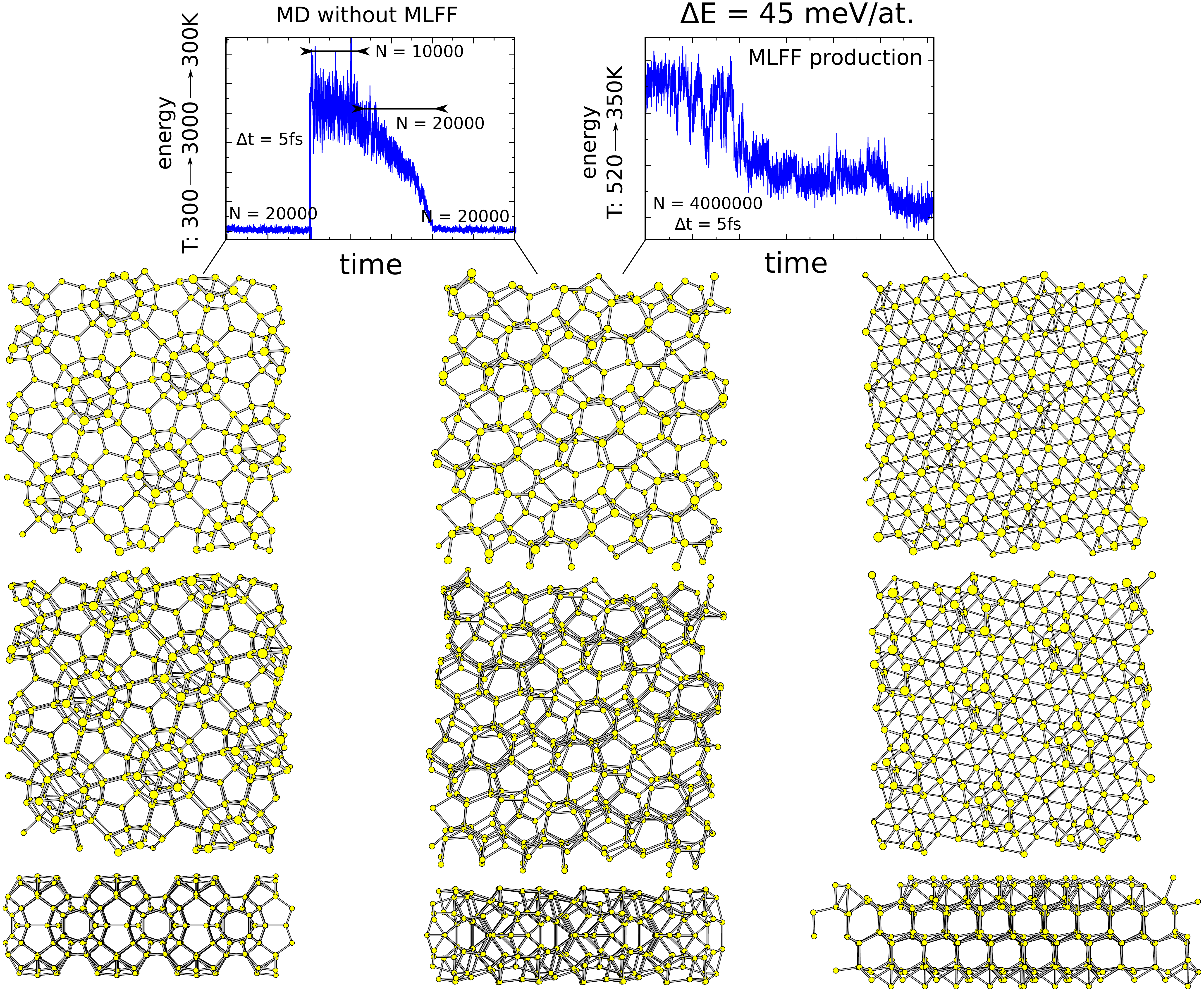}
	\begin{tabular}[h!]{p{15.6cm}}
		\vspace{-0.0cm}
		\label{top1}Figure S6.1: {\small{Spontaneous re-crystallization of free-standing initially symmetrical Ge film, sample designated as Ge.equi-II-3.5 in the main text, see Subsubsection~\hyperlink{s9.2.4}{S9.2.4} for the atomic coordinates.}}
	\end{tabular}
\end{figure}

\begin{center}
	\hypertarget{top2}{}
	\includegraphics[width=15.8cm]{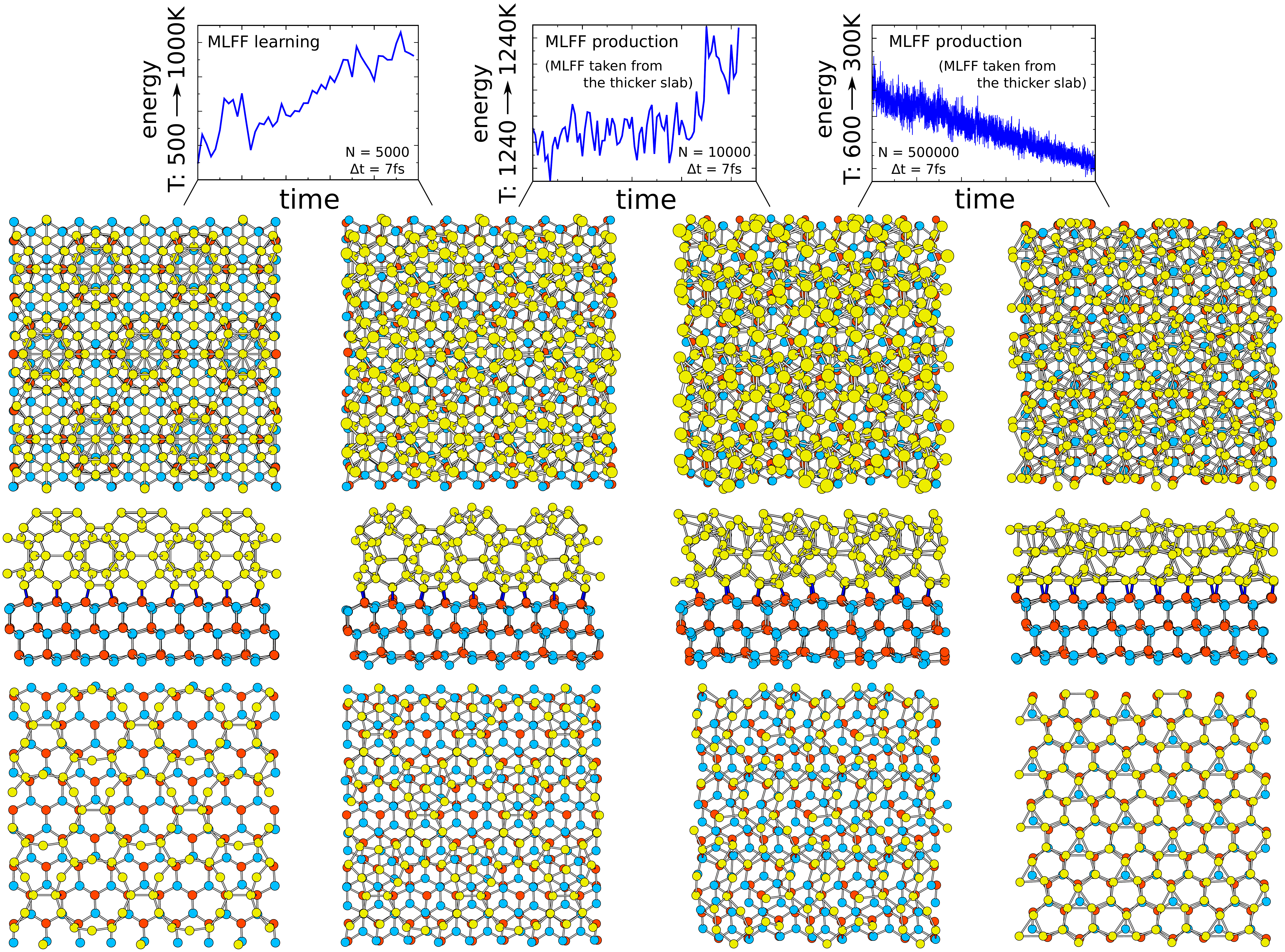}
	\begin{tabular}[h!]{p{15.6cm}}
		\vspace{-0.0cm}
		Figure S6.2: {\small{Spontaneous re-crystallization of Ge films on InN. Starting from~the thinnest clathrate film with~6-caps, see Subsubsection~\hyperlink{s9.4.1}{S9.4.1} for the atomic coordinates.}}
	\end{tabular}
\end{center}

\begin{center}
	\includegraphics[width=15.8cm]{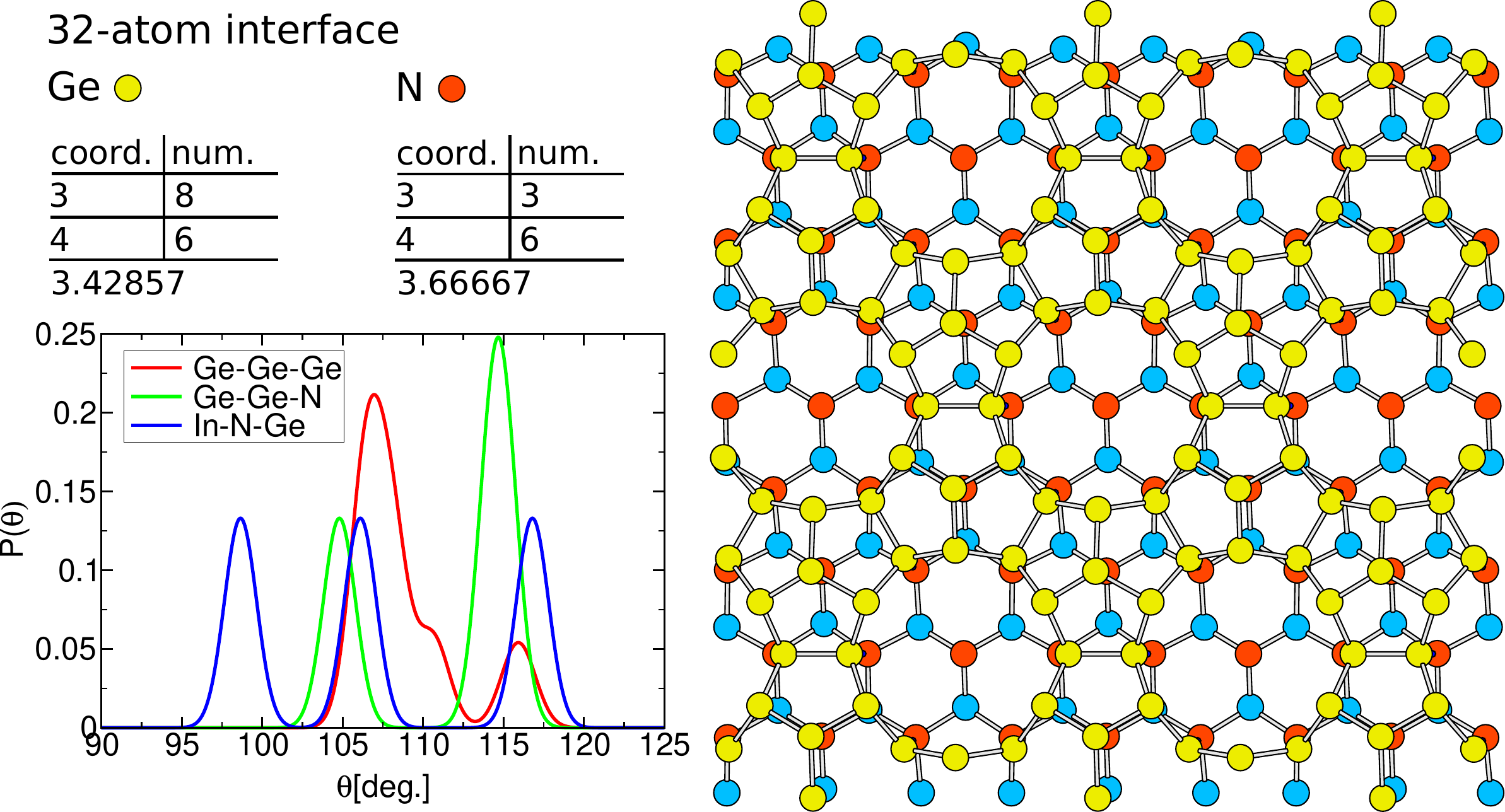}
	\begin{tabular}[h!]{p{15.6cm}}
		\vspace{-0.0cm}
		\label{top3}Figure S6.3: {\small{Coherent clathrate/diamond interface made up of Ge/InN.}}
	\end{tabular}
\end{center}

\begin{center}
	\includegraphics[width=15.8cm]{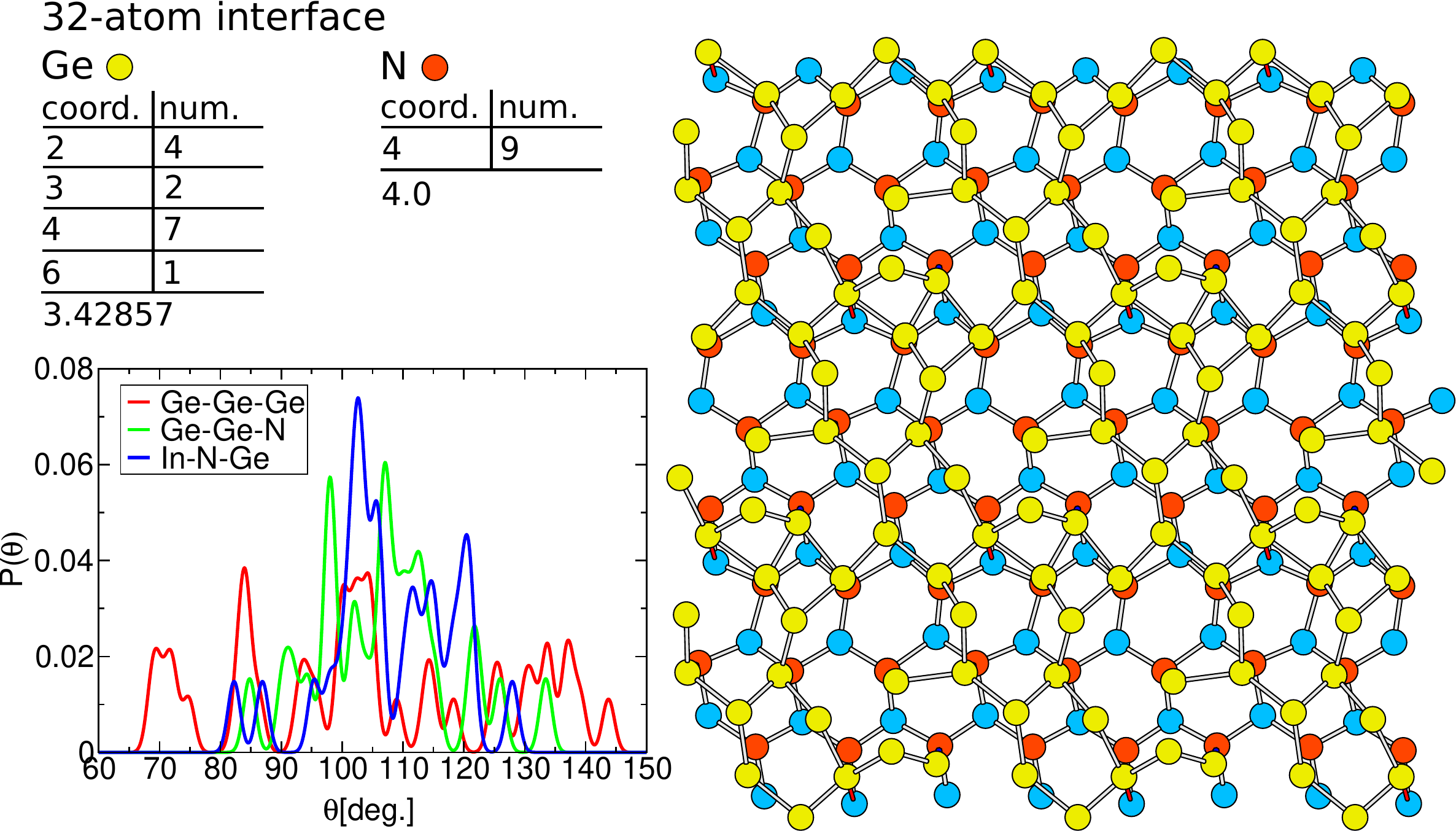}
	\begin{tabular}[h!]{p{15.6cm}}
		\vspace{-0.0cm}
		\label{top4}Figure S6.4: {\small{Interface between diamond and clathrate melt after $N = 10000$ $\mathrm{d}t = 7\:\mathrm{fs}$-steps at $1240\:\mathrm{K}$, according to roadmap~\hyperlink{top2}{S6.2}.}}
	\end{tabular}
\end{center}

\begin{center}
	\includegraphics[width=15.8cm]{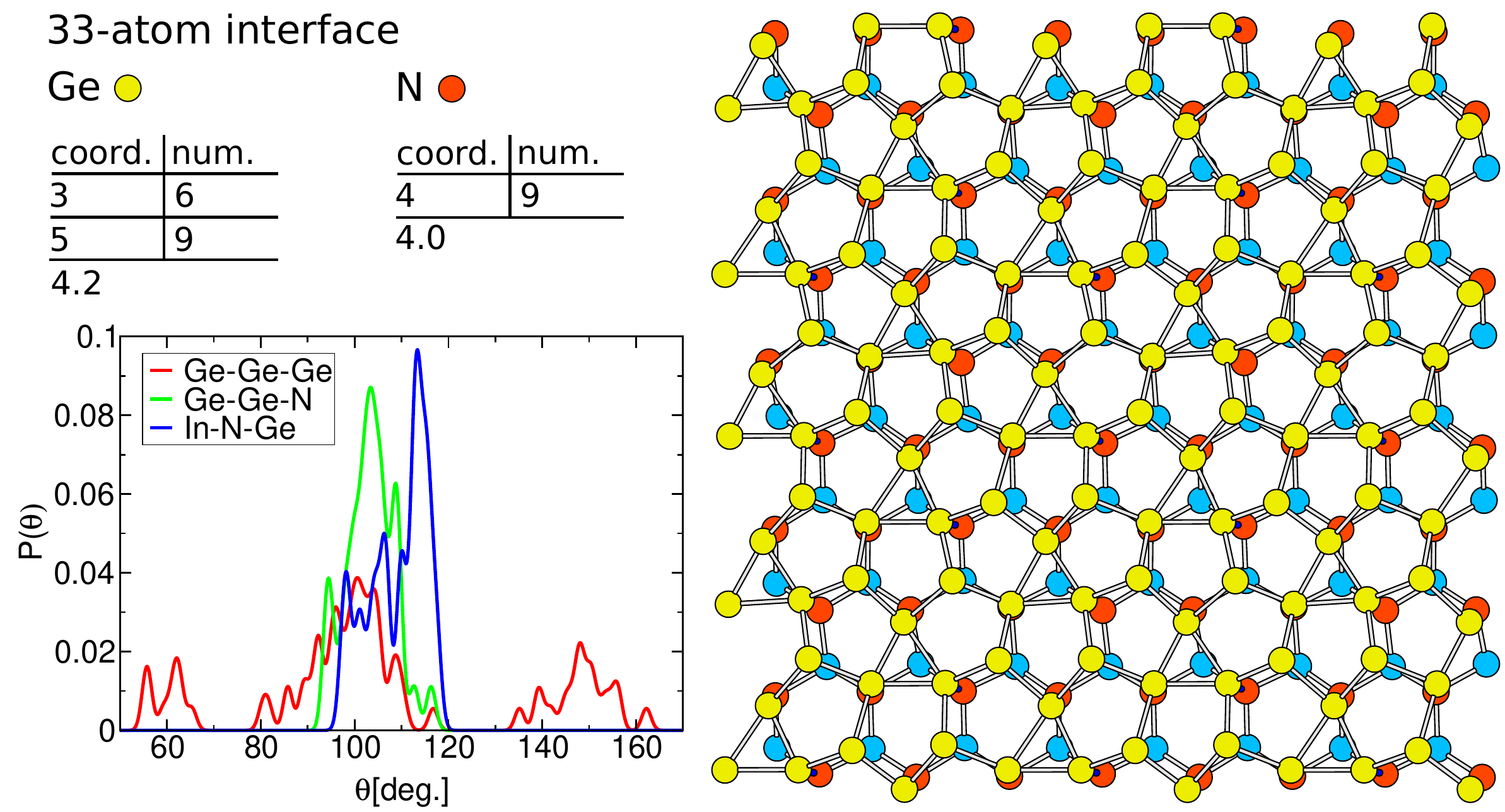}
	\begin{tabular}[h!]{p{15.6cm}}
		\vspace{-0.0cm}
		\label{top5}Figure S6.5: {\small{Interface between diamond and re-crystallized clathrate melt according to Figure~\hyperlink{top2}{S6.2}.}}
	\end{tabular}
\end{center}


\begin{center}
	\hypertarget{top6}{}
	\includegraphics[width=15.8cm]{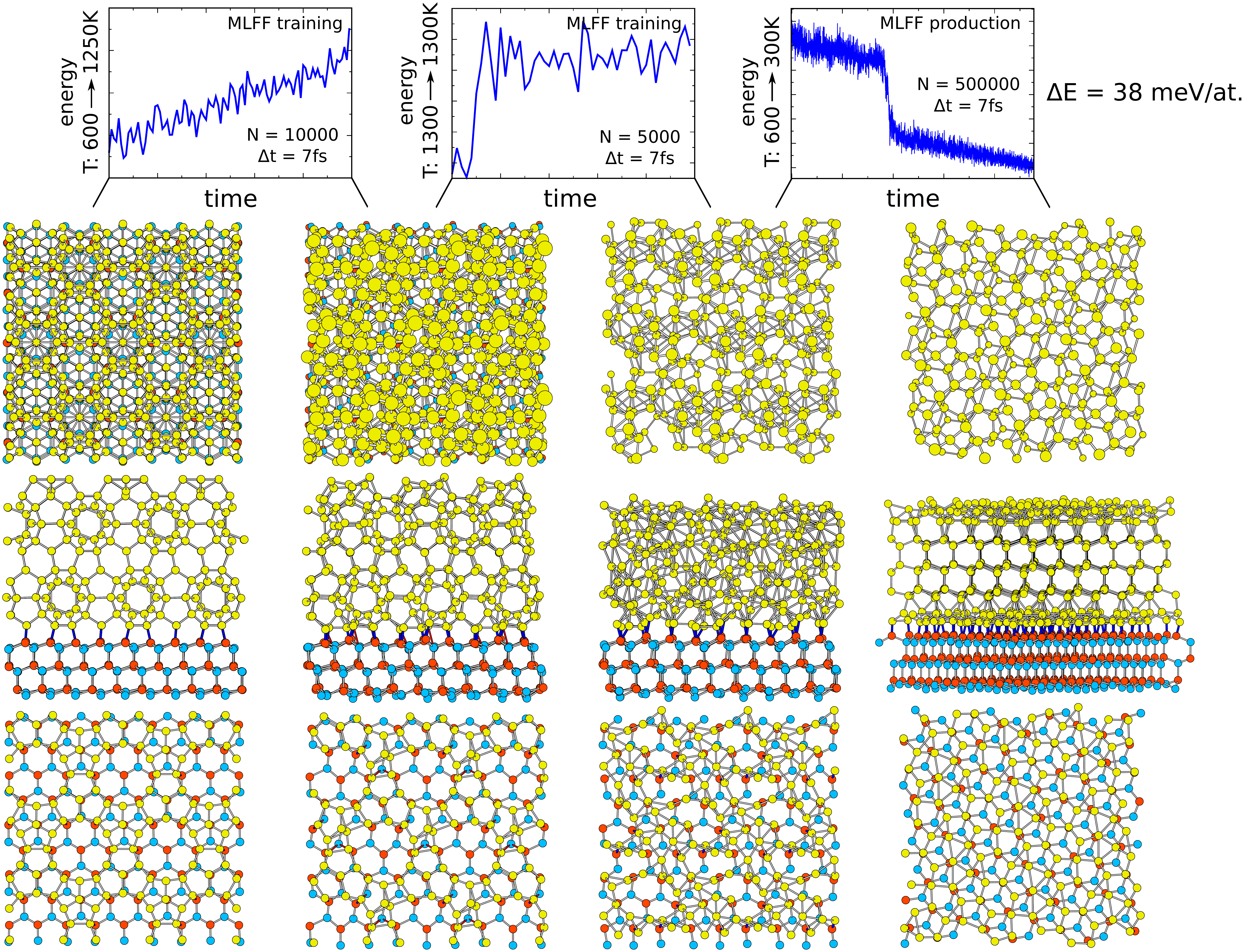}
	\begin{tabular}[h!]{p{15.6cm}}
		\vspace{-0.0cm}
	    Figure S6.6: {\small{Spontaneous re-crystallization of Ge films on InN. Starting from~thicker clathrate film with~6-caps, see Subsubsection~\hyperlink{s9.4.2}{S9.4.2} for the atomic coordinates.}}
	\end{tabular}
\end{center}

\begin{center}
	\includegraphics[width=15.8cm]{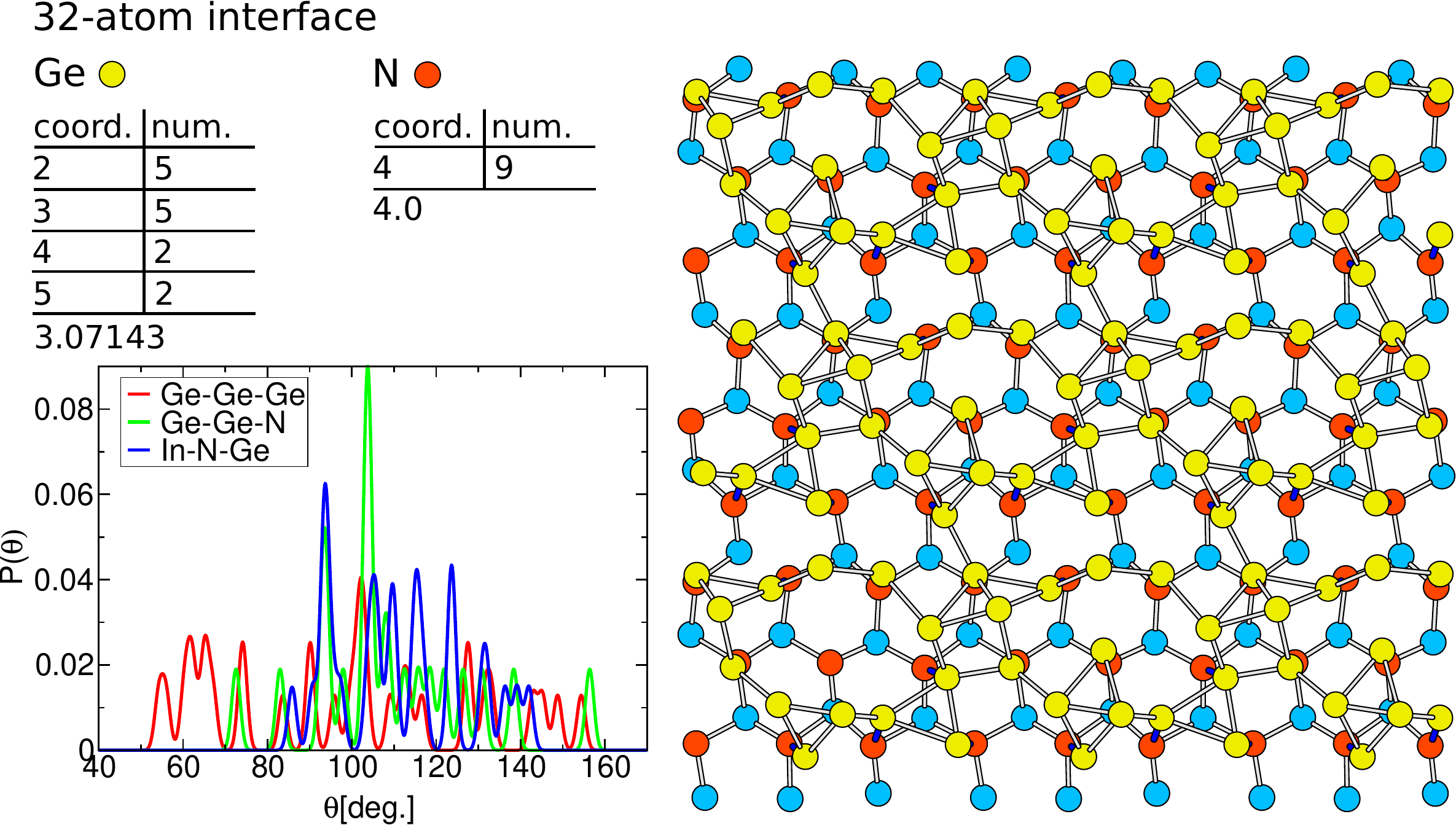}
	\begin{tabular}[h!]{p{15.6cm}}
		\vspace{-0.0cm}
		\label{top7}Figure S6.7: {\small{Interface between diamond and clathrate melt after $N = 5000$ $\mathrm{d}t = 7\:\mathrm{fs}$-steps at $1300\:\mathrm{K}$, according to Figure~\hyperlink{top6}{S6.6}.}}
	\end{tabular}
\end{center}

\begin{center}
	\includegraphics[width=15.8cm]{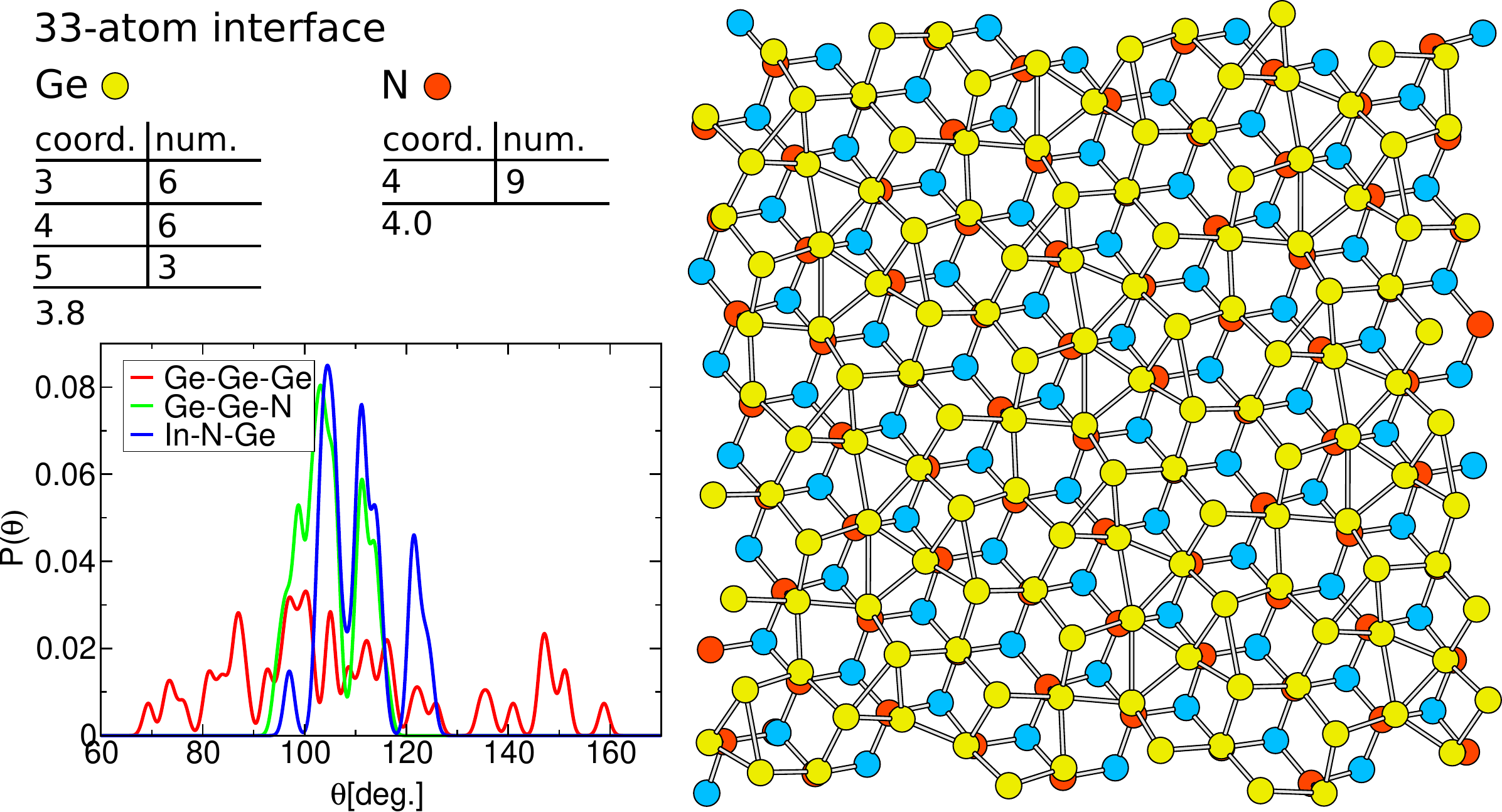}
	\begin{tabular}[h!]{p{15.6cm}}
		\vspace{-0.0cm}
		\label{top8}Figure S6.8: {\small{Interface between diamond and re-crystallized clathrate melt according to Figure~\hyperlink{top6}{S6.6}.}}
	\end{tabular}
\end{center}


\newpage
\hypertarget{s7}{}
\noindent {\Large{\textbf{S7. van der Waals Dispersion Correction of G4 Slabs}}} \\
\label{sec:vdW}

\begin{figure}[h!]
	\centering
	\includegraphics[width = 15.8cm]{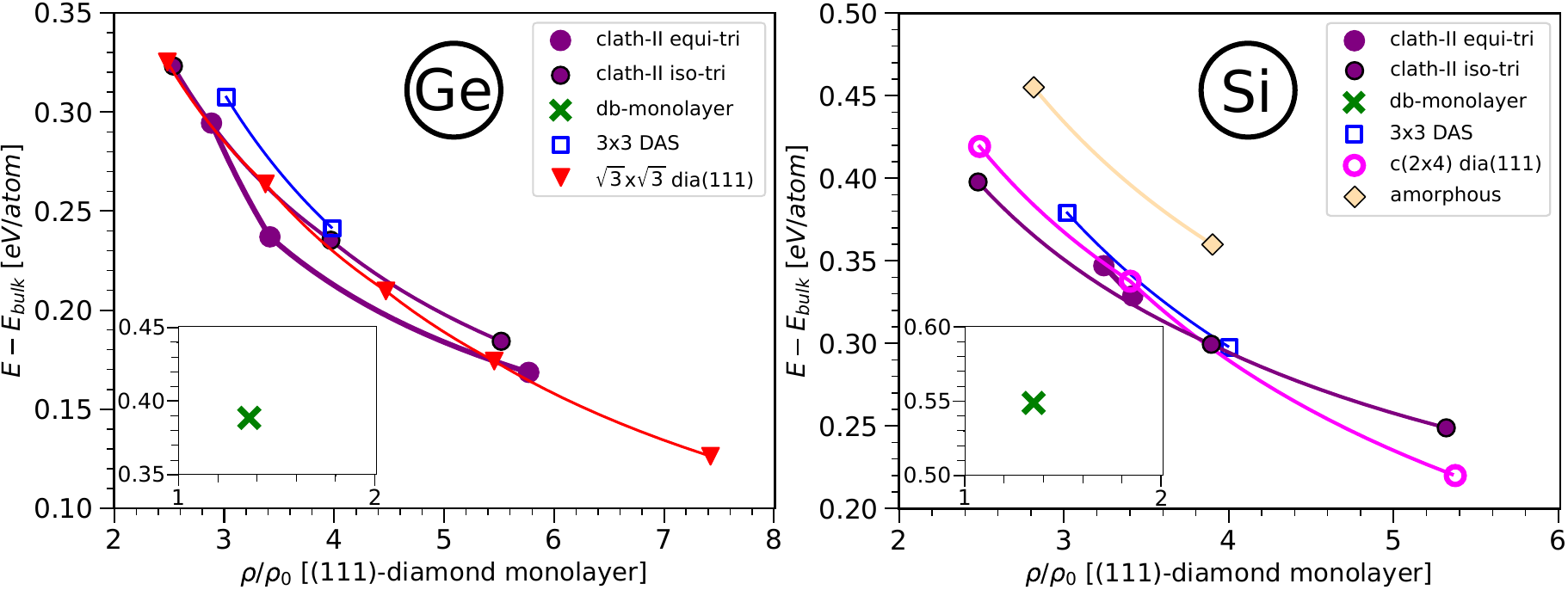}
	\begin{tabular}[h!]{p{15.8cm}}
		\vspace{-0.0cm}
	\label{vdw-pic}Figure S7.1: {\small{Change of the diamond-clathrate stability competition in~response to the van der Waals dispersion correction PBE-D3 (\texttt{ivdw=12}), visit Subsection 6.4 in the main text.}}
   \end{tabular}
\end{figure}


\newpage
\hypertarget{s8}{}
\noindent {\Large{\textbf{S8. Electronic Density of States: Clathrate Slabs \\ \& Bulk}}}
\label{sec:edos}

\vspace{1.0cm}

\begin{figure}[h!]
	\centering
	\includegraphics[width = 12.2cm]{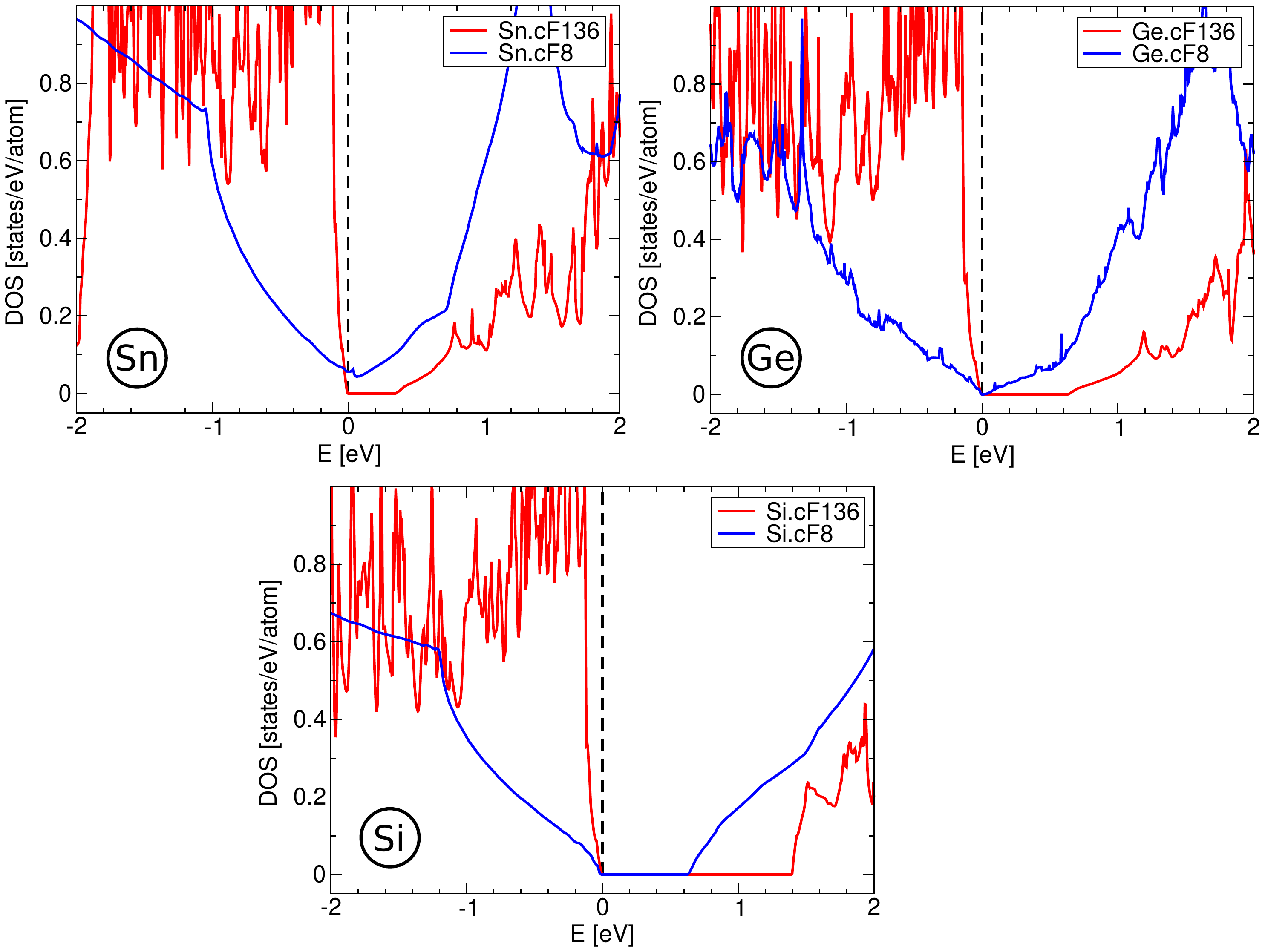}
	\begin{tabular}[h!]{p{15.8cm}}
		\vspace{-0.0cm}
		\label{edos-bulk}Figure S8.1: {\small{Electronic Density of States, eDOS, of the bulk type-II clathrates (cF136) compared with bulk diamond (cF8) for Sn, Ge and Si. Fermi energy, $E_{F}$, is designated by the vertical thick dashed line.}}
	\end{tabular}
\end{figure}

\begin{figure}[h!]
	\centering
	\hspace{0.0cm}\includegraphics[width = 12.2cm]{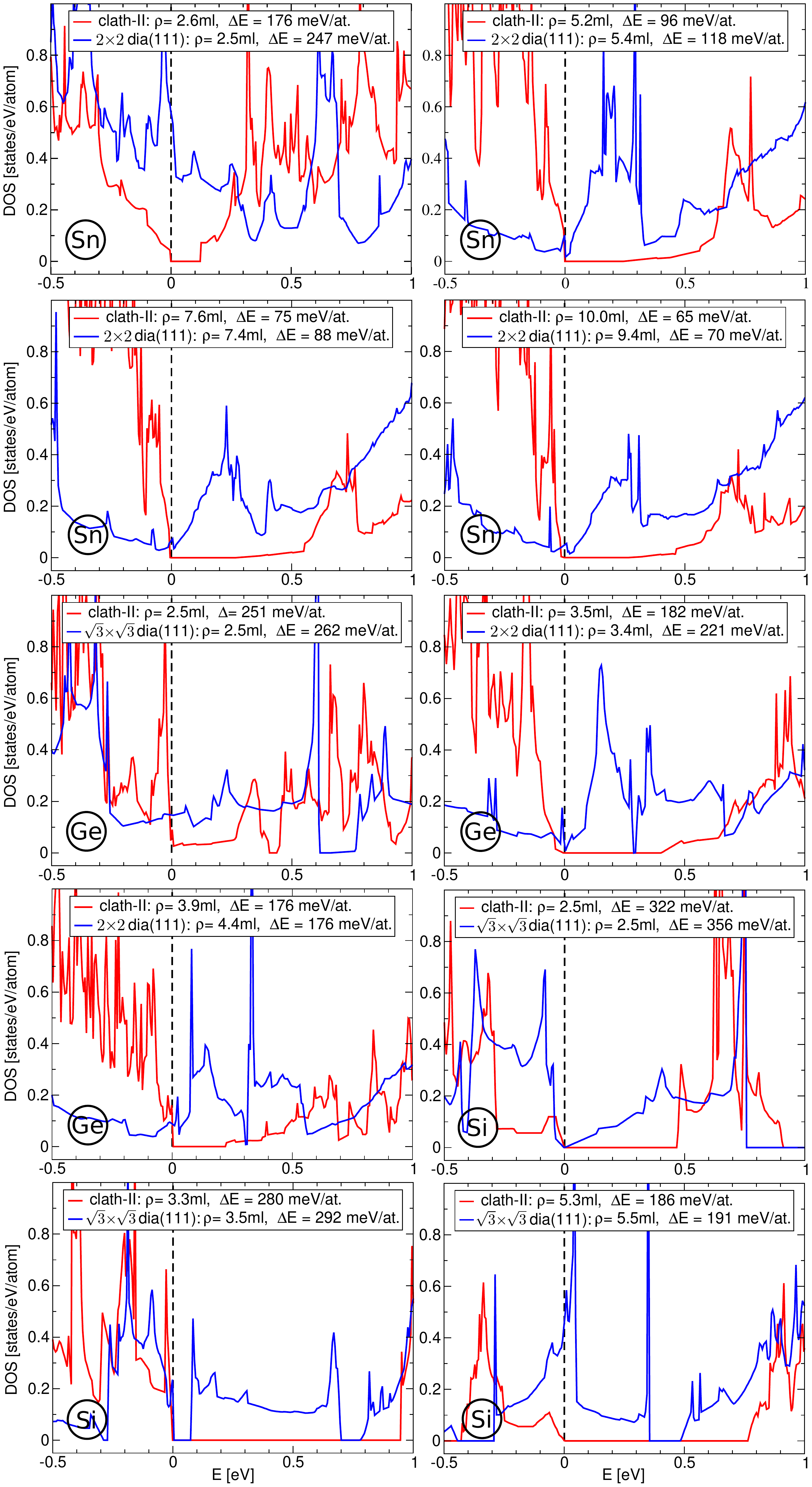}
	\begin{tabular}[h!]{p{16.2cm}}
	\vspace{-0.35cm}
	\label{dos}Figure S8.2: {\small{Electronic Density of States, eDOS, of the most stable clathrate slabs, compared with that of the competitive diamond slabs of approximately the same thickness $\rho$. Fermi energy, $E_{F}$, is designated by vertical thick dashed line. Visit Subsection 5.4 in the main text.}}
    \end{tabular}
    \vspace{-0.35cm}
\end{figure}


\newpage
\hypertarget{s9}{}
\noindent {\Large{\textbf{S9. xyz Files}}} \\
\label{sec:xyz}


\hypertarget{s9.1}{}
\noindent {\large{\textbf{S9.1 Tin-Slab xyz Files}}} \\
\label{sec:xyz.Sn}


\hypertarget{s9.1.1}{}
\noindent {\textbf{S9.1.1 Sn9-web}} \\
\noindent $\rho = 0.1749\:\AA^{-2} = 1.667\:\mathrm{ML}$, $E - E_{\mathrm{bulk}} = 216\:\mathrm{meV/at.}$ \\

\begin{tabular}{llllll}
	23.9138  &   0.00000  &   0.00000  &     &     &      \\ 
	0.00000  &   23.9138  &   0.00000  &     &     &      \\ 
	0.00000  &   0.00000  &   23.9138  &     &     &      \\ 
	100 atoms    &     &     &     &     &      \\ 
	0.96713  &   0.89507  &   0.02979  &   50  &   1  &   1 \\ 
	0.18163  &   0.30665  &   0.95156  &   50  &   2  &   2 \\ 
	0.00243  &   0.36734  &   0.04387  &   50  &   3  &   3 \\ 
	0.11868  &   0.91534  &   0.92111  &   50  &   4  &   4 \\
\end{tabular}

\begin{tabular}{llllll} 
0.69423  &   0.89719  &   0.96087  &   50  &   36  &   36 \\ 
0.50253  &   0.38736  &   0.05731  &   50  &   37  &   37 \\ 
0.45773  &   0.04332  &   0.91417  &   50  &   38  &   38 \\ 
0.82545  &   0.91695  &   0.04931  &   50  &   39  &   39 \\ 
0.55073  &   0.81738  &   0.08054  &   50  &   40  &   40 \\ 
0.06642  &   0.18467  &   0.94560  &   50  &   41  &   41 \\ 
0.34977  &   0.85690  &   0.04637  &   50  &   42  &   42 \\ 
0.10212  &   0.43689  &   0.07187  &   50  &   43  &   43 \\
\end{tabular}

\begin{tabular}{llllll}
0.33400  &   0.29204  &   0.99652  &   50  &   86  &   86 \\ 
0.67111  &   0.52100  &   0.93956  &   50  &   87  &   87 \\ 
0.41257  &   0.58602  &   0.10125  &   50  &   88  &   88 \\ 
0.06569  &   0.80185  &   0.94906  &   50  &   89  &   89 \\ 
0.04873  &   0.31091  &   0.93358  &   50  &   90  &   90 \\ 
0.33098  &   0.02608  &   0.97160  &   50  &   91  &   91 \\ 
0.40518  &   0.46092  &   0.08388  &   50  &   92  &   92 \\ 
0.45882  &   0.30776  &   0.96075  &   50  &   93  &   93 \\ 
0.43056  &   0.16401  &   0.06316  &   50  &   94  &   94 \\ 
0.61008  &   0.61374  &   0.87023  &   50  &   95  &   95 \\ 
0.32467  &   0.16033  &   0.98493  &   50  &   96  &   96 \\ 
0.53509  &   0.53990  &   0.93945  &   50  &   97  &   97 \\ 
0.53490  &   0.58237  &   0.06596  &   50  &   98  &   98 \\ 
0.15668  &   0.13509  &   0.87445  &   50  &   99  &   99 \\ 
0.59489  &   0.47624  &   0.03422  &   50  &   100  &   100 \\ 
\end{tabular}

\bigskip

\hypertarget{s9.1.2}{}
\noindent {\textbf{S9.1.2 triangles}} \\
\noindent $\rho = 0.2225\:\AA^{-2} = 2.121\:\mathrm{ML}$, $E - E_{\mathrm{bulk}} = 198\:\mathrm{meV/at.}$ \\

\begin{tabular}{llllll}
 3.55789 & -0.00806 & 0.00000 &        &         &        \\
-3.56552 &  4.68674 & 0.00000 &        &         &        \\
 0.00000 &  0.00000 & 28.0000 &        &         &        \\
4 atoms  &          &         &        &         &        \\
0        &  0.25    & 0.07803 &   50   &    1    &    1   \\
0        &  0.75    & 0.92197 &   50   &    2    &    2   \\
0        &  0.25    & 0.96739 &   50   &    3    &    3   \\
0        &  0.75    & 0.03261 &   50   &    4    &    4   \\
\end{tabular}

\bigskip

\hypertarget{s9.1.3}{}
\noindent {\textbf{S9.1.3 squares}} \\
\noindent $\rho = 0.2448\:\AA^{-2} = 2.333\:\mathrm{ML}$, $E - E_{\mathrm{bulk}} = 158\:\mathrm{meV/at.}$ \\

\begin{tabular}{llllll}
5.72637 & 0.00000 & 0.00000 &           &           &           \\
0.00000 & 5.72637 & 0.00000 &           &           &           \\
0.00000 & 0.00000 & 20.5771 &           &           &           \\
8 atoms &         &         &           &           &           \\
0       &  0      & 0.56989 &    50     &     1     &     1     \\
0.5     &  0.5    & 0.56989 &    50     &     2     &     2     \\
0       &  0      & 0.41700 &    50     &     3     &     3     \\
0.5     &  0.5    & 0.41700 &    50     &     4     &     4     \\
0       &  0.5    & 0.62798 &    50     &     5     &     5     \\
0.5     &  0      & 0.35890 &    50     &     6     &     6     \\
0       &  0.5    & 0.47756 &    50     &     7     &     7     \\
0.5     &  0      & 0.50932 &    50     &     8     &     8     \\
\end{tabular}

\bigskip

\hypertarget{s9.1.4}{}
\noindent {\textbf{S9.1.4 Sn.equi-II-3}} \\
\noindent $\rho = 0.3428\:\AA^{-2} = 3.268\:\mathrm{ML}$, $E - E_{\mathrm{bulk}} = 142\:\mathrm{meV/at.}$ \\

\begin{tabular}{llllll}
	12.44832  &   -0.00000  &   -0.00008  &     &     &      \\ 
	-6.22416  &   10.78059  &   -0.00011  &     &     &      \\ 
	-0.00022  &   -0.00049  &   34.25425  &     &     &      \\ 
	46 atoms    &     &     &     &     &      \\ 
	0.33338  &   0.05896  &   0.26941  &   50  &   1  &   1 \\ 
	0.94107  &   0.27436  &   0.26942  &   50  &   2  &   2 \\ 
	0.72567  &   0.66667  &   0.26941  &   50  &   3  &   3 \\ 
	0.70092  &   0.03421  &   0.26532  &   50  &   4  &   4 \\
\end{tabular}

\bigskip

\hypertarget{s9.1.5}{}
\noindent {\textbf{S9.1.5 Sn.equi-II-5}} \\
\noindent $\rho = 0.5501\:\AA^{-2} = 5.244\:\mathrm{ML}$, $E - E_{\mathrm{bulk}} = 96\:\mathrm{meV/at.}$ \\

\begin{tabular}{llllll}
	12.46276  &   0.00004  &   0.00003  &     &     &      \\ 
	-6.23135  &   10.79309  &   -0.00004  &     &     &      \\ 
	0.00009  &   -0.00009  &   41.74180  &     &     &      \\ 
	74 atoms    &     &     &     &     &      \\ 
	0.33759  &   0.05312  &   0.21120  &   50  &   1  &   1 \\ 
	0.94690  &   0.28451  &   0.21120  &   50  &   2  &   2 \\ 
	0.71549  &   0.66242  &   0.21120  &   50  &   3  &   3 \\ 
	0.71549  &   0.05310  &   0.21120  &   50  &   4  &   4 \\
\end{tabular}

\begin{tabular}{llllll}
0.66665  &   0.33335  &   0.08417  &   50  &   41  &   41 \\ 
0.33332  &   0.66668  &   0.33051  &   50  &   42  &   42 \\ 
0.46549  &   0.53451  &   0.11781  &   50  &   43  &   43 \\ 
0.46549  &   0.93101  &   0.11781  &   50  &   44  &   44 \\ 
0.06899  &   0.53451  &   0.11781  &   50  &   45  &   45 \\ 
0.13203  &   0.86797  &   0.36675  &   50  &   46  &   46 \\ 
0.13204  &   0.26412  &   0.36675  &   50  &   47  &   47 \\ 
0.73588  &   0.86797  &   0.36675  &   50  &   48  &   48 \\
\end{tabular}

\bigskip

\hypertarget{s9.1.6}{}
\noindent {\textbf{S9.1.6 Sn.equi-II-7.5}} \\
\noindent $\rho = 0.7981\:\AA^{-2} = 7.608\:\mathrm{ML}$, $E - E_{\mathrm{bulk}} = 75\:\mathrm{meV/at.}$ \\

\begin{tabular}{llllll}
	12.50059  &   -0.00001  &   0.00007  &     &     &      \\ 
	-6.25030  &   10.82582  &   -0.00009  &     &     &      \\ 
	0.00033  &   -0.00026  &   52.72611  &     &     &      \\ 
	108 atoms    &     &     &     &     &      \\ 
	0.33844  &   0.05386  &   0.16954  &   50  &   1  &   1 \\ 
	0.94615  &   0.28458  &   0.16954  &   50  &   2  &   2 \\ 
	0.71543  &   0.66156  &   0.16954  &   50  &   3  &   3 \\ 
	0.29003  &   0.95257  &   0.82048  &   50  &   4  &   4 \\ 
	0.66258  &   0.70993  &   0.82048  &   50  &   5  &   5 \\ 
\end{tabular}

\begin{tabular}{llllll}
0.04743  &   0.33741  &   0.82048  &   50  &   6  &   6 \\ 
0.66258  &   0.95257  &   0.82048  &   50  &   7  &   7 \\ 
0.04742  &   0.70996  &   0.82048  &   50  &   8  &   8 \\ 
0.29006  &   0.33742  &   0.82048  &   50  &   9  &   9 \\ 
0.71542  &   0.05385  &   0.16954  &   50  &   10  &   10 \\ 	
0.33844  &   0.28458  &   0.16954  &   50  &   11  &   11 \\ 
0.94614  &   0.66156  &   0.16954  &   50  &   12  &   12 \\ 
0.95126  &   0.27947  &   0.02386  &   50  &   13  &   13 \\
\end{tabular}

\begin{tabular}{llllll}
0.45879  &   0.54120  &   0.85571  &   50  &   56  &   56 \\ 
0.75388  &   0.87693  &   0.05385  &   50  &   57  &   57 \\ 
0.12307  &   0.24612  &   0.05385  &   50  &   58  &   58 \\ 
0.12307  &   0.87693  &   0.05385  &   50  &   59  &   59 \\ 
0.87645  &   0.12355  &   0.94334  &   50  &   60  &   60 \\ 
0.87645  &   0.75286  &   0.94334  &   50  &   61  &   61 \\ 
0.24714  &   0.12355  &   0.94334  &   50  &   62  &   62 \\ 
0.41952  &   0.20977  &   0.25007  &   50  &   63  &   63 \\
\end{tabular}

\begin{tabular}{llllll}
0.49994  &   0.50006  &   0.32131  &   50  &   106  &   106 \\ 
0.00013  &   0.50007  &   0.32132  &   50  &   107  &   107 \\ 
0.49993  &   0.99988  &   0.32132  &   50  &   108  &   108 \\ 
\end{tabular}

\bigskip

\hypertarget{s9.1.7}{}
\noindent {\textbf{S9.1.7 Sn.equi-II-10}} \\
\noindent $\rho = 1.047\:\AA^{-2} = 9.978\:\mathrm{ML}$, $E - E_{\mathrm{bulk}} = 65\:\mathrm{meV/at.}$ \\

\begin{tabular}{llllll}
	12.51618  &   0.00000  &   0.00000  &     &     &      \\ 
	-6.25809  &   10.83933  &   0.00000  &     &     &      \\ 
	0.00000  &   0.00000  &   62.18220  &     &     &      \\ 
	142 atoms    &     &     &     &     &      \\ 
	0.33865  &   0.05400  &   0.14502  &   50  &   1  &   1 \\ 
	0.94600  &   0.28466  &   0.14502  &   50  &   2  &   2 \\ 
	0.71534  &   0.66135  &   0.14502  &   50  &   3  &   3 \\ 
	0.28466  &   0.94600  &   0.85498  &   50  &   4  &   4 \\
\end{tabular}

\begin{tabular}{llllll}
0.27916  &   0.32776  &   0.97854  &   50  &   36  &   36 \\ 
0.54119  &   0.45881  &   0.04431  &   50  &   37  &   37 \\ 
0.54119  &   0.08239  &   0.04431  &   50  &   38  &   38 \\ 
0.91761  &   0.45881  &   0.04431  &   50  &   39  &   39 \\ 
0.08239  &   0.54119  &   0.95569  &   50  &   40  &   40 \\ 
0.45881  &   0.91761  &   0.95569  &   50  &   41  &   41 \\ 
0.45881  &   0.54119  &   0.95569  &   50  &   42  &   42 \\ 
0.20780  &   0.79220  &   0.21105  &   50  &   43  &   43 \\
\end{tabular}

\begin{tabular}{llllll}
0.46511  &   0.93022  &   0.08246  &   50  &   86  &   86 \\ 
0.06978  &   0.53489  &   0.08246  &   50  &   87  &   87 \\ 
0.93022  &   0.46511  &   0.91754  &   50  &   88  &   88 \\ 
0.53489  &   0.06978  &   0.91754  &   50  &   89  &   89 \\ 
0.53489  &   0.46511  &   0.91754  &   50  &   90  &   90 \\ 
0.13189  &   0.86811  &   0.24930  &   50  &   91  &   91 \\ 
0.13189  &   0.26378  &   0.24930  &   50  &   92  &   92 \\ 
0.73622  &   0.86811  &   0.24930  &   50  &   93  &   93 \\
\end{tabular}

\begin{tabular}{llllll}
0.73604  &   0.86802  &   0.74940  &   50  &   136  &   136 \\ 
0.50005  &   0.49995  &   0.72624  &   50  &   137  &   137 \\ 
0.99990  &   0.49995  &   0.72624  &   50  &   138  &   138 \\ 
0.50005  &   0.00010  &   0.72624  &   50  &   139  &   139 \\ 
0.49995  &   0.50005  &   0.27376  &   50  &   140  &   140 \\ 
0.00010  &   0.50005  &   0.27376  &   50  &   141  &   141 \\ 
0.49995  &   0.99990  &   0.27376  &   50  &   142  &   142 \\ 
\end{tabular}


\newpage
\hypertarget{s9.2}{}
\noindent {\large{\textbf{S9.2 Germanium-Slab xyz Files}}} \\
\label{sec:xyz.Ge}


\hypertarget{s9.2.1}{}
\noindent {\textbf{S9.2.1 Ge.db-monolayer (oP10)}} \\
\noindent $\rho = 0.1884\:\AA^{-2} = 1.359\:\mathrm{ML}$, $E - E_{\mathrm{bulk}} = 308\:\mathrm{meV/at.}$ \\

\begin{tabular}{llllll}
	7.84310  &   0.00000  &   0.00000  &     &     &      \\ 
	0.00000  &   38.86961  &   0.00000  &     &     &      \\ 
	0.00000  &   0.00000  &   6.76570  &     &     &      \\ 
	10 atoms    &     &     &     &     &      \\ 
	0.02034  &   0.00000  &   0.81941  &   32  &   1  &   1 \\ 
	0.47966  &   0.00000  &   0.81941  &   32  &   2  &   2 \\ 
	0.52034  &   0.00000  &   0.18059  &   32  &   3  &   3 \\ 
	0.97966  &   0.00000  &   0.18059  &   32  &   4  &   4 \\ 
	0.25000  &   0.03815  &   0.67495  &   32  &   5  &   5 \\ 
	0.25000  &   0.96185  &   0.67495  &   32  &   6  &   6 \\ 
	0.75000  &   0.96185  &   0.32505  &   32  &   7  &   7 \\ 
	0.75000  &   0.03815  &   0.32505  &   32  &   8  &   8 \\ 
	0.25000  &   0.00000  &   0.37240  &   32  &   9  &   9 \\ 
	0.75000  &   0.00000  &   0.62760  &   32  &   10  &   10 \\ 
\end{tabular}

\bigskip

\hypertarget{s9.2.2}{}
\noindent {\textbf{S9.2.2 Ge.iso-II-2.5}} \\
\noindent $\rho = 0.3534\:\AA^{-2} = 2.548\:\mathrm{ML}$, $E - E_{\mathrm{bulk}} = 251\:\mathrm{meV/at.}$ \\

\begin{tabular}{llllll}
	15.67423  &   0.00000  &   0.00000  &     &     &      \\ 
	0.00000  &   11.19385  &   0.00000  &     &     &      \\ 
	0.00000  &   0.00000  &   28.11850  &     &     &      \\ 
	62 atoms    &     &     &     &     &      \\ 
	0.00001  &   0.00000  &   0.00000  &   32  &   1  &   1 \\ 
	0.36467  &   0.00000  &   0.10991  &   32  &   2  &   2 \\ 
	0.36467  &   0.00000  &   0.89009  &   32  &   3  &   3 \\ 
	0.50130  &   0.00000  &   0.15808  &   32  &   4  &   4 \\ 
	0.50130  &   0.00000  &   0.84192  &   32  &   5  &   5 \\ 
	0.76010  &   0.00000  &   0.04475  &   32  &   6  &   6 \\ 
	0.76010  &   0.00000  &   0.95525  &   32  &   7  &   7 \\ 
	0.91225  &   0.00000  &   0.07406  &   32  &   8  &   8 \\ 
	0.91225  &   0.00000  &   0.92595  &   32  &   9  &   9 \\ 
	0.49329  &   0.00000  &   0.05356  &   32  &   10  &   10 \\ 
	0.49329  &   0.00000  &   0.94644  &   32  &   11  &   11 \\ 
	0.24297  &   0.11297  &   0.00000  &   32  &   12  &   12 \\ 
	0.24297  &   0.88703  &   0.00000  &   32  &   13  &   13 \\ 
	0.55267  &   0.16644  &   0.10305  &   32  &   14  &   14 \\ 
	0.55267  &   0.83356  &   0.89694  &   32  &   15  &   15 \\ 
	0.55267  &   0.16644  &   0.89694  &   32  &   16  &   16 \\ 
	0.55267  &   0.83356  &   0.10305  &   32  &   17  &   17 \\ 
	0.30975  &   0.18824  &   0.07465  &   32  &   18  &   18 \\ 
	0.30975  &   0.81176  &   0.92535  &   32  &   19  &   19 \\ 
	0.30975  &   0.18824  &   0.92535  &   32  &   20  &   20 \\ 
\end{tabular}

\begin{tabular}{llllll}
0.30975  &   0.81176  &   0.07465  &   32  &   21  &   21 \\ 
0.95601  &   0.17575  &   0.12814  &   32  &   22  &   22 \\ 
0.95601  &   0.82425  &   0.87186  &   32  &   23  &   23 \\ 
0.95601  &   0.17575  &   0.87186  &   32  &   24  &   24 \\ 
0.95601  &   0.82425  &   0.12814  &   32  &   25  &   25 \\ 
0.08999  &   0.18182  &   0.00000  &   32  &   26  &   26 \\ 
0.08999  &   0.81818  &   0.00000  &   32  &   27  &   27 \\ 
0.69977  &   0.19543  &   0.07184  &   32  &   28  &   28 \\
\end{tabular}

\bigskip

\hypertarget{s9.2.3}{}
\noindent {\textbf{S9.2.3 Ge.equi-II-3}} \\
\noindent $\rho = 0.4037\:\AA^{-2} = 2.911\:\mathrm{ML}$, $E - E_{\mathrm{bulk}} = 222\:\mathrm{meV/at.}$ \\

\begin{tabular}{llllll}
10.88372  &   -0.00000  &   0.00204  &     &     &      \\ 
-5.44186  &   9.33195  &   -0.00089  &     &     &      \\ 
0.00667  &   0.00048  &   34.96546  &     &     &      \\ 
41 atoms    &     &     &     &     &      \\ 
0.33470  &   0.04609  &   0.25061  &   50  &   1  &   1 \\ 
0.95527  &   0.28723  &   0.25061  &   50  &   2  &   2 \\ 
0.72578  &   0.66666  &   0.24872  &   50  &   3  &   3 \\ 
0.71201  &   0.04620  &   0.25327  &   50  &   4  &   4 \\
\end{tabular}

\bigskip

\hypertarget{s9.2.4}{}
\noindent {\textbf{S9.2.4 Ge.equi-II-3.5}} \\
\noindent $\rho = 0.4781\:\AA^{-2} = 3.448\:\mathrm{ML}$, $E - E_{\mathrm{bulk}} = 183\:\mathrm{meV/at.}$

\begin{tabular}{llllll}
	10.71370  &   -0.01100  &   -0.00829  &     &     &      \\ 
	-5.49272  &   9.36400  &   -0.25389  &     &     &      \\ 
	-0.21007  &   -0.40637  &   30.60224  &     &     &      \\ 
	48 atoms    &     &     &     &     &      \\ 
	0.03938  &   0.81651  &   0.19581  &   32  &   1  &   1 \\ 
	0.04665  &   0.82625  &   0.27601  &   32  &   2  &   2 \\ 
	0.11219  &   0.35595  &   0.23590  &   32  &   3  &   3 \\ 
	0.13361  &   0.25074  &   0.05846  &   32  &   4  &   4 \\
\end{tabular}

\begin{tabular}{llllll}
0.57690  &   0.89760  &   0.16922  &   32  &   45  &   45 \\ 
0.59430  &   0.91642  &   0.30314  &   32  &   46  &   46 \\ 
0.36517  &   0.44353  &   0.12499  &   32  &   47  &   47 \\ 
0.38523  &   0.53138  &   0.34682  &   32  &   48  &   48 \\ 
\end{tabular}

\bigskip

\hypertarget{s9.2.5}{}
\noindent {\textbf{S9.2.5 Ge.iso-II-4}} \\
\noindent $\rho = 0.5461\:\AA^{-2} = 3.939\:\mathrm{ML}$, $E - E_{\mathrm{bulk}} = 176\:\mathrm{meV/at.}$ \\

\begin{tabular}{llllll}
	-15.82209  &   15.82209  &   0.00000  &     &     &      \\ 
	-0.00007  &   -0.00007  &   15.71212  &     &     &      \\ 
	47.38157  &   47.38157  &   -0.00043  &     &     &      \\ 
	192 atoms    &     &     &     &     &      \\ 
	0.00000  &   0.88780  &   0.12974  &   32  &   1  &   1 \\ 
	0.00000  &   0.50254  &   0.93508  &   32  &   2  &   2 \\ 
	0.94334  &   0.00665  &   0.08217  &   32  &   3  &   3 \\ 
	0.94333  &   0.24350  &   0.00107  &   32  &   4  &   4 \\
\end{tabular}

\begin{tabular}{llllll}
0.75000  &   0.99488  &   0.02032  &   32  &   36  &   36 \\ 
0.75000  &   0.99855  &   0.14880  &   32  &   37  &   37 \\ 
0.74999  &   0.84154  &   0.97019  &   32  &   38  &   38 \\ 
0.75000  &   0.84207  &   0.03112  &   32  &   39  &   39 \\ 
0.75000  &   0.86023  &   0.12929  &   32  &   40  &   40 \\ 
0.75000  &   0.75066  &   0.00081  &   32  &   41  &   41 \\ 
0.75000  &   0.49909  &   0.08258  &   32  &   42  &   42 \\ 
0.75000  &   0.25199  &   0.93431  &   32  &   43  &   43 \\
\end{tabular}

\begin{tabular}{llllll}
0.50000  &   0.24902  &   0.08172  &   32  &   86  &   86 \\ 
0.50001  &   0.36206  &   0.95334  &   32  &   87  &   87 \\ 
0.50000  &   0.34292  &   0.05197  &   32  &   88  &   88 \\ 
0.50000  &   0.49621  &   0.06306  &   32  &   89  &   89 \\ 
0.50000  &   0.49435  &   0.10075  &   32  &   90  &   90 \\ 
0.99998  &   0.90791  &   0.97172  &   32  &   91  &   91 \\ 
0.00000  &   0.90693  &   0.03132  &   32  &   92  &   92 \\ 
0.99999  &   0.75551  &   0.98260  &   32  &   93  &   93 \\
\end{tabular}

\begin{tabular}{llllll}
0.25000  &   0.99488  &   0.02032  &   32  &   136  &   136 \\ 
0.25000  &   0.99855  &   0.14880  &   32  &   137  &   137 \\ 
0.24999  &   0.84154  &   0.97019  &   32  &   138  &   138 \\ 
0.25000  &   0.84207  &   0.03112  &   32  &   139  &   139 \\ 
0.25000  &   0.86023  &   0.12929  &   32  &   140  &   140 \\ 
0.25000  &   0.75066  &   0.00081  &   32  &   141  &   141 \\ 
0.25000  &   0.49909  &   0.08258  &   32  &   142  &   142 \\ 
0.25000  &   0.25199  &   0.93431  &   32  &   143  &   143 \\
\end{tabular}

\begin{tabular}{llllll}
0.00001  &   0.48841  &   0.97807  &   32  &   185  &   185 \\ 
0.75000  &   0.26178  &   0.97745  &   32  &   186  &   186 \\ 
0.25000  &   0.26178  &   0.97745  &   32  &   187  &   187 \\ 
0.50001  &   0.48841  &   0.97807  &   32  &   188  &   188 \\ 
0.75000  &   0.98864  &   0.10562  &   32  &   189  &   189 \\ 
1.00000  &   0.76126  &   0.10506  &   32  &   190  &   190 \\ 
0.25000  &   0.98864  &   0.10562  &   32  &   191  &   191 \\ 
0.50000  &   0.76126  &   0.10506  &   32  &   192  &   192 \\ 
\end{tabular}

\bigskip

\hypertarget{s9.2.6}{}
\noindent {\textbf{S9.2.6 Ge.equi-II-6}} \\
\noindent $\rho = 0.8042\:\AA^{-2} = 5.800\:\mathrm{ML}$, $E - E_{\mathrm{bulk}} = 120\:\mathrm{meV/at.}$ \\

\begin{tabular}{llllll}
	10.86109  &   -0.01836  &   -0.00140  &     &     &      \\ 
	-5.44644  &   9.39675  &   -0.00201  &     &     &      \\ 
	-0.00506  &   -0.01138  &   38.26388  &     &     &      \\ 
	82 atoms    &     &     &     &     &      \\ 
	0.34308  &   0.99782  &   0.20027  &   32  &   1  &   1 \\ 
	0.95338  &   0.22993  &   0.20030  &   32  &   2  &   2 \\ 
	0.72149  &   0.60798  &   0.20028  &   32  &   3  &   3 \\ 
	0.72123  &   0.99788  &   0.20033  &   32  &   4  &   4 \\
\end{tabular}

\begin{tabular}{llllll}
0.92028  &   0.40228  &   0.16462  &   32  &   33  &   33 \\ 
0.75975  &   0.82036  &   0.06012  &   32  &   34  &   34 \\ 
0.12706  &   0.18979  &   0.06017  &   32  &   35  &   35 \\ 
0.13048  &   0.82432  &   0.05992  &   32  &   36  &   36 \\ 
0.42506  &   0.15537  &   0.29660  &   32  &   37  &   37 \\ 
0.79642  &   0.52673  &   0.29667  &   32  &   38  &   38 \\ 
0.79584  &   0.15530  &   0.29662  &   32  &   39  &   39 \\ 
0.00578  &   0.94493  &   0.14701  &   32  &   40  &   40 \\
\end{tabular}


\newpage
\hypertarget{s9.3}{}
\noindent {\large{\textbf{S9.3 Silicon-Slab xyz Files}}} \\
\label{sec:xyz.Si}


\hypertarget{s9.3.1}{}
\noindent {\textbf{S9.3.1 Si.db-monolayer (oP10)}} \\
\noindent $\rho = 0.2110\:\AA^{-2} = 1.364\:\mathrm{ML}$, $E - E_{\mathrm{bulk}} = 430\:\mathrm{meV/at.}$ \\

\begin{tabular}{llllll}
	7.40937  &   0.00000  &   0.00000  &     &     &      \\ 
	0.00000  &   35.75410  &   0.00000  &     &     &      \\ 
	0.00000  &   0.00000  &   6.39543  &     &     &      \\ 
	10 atoms    &     &     &     &     &      \\ 
	0.01913  &   0.00000  &   0.81934  &   14  &   1  &   1 \\ 
	0.48087  &   0.00000  &   0.81934  &   14  &   2  &   2 \\ 
	0.51913  &   0.00000  &   0.18066  &   14  &   3  &   3 \\ 
	0.98087  &   0.00000  &   0.18066  &   14  &   4  &   4 \\ 
	0.25000  &   0.03788  &   0.67701  &   14  &   5  &   5 \\ 
	0.25000  &   0.96212  &   0.67701  &   14  &   6  &   6 \\ 
	0.75000  &   0.96212  &   0.32299  &   14  &   7  &   7 \\ 
	0.75000  &   0.03788  &   0.32299  &   14  &   8  &   8 \\ 
	0.25000  &   0.00000  &   0.37407  &   14  &   9  &   9 \\ 
	0.75000  &   0.00000  &   0.62593  &   14  &   10  &   10 \\ 
\end{tabular}

\bigskip

\hypertarget{s9.3.2}{}
\noindent {\textbf{S9.3.2 Si.iso-II-2.5}} \\
\noindent $\rho = 0.3851\:\AA^{-2} = 2.489\:\mathrm{ML}$, $E - E_{\mathrm{bulk}} = 322\:\mathrm{meV/at.}$ \\

\begin{tabular}{llllll}
	15.01360  &   0.00000  &   0.00000  &     &     &      \\ 
	0.00000  &   10.72374  &   0.00000  &     &     &      \\ 
	0.00000  &   0.00000  &   24.81168  &     &     &      \\ 
	62 atoms    &     &     &     &     &      \\ 
	0.00530  &   0.00000  &   0.00000  &   14  &   1  &   1 \\ 
	0.36693  &   0.00000  &   0.11410  &   14  &   2  &   2 \\ 
	0.36693  &   0.00000  &   0.88590  &   14  &   3  &   3 \\ 
	0.50034  &   0.00000  &   0.16477  &   14  &   4  &   4 \\ 
	0.50034  &   0.00000  &   0.83523  &   14  &   5  &   5 \\ 
	0.76091  &   0.00000  &   0.04838  &   14  &   6  &   6 \\ 
	0.76091  &   0.00000  &   0.95162  &   14  &   7  &   7 \\ 
	0.91223  &   0.00000  &   0.07669  &   14  &   8  &   8 \\ 
	0.91223  &   0.00000  &   0.92331  &   14  &   9  &   9 \\ 
	0.49346  &   0.00000  &   0.05520  &   14  &   10  &   10 \\ 
	0.49346  &   0.00000  &   0.94480  &   14  &   11  &   11 \\ 
	0.24599  &   0.11233  &   0.00000  &   14  &   12  &   12 \\ 
	0.24599  &   0.88767  &   0.00000  &   14  &   13  &   13 \\ 
	0.55251  &   0.16366  &   0.10837  &   14  &   14  &   14 \\ 
	0.55251  &   0.83634  &   0.89163  &   14  &   15  &   15 \\ 
	0.55251  &   0.16366  &   0.89163  &   14  &   16  &   16 \\ 
	0.55251  &   0.83634  &   0.10837  &   14  &   17  &   17 \\ 
	0.31244  &   0.18904  &   0.07990  &   14  &   18  &   18 \\ 
	0.31244  &   0.81096  &   0.92010  &   14  &   19  &   19 \\ 
	0.31244  &   0.18904  &   0.92010  &   14  &   20  &   20 \\ 
\end{tabular}

\begin{tabular}{llllll}
0.31244  &   0.81096  &   0.07990  &   14  &   21  &   21 \\ 
0.94147  &   0.19531  &   0.11305  &   14  &   22  &   22 \\ 
0.94147  &   0.80469  &   0.88695  &   14  &   23  &   23 \\ 
0.94147  &   0.19531  &   0.88695  &   14  &   24  &   24 \\ 
0.94147  &   0.80469  &   0.11305  &   14  &   25  &   25 \\ 
0.09512  &   0.18183  &   0.00000  &   14  &   26  &   26 \\ 
0.09512  &   0.81817  &   0.00000  &   14  &   27  &   27 \\ 
0.69757  &   0.19292  &   0.07633  &   14  &   28  &   28 \\
\end{tabular}

\bigskip

\hypertarget{s9.3.3}{}
\noindent {\textbf{S9.3.3 Si.equi-II-3}} \\
\noindent $\rho = 0.5032\:\AA^{-2} = 3.253\:\mathrm{ML}$, $E - E_{\mathrm{bulk}} = 280\:\mathrm{meV/at.}$ \\

\begin{tabular}{llllll}
	10.27433  &   -0.00034  &   0.00000  &     &     &      \\ 
	-5.13746  &   8.89766  &   -0.00000  &     &     &      \\ 
	0.00001  &   -0.00000  &   26.09222  &     &     &      \\ 
	46 atoms    &     &     &     &     &      \\ 
	0.33334  &   0.04309  &   0.27406  &   14  &   1  &   1 \\ 
	0.95693  &   0.29027  &   0.27407  &   14  &   2  &   2 \\ 
	0.70976  &   0.66667  &   0.27406  &   14  &   3  &   3 \\ 
	0.71884  &   0.05218  &   0.27250  &   14  &   4  &   4 \\
\end{tabular}

\bigskip

\hypertarget{s9.3.4}{}
\noindent {\textbf{S9.3.4 Si.iso-II-4}} \\
\noindent $\rho = 0.6056\:\AA^{-2} = 3.915\:\mathrm{ML}$, $E - E_{\mathrm{bulk}} = 232\:\mathrm{meV/at.}$ \\

\begin{tabular}{llllll}
	10.62542  &   0.00000  &   0.00000  &     &     &      \\ 
	0.00000  &   14.91779  &   0.00000  &     &     &      \\ 
	0.00000  &   0.00000  &   34.35054  &     &     &      \\ 
	96 atoms    &     &     &     &     &      \\ 
	0.25000  &   0.51535  &   0.23141  &   14  &   1  &   1 \\ 
	0.75000  &   0.48465  &   0.23141  &   14  &   2  &   2 \\ 
	0.25000  &   0.11281  &   0.95396  &   14  &   3  &   3 \\ 
	0.75000  &   0.88719  &   0.95396  &   14  &   4  &   4 \\
\end{tabular}

\begin{tabular}{llllll}
0.43809  &   0.93304  &   0.05602  &   14  &   41  &   41 \\ 
0.56191  &   0.06696  &   0.05602  &   14  &   42  &   42 \\ 
0.05523  &   0.93739  &   0.22893  &   14  &   43  &   43 \\ 
0.94477  &   0.06261  &   0.22893  &   14  &   44  &   44 \\ 
0.44477  &   0.93739  &   0.22893  &   14  &   45  &   45 \\ 	
0.55523  &   0.06261  &   0.22893  &   14  &   46  &   46 \\ 
0.05633  &   0.31849  &   0.94166  &   14  &   47  &   47 \\ 
0.94367  &   0.68151  &   0.94166  &   14  &   48  &   48 \\
\end{tabular}

\begin{tabular}{llllll}
0.75000  &   0.88113  &   0.18105  &   14  &   90  &   90 \\ 
0.25000  &   0.98466  &   0.91187  &   14  &   91  &   91 \\ 
0.75000  &   0.01534  &   0.91187  &   14  &   92  &   92 \\ 
0.25000  &   0.96752  &   0.09113  &   14  &   93  &   93 \\ 
0.75000  &   0.03248  &   0.09113  &   14  &   94  &   94 \\ 
0.25000  &   0.96728  &   0.20182  &   14  &   95  &   95 \\ 
0.75000  &   0.03272  &   0.20182  &   14  &   96  &   96 \\ 
\end{tabular}

\bigskip

\hypertarget{s9.3.5}{}
\noindent {\textbf{S9.3.5 Si.iso-II-5}} \\
\noindent $\rho = 0.8272\:\AA^{-2} = 5.347\:\mathrm{ML}$, $E - E_{\mathrm{bulk}} = 186\:\mathrm{meV/at.}$ \\

\begin{tabular}{llllll}
	10.57224  &   0.00000  &   0.00000  &     &     &      \\ 
	0.00000  &   14.86573  &   0.00000  &     &     &      \\ 
	0.00000  &   0.00000  &   43.59367  &     &     &      \\ 
	130 atoms    &     &     &     &     &      \\ 
	0.25000  &   0.62159  &   0.21259  &   14  &   1  &   1 \\ 
	0.25000  &   0.62159  &   0.78741  &   14  &   2  &   2 \\ 
	0.75000  &   0.37841  &   0.78741  &   14  &   3  &   3 \\ 
	0.75000  &   0.37841  &   0.21259  &   14  &   4  &   4 \\
\end{tabular}

\begin{tabular}{llllll}
0.41735  &   0.67456  &   0.81844  &   14  &   33  &   33 \\ 
0.08265  &   0.67456  &   0.18156  &   14  &   34  &   34 \\
0.91735  &   0.32544  &   0.18156  &   14  &   35  &   35 \\ 
0.58265  &   0.32544  &   0.81844  &   14  &   36  &   36 \\ 
0.56182  &   0.56654  &   0.16429  &   14  &   37  &   37 \\ 
0.93818  &   0.56654  &   0.83571  &   14  &   38  &   38 \\ 
0.06182  &   0.43346  &   0.83571  &   14  &   39  &   39 \\ 
0.43818  &   0.43346  &   0.16429  &   14  &   40  &   40 \\
\end{tabular}

\begin{tabular}{llllll}
0.05344  &   0.81791  &   0.92634  &   14  &   83  &   83 \\ 
0.44656  &   0.81791  &   0.07366  &   14  &   84  &   84 \\ 
0.44656  &   0.81791  &   0.92634  &   14  &   85  &   85 \\ 
0.05344  &   0.81791  &   0.07366  &   14  &   86  &   86 \\ 
0.94656  &   0.18209  &   0.07366  &   14  &   87  &   87 \\ 
0.55344  &   0.18209  &   0.92634  &   14  &   88  &   88 \\ 
0.55322  &   0.31790  &   0.04458  &   14  &   89  &   89 \\ 
0.94678  &   0.31790  &   0.95542  &   14  &   90  &   90 \\
\end{tabular}


\newpage
\hypertarget{s9.4}{}
\noindent {\large{\textbf{S9.4 Ge-Clathrate/InN}}} \\
\label{sec:xyz.Ge.InN}

\hypertarget{s9.4.1}{}
\noindent {\textbf{S9.4.1 Thinner Ge Clathrate Film on InN (initial configuration)}} \\
\noindent See Figure \hyperlink{top2}{S6.2}. \\

\begin{tabular}{llllll}
	10.89643  &   0.00000  &   -0.00000  &     &     &      \\ 
	-5.44821  &   9.43658  &   0.00000  &     &     &      \\ 
	0.00000  &   0.00000  &   26.29850  &     &     &      \\ 
	95 atoms    &     &     &     &     &      \\ 
	0.26262  &   0.13131  &   0.17022  &   32  &   1  &   1 \\ 
	0.86869  &   0.13131  &   0.17022  &   32  &   2  &   2 \\ 
	0.86869  &   0.73738  &   0.17022  &   32  &   3  &   3 \\ 
	0.66667  &   0.33333  &   0.21704  &   32  &   4  &   4 \\
\end{tabular}

\begin{tabular}{llllll} 
-0.03003  &   0.76052  &   0.35200  &   32  &   39  &   39 \\ 
-0.03003  &   0.23956  &   0.35200  &   32  &   40  &   40 \\ 
0.00000  &   0.00000  &   0.30200  &   32  &   41  &   41 \\ 
0.66667  &   0.33333  &   0.73784  &   49  &   42  &   42 \\  
0.33333  &   0.66667  &   0.73561  &   49  &   43  &   43 \\ 
0.00000  &   0.00000  &   0.73969  &   49  &   44  &   44 \\
0.00385  &   0.33274  &   0.72373  &   49  &   45  &   45 \\ 
0.00385  &   0.67111  &   0.72373  &   49  &   46  &   46 \\
\end{tabular}

\begin{tabular}{llllll}
0.33884  &   0.33860  &   0.97522  &   7  &   88  &   88 \\ 
0.66140  &   0.00024  &   0.97522  &   7  &   89  &   89 \\ 
0.66140  &   0.66116  &   0.97522  &   7  &   90  &   90 \\ 
0.99976  &   0.33860  &   0.97522  &   7  &   91  &   91 \\ 
0.99976  &   0.66116  &   0.97522  &   7  &   92  &   92 \\ 
0.00000  &   0.00000  &   0.98093  &   7  &   93  &   93 \\ 
0.66667  &   0.33333  &   0.97516  &   7  &   94  &   94 \\ 
0.33333  &   0.66667  &   0.98001  &   7  &   95  &   95 \\ 
\end{tabular}

\bigskip

\hypertarget{s9.4.2}{}
\noindent {\textbf{S9.4.2 Thicker Ge Clathrate Film on InN (initial configuration)}} \\
\noindent See Figure \hyperlink{top6}{S6.6}. \\

\begin{tabular}{llllll}
	10.89643  &   0.00000  &   -0.00000  &     &     &      \\ 
	-5.44821  &   9.43658  &   0.00000  &     &     &      \\ 
	0.00000  &   0.00000  &   50.00000  &     &     &      \\ 
	129 atoms    &     &     &     &     &      \\ 
	1.00700  &   1.00700  &   0.36991  &   32  &   1  &   1 \\ 
	0.34644  &   0.06187  &   0.16999  &   32  &   2  &   2 \\ 
	-0.04787  &   0.29157  &   0.16999  &   32  &   3  &   3 \\ 
	-0.27757  &   -0.33244  &   0.16999  &   32  &   4  &   4 \\
\end{tabular}

\begin{tabular}{llllll}
	-0.32633  &   0.34033  &   0.12403  &   32  &   31  &   31 \\ 
-0.44907  &   0.46307  &   0.14127  &   32  &   32  &   32 \\ 
-0.44907  &   0.09486  &   0.14127  &   32  &   33  &   33 \\ 
-0.08086  &   0.46307  &   0.14127  &   32  &   34  &   34 \\ 
-0.23847  &   -0.11574  &   0.05998  &   32  &   35  &   35 \\ 
0.12974  &   0.25247  &   0.05998  &   32  &   36  &   36 \\ 
0.12974  &   -0.11574  &   0.05998  &   32  &   37  &   37 \\ 
0.42819  &   0.21760  &   0.24723  &   32  &   38  &   38 \\
\end{tabular}

\begin{tabular}{llllll}
0.32889  &   0.33274  &   -0.14531  &   49  &   81  &   81 \\ 
0.32889  &   0.99615  &   -0.14531  &   49  &   82  &   82 \\ 
0.66726  &   0.67111  &   -0.14531  &   49  &   83  &   83 \\ 
0.66726  &   0.99615  &   -0.14531  &   49  &   84  &   84 \\ 
0.43857  &   0.21928  &   -0.08586  &   49  &   85  &   85 \\ 
0.78072  &   0.21928  &   -0.08586  &   49  &   86  &   86 \\ 
0.78072  &   0.56143  &   -0.08586  &   49  &   87  &   87 \\ 
0.11075  &   0.22151  &   -0.08291  &   49  &   88  &   88 \\
\end{tabular}


\newpage
\hypertarget{s10}{}
\noindent {\Large{\textbf{S10. Tabulated $\bm{E(\rho)}$ Dependencies}}} \\
\label{sec:e-rho}

\hypertarget{s10.1}{}
\noindent {\large{\textbf{S10.1 $\bm{E(\rho)}$ of Sn}}} \\
\label{sec:e-rho-Sn}

\begin{table}[h!]
	\centering
	\begin{tabular}{c|c|c|c|c|l}
		$\rho$ &  $\rho$ & $E_{\mathrm{KS}}$ & $E - E_{\mathrm{bulk}}$ &  $E_{\mathrm{g}}$ & sample \\
		$[\mathrm{at./\AA^{2}}]$ &  $[\mathrm{ML}]$ & $[\mathrm{eV/at.}]$ & $[\mathrm{meV/at.}]$ & $[\mathrm{eV}]$ & designation \\
		\hline
		\hline
		0.1422 &   1.356      &  -3.6043   &   265    &  0.038  & db-monolayer (oP10)                      \\
		\hline
		0.1741 &   1.660      &  -3.6159   &   253    &  0.000  & db-monolayer (hP11)                      \\
		\hline
		0.1425 &   1.358      &  -3.6268   &   242    &  0.065  & Sn9-web (40 atoms)                       \\  
		\hline       
		0.1749 &   1.667      &  -3.6530   &   216    &  0.000  & Sn9-web (100 atoms)                      \\  
		\hline
		0.2224 &   2.120      &  -3.6707   &   198    &  0.000  & triangles (bilayer)                      \\ 
		\hline        
		0.4449 &   4.241      &  -3.7092   &   160    &  0.000  & triangles ($2\times$bilayer)             \\
\end{tabular}
\end{table}

\vspace{-0.35cm}
\noindent\label{erho-Sn}Table S10.1: {\small{\textit{Tabulated content of Figure~6 (main text) for Sn thin films. ``KS'' denotes total Kohn-Sham energy as a direct output of VASP (without dispersion vdW correction). For~unit conversion: thickness of $1\:\mathrm{ML}$ of~(111)-diamond cF8 Sn is $\rho_{0} = 0.10490\:\mathrm{at./\AA^{2}}$ and the respective bulk Kohn-Sham per-atom energy $E_{\mathrm{bulk}} = -3.8688\:\mathrm{eV/at.}$ $E_{\mathrm{g}}$ stands for the width of electronic band gap.}}} \\

\newpage

\hypertarget{s10.2}{}
\noindent {\large{\textbf{S10.2 $\bm{E(\rho)}$ of Ge}}} \\
\label{sec:e-rho-Ge}

\begin{table}[h!]
	\centering
	\begin{tabular}{c|c|c|c|c|l}
		$\rho$ &  $\rho$ & $E_{\mathrm{KS}}$ & $E - E_{\mathrm{bulk}}$ & $E_{\mathrm{g}}$ & sample \\
		$[\mathrm{at./\AA^{2}}]$ &  $[\mathrm{ML}]$ & $[\mathrm{eV/at.}]$ & $[\mathrm{meV/at.}]$ & $[\mathrm{eV}]$ & designation \\
		\hline
		\hline
	    0.1884 &    1.359     &  -4.2083   &   308    &  0.100  & db-monolayer (oP10)                      \\
	    \hline       
       	0.2367 &    1.707     &  -4.1636   &   352    &  0.000  & GS-bilayer                               \\
       	\hline
	    0.3534 &    2.549     &  -4.2648   &   251    &  0.256  & Ge.iso-II-2.5                            \\  
	    \hline       
	    0.4037 &    2.911     &  -4.2941   &   222    &  0.400  & Ge.equi-II-3                             \\ 
	    \hline        
	    0.4781 &    3.448     &  -4.3328   &   183    &  0.428  & Ge.equi-II-3.5                           \\ 
	    \hline         
	    0.5461 &    3.938     &  -4.3404   &   176    &  0.217  & Ge.iso-II-4                              \\
	    \hline       
	    0.8042 &    5.800     &  -4.3957   &   120    &  0.517  & Ge.equi-II-6                             \\
	    \hline       
	    0.3455 &    2.492     &  -4.2535   &   262    &  0.000  & ($\sqrt{3}\times\sqrt{3}$)-dia (bilayer) \\
	    \hline       
	    0.3331 &    2.402     &  -4.2252   &   291    &  0.133  & ($2\times2$)-dia (bilayer)               \\ 
	    \hline        
	    0.4683 &    3.377     &  -4.2950   &   221    &  0.000  & ($2\times2$)-dia (trilayer)              \\ 
	    \hline        
	    0.6059 &    4.370     &  -4.3404   &   176    &  0.000  & ($2\times2$)-dia (fourlayer)             \\ 
	    \hline         
	    0.7440 &    5.366     &  -4.3736   &   142    &  0.000  & ($2\times2$)-dia (fivelayer)             \\
	    \hline        
	    0.8822 &    6.362     &  -4.3948   &   121    &  0.000  & ($2\times2$)-dia (sixlayer)              \\ 
	    \hline        
	    1.0208 &    7.362     &  -4.4118   &   104    &  0.022  & ($2\times2$)-dia (sevenlayer)            \\ 
	    \hline         
	    1.1585 &    8.355     &  -4.4245   &   91     &  0.057  & ($2\times2$)-dia (eightlayer)            \\
	    \hline         
	    1.2982 &    9.362     &  -4.4343   &   82     &  0.108  & ($2\times2$)-dia (ninelayer)             \\   
	\end{tabular}
\end{table}

\vspace{-0.35cm}
\noindent\label{erho-Ge-1}Table S10.2.1: {\small{\textit{Tabulated content of Figure~6 (main text) for Ge thin films. ``KS'' denotes total Kohn-Sham energy as a direct output of VASP (without dispersion vdW correction). For~unit conversion: thickness of $1\:\mathrm{ML}$ of~(111)-diamond cF8 Ge is $\rho_{0} = 0.13866\:\mathrm{at./\AA^{2}}$ and the respective bulk Kohn-Sham per-atom energy $E_{\mathrm{bulk}} = -4.5159\:\mathrm{eV/at.}$ $E_{\mathrm{g}}$ stands for the width of electronic band gap.}}} \\

\medskip

\begin{table}[h!]
	\centering
	\begin{tabular}{c|c|c|c|c|l}
		$\rho$ &  $\rho$ & $E_{\mathrm{KS}}$ & $E - E_{\mathrm{bulk}}$ &  $E_{\mathrm{g}}$ & sample\\
		$[\mathrm{at./\AA^{2}}]$ &  $[\mathrm{ML}]$ & $[\mathrm{eV/at.}]$ & $[\mathrm{meV/at.}]$ &  $[\mathrm{eV}]$ & designation \\
		\hline
		\hline
		0.1924 &    1.361     &   -4.3810   &   389    &  0.000  &  db-monolayer (oP10)                          \\
		\hline
		0.3591 &    2.540     &   -4.4461   &   323    &  0.000  &  Ge.iso-II-2.5                                \\  
		\hline       
	 	0.5612 &    3.970     &   -4.5343   &   235    &  0.255  &  Ge.iso-II-4                                  \\ 
		\hline        
		0.7908 &    5.595     &   -4.5855   &   184    &  0.435  &  Ge.iso-II-5.5                                \\ 
		\hline         
		0.4107 &    2.905     &   -4.4844   &   285    &  0.455  &  Ge.equi-II-3                                 \\
	    \hline         
		0.4912 &    3.475     &   -4.5313   &   238    &  0.490  &  Ge.equi-II-3.5                               \\    
		\hline       
		0.8137 &    5.756     &   -4.5952   &   174    &  0.450  &  Ge.equi-II-6                                 \\
		\hline       
		0.3510 &    2.483     &   -4.4444   &   325    &  0.000  &  ($\sqrt{3}\times\sqrt{3}$)-dia (bilayer)     \\
		\hline       
		0.4771 &    3.374     &   -4.5069   &   263    &  0.000  &  ($2\times2$)-dia (trilayer)    \\ 
		\hline        
	 	0.6320 &    4.471     &   -4.5609   &   209    &  0.000  &  ($\sqrt{3}\times\sqrt{3}$)-dia (fourlayer)   \\ 
		\hline        
		0.7712 &    5.456     &   -4.5959   &   174    &  0.000  &  ($\sqrt{3}\times\sqrt{3}$)-dia (fivelayer)   \\ 
		\hline         
	 	1.0492 &    7.422     &   -4.6433   &   126    &  0.000  &  ($\sqrt{3}\times\sqrt{3}$)-dia (sevenlayer)  \\ 
	 	\hline         
	 	0.4276 &    3.025     &   -4.4632   &   307    &  0.000  &  ($3\times3$)-DAS (trilayer)                  \\ 
	 	\hline         
	 	0.5650 &    3.997     &   -4.5296   &   240    &  0.000  &  ($3\times3$)-DAS (fourlayer)                 \\ 
	\end{tabular}
\end{table}

\vspace{-0.35cm}
\noindent\label{erho-Ge-2}Table S10.2.2: {\small{\textit{Tabulated content of \hyperlink{s7}{Figure~S7.1} (SM) for Ge thin films. ``KS'' denotes total Kohn-Sham energy as a direct output of VASP, \emph{with} dispersion vdW correction (PBE-D3, \texttt{IVDW=12}). For~unit conversion: thickness of $1\:\mathrm{ML}$ of~(111)-diamond cF8 Ge is $\rho_{0} = 0.141357\:\mathrm{at./\AA^{2}}$ and the respective bulk Kohn-Sham per-atom energy $E_{\mathrm{bulk}} = -4.7696\:\mathrm{eV/at.}$ $E_{\mathrm{g}}$ stands for the width of electronic band gap.}}} \\

\bigskip

\hypertarget{s10.3}{}
\noindent {\large{\textbf{S10.3 $\bm{E(\rho)}$ of Si}}} \\
\label{sec:e-rho-Si}

\begin{table}[h!]
	\centering
	\begin{tabular}{c|c|c|c|c|l}
		$\rho$ &  $\rho$ & $E_{\mathrm{KS}}$ & $E - E_{\mathrm{bulk}}$ & $E_{\mathrm{g}}$ & sample \\
		$[\mathrm{at./\AA^{2}}]$ & $[\mathrm{ML}]$ & $[\mathrm{eV/at.}]$ & $[\mathrm{meV/at.}]$ & $[\mathrm{eV}]$ & designation \\
		\hline
		\hline
		0.2110 &    1.364     &  -4.9906   &   430    &  0.650  & db-monolayer (oP10)                      \\
		\hline       
	    0.2709 &    1.751     &  -4.9566   &   464    &  0.000  & GS-bilayer                               \\
	    \hline
	    0.3851 &    2.489     &  -5.0987   &   322    &  0.467  & Si.iso-II-2.5                            \\  
	    \hline       
	    0.5032 &    3.253     &  -5.1399   &   280    &  0.950  & Si.equi-II-3                             \\  
	    \hline       
	    0.6056 &    3.915     &  -5.1885   &   232    &  0.837  & Si.iso-II-4                              \\   
	    \hline       
	    0.8272 &    5.347     &  -5.2338   &   186    &  0.762  & Si.iso-II-5                              \\
	    \hline       
	    0.3866 &    2.499     &  -5.0643   &   356    &  0.138  & ($\sqrt{3}\times\sqrt{3}$)-dia (bilayer) \\
	    \hline       
		0.3690 &    2.385     &  -5.0080   &   412    &  0.000  & ($2\times2$)-dia (bilayer)               \\  
	    \hline        
		0.5263 &    3.402     &  -5.1245   &   296    &  0.450  & ($2\times2$)-dia (trilayer)              \\ 
        \hline        
        0.6788 &    4.388     &  -5.1834   &   237    &  0.300  & ($2\times2$)-dia (fourlayer)             \\ 
        \hline         
        0.8339 &    5.391     &  -5.2279   &   192    &  0.465  & ($2\times2$)-dia (fivelayer)             \\
        \hline        
        0.9874 &    6.383     &  -5.2561   &   164    &  0.436  & ($2\times2$)-dia (sixlayer)              \\ 
        \hline        
        1.1423 &    7.384     &  -5.2792   &   141    &  0.450  & ($2\times2$)-dia (sevenlayer)            \\ 
        \hline         
        1.2969 &    8.384     &  -5.2961   &   124    &  0.473  & ($2\times2$)-dia (eightlayer)            \\
        \hline         
        1.4515 &    9.383     &  -5.3094   &   111    &  0.480  & ($2\times2$)-dia (ninelayer)             \\   
\end{tabular}
\end{table}

\vspace{-0.35cm}
\noindent\label{erho-Si-1}Table S10.3.1: {\small{\textit{Tabulated content of Figure~6 (main text) for Si thin films. ``KS'' denotes total Kohn-Sham energy as a direct output of VASP (without dispersion vdW correction). For~unit conversion: thickness of $1\:\mathrm{ML}$ of~(111)-diamond cF8 Si is $\rho_{0} = 0.15469\:\mathrm{at./\AA^{2}}$ and the respective bulk Kohn-Sham per-atom energy $E_{\mathrm{bulk}} = -5.4202\:\mathrm{eV/at.}$ $E_{\mathrm{g}}$ stands for the width of electronic band gap.}}}

\medskip

\begin{table}[h!]
	\centering
	\begin{tabular}{c|c|c|c|c|l}
		$\rho$ &  $\rho$ & $E_{\mathrm{KS}}$ & $E - E_{\mathrm{bulk}}$ & $E_{\mathrm{g}}$ & sample \\
		$[\mathrm{at./\AA^{2}}]$ &  $[\mathrm{ML}]$ & $[\mathrm{eV/at.}]$ & $[\mathrm{meV/at.}]$ & $[\mathrm{eV}]$ & designation \\
		\hline
		\hline
		0.2127 &    1.351     &  -5.1751   &   550    &  0.508  &  db-monolayer (oP10)                      \\
		\hline
		0.3903 &    2.479     &  -5.3273   &   398    &  0.371  &  Si.iso-II-2.5                            \\  
		\hline       
		0.6131 &    3.894     &  -5.4256   &   300    &  0.787  &  Si.iso-II-4                              \\  
		\hline       
		0.8376 &    5.320     &  -5.4762   &   249    &  0.731  &  Si.iso-II-5.5                            \\   
		\hline       
		0.5105 &    3.242     &  -5.3773   &   348    &  0.950  &  Si.equi-II-3                             \\
	    \hline       
		0.5381 &    3.417     &  -5.3957   &   329    &  0.847  &  Si.equi-II-3.5                           \\
		\hline       
		0.3919 &    2.488     &  -5.3044   &   421    &  0.000  &  ($\sqrt{3}\times\sqrt{3}$)-dia (bilayer) \\
		\hline       
		0.5356 &    3.402     &  -5.3874   &   338    &  0.420  &  $(2\times2)$-dia (trilayer)              \\  
		\hline	   
		0.8462 &    5.374     &  -5.5051   &   220    &  0.334  &  $c(2\times4)$-dia (fivelayer)            \\  
		\hline
	 	0.4755 &    3.020     &  -5.3460   &   379    &  0.000  &  ($3\times3$)-DAS (trilayer)              \\ 
        \hline         
        0.6305 &    4.004     &  -5.4274   &   298    &  0.000  &  ($3\times3$)-DAS (fourlayer)             \\
        \hline
        0.4435 &    2.817     &  -5.2692   &   456    &  0.034  &  a-Si (trilayer)                          \\ 
        \hline         
        0.6143 &    3.901     &  -5.3652   &   360    &  0.070  &  a-Si (fourlayer, cF8 nuclei)             \\
	\end{tabular}
\end{table}

\vspace{-0.0cm}
\noindent\label{erho-Si-2}Table S10.3.2: {\small{\textit{Tabulated content of \hyperlink{s7}{Figure~S7.1} (SM) for Si thin films. ``KS'' denotes total Kohn-Sham energy as a direct output of VASP, \emph{with} dispersion vdW correction (PBE-D3, \texttt{IVDW=12}). For~unit conversion: thickness of $1\:\mathrm{ML}$ of~(111)-diamond cF8 Si is $\rho_{0} = 0.157457\:\mathrm{at./\AA^{2}}$ and the respective bulk Kohn-Sham per-atom energy $E_{\mathrm{bulk}} = -5.7251\:\mathrm{eV/at.}$ $E_{\mathrm{g}}$ stands for the width of electronic band gap.}}}


\newpage
\thispagestyle{empty}
\def\thispagestyle#1{}

\renewcommand\bibpreamble{\vspace{-1.0\baselineskip}} 

\titleformat{\chapter}[hang]{\flushleft\fontseries{b}\fontsize{80}{100}\selectfont}{\fontseries{b}\fontsize{100}{130}\selectfont \textcolor{white}\thechapter\hsp}{0pt}{${}$ \\\Huge\bfseries}[]

\clearpage
\phantomsection
\bibliographystyle{unsrt}
\phantomsection
\addcontentsline{toc}{chapter}{Bibliography}
